\documentclass[useAMS,usenatbib]{mn2e}
\usepackage{graphicx, rotating, color}

\usepackage{aas_macros}  
\usepackage{graphicx}
\usepackage{amssymb}
\usepackage{rotating}
\usepackage{natbib}
\usepackage{url}
\usepackage{multirow}
\usepackage{fmtcount}
\usepackage[colorlinks=true,citecolor=black,linkcolor=black,urlcolor=blue,bookmarks=true]{hyperref}
\usepackage{microtype}

\newcommand{\kms}{km\,s$^{-1}$}
\newcommand{\reff}{r$_\mathrm{eff}$}
\newcommand{\aeff}{a$_\mathrm{eff}$}

\newcommand{\ulyss}{ULySS}

\begin{document}
\title[Population gradients in early-types galaxies]{
Age and metallicity gradients in early-type galaxies: A dwarf to giant sequence.
} 
\author[Mina Koleva et
  al.]{Mina Koleva$^{1,2,3}$\thanks{E-mail:
    koleva@iac.es},
  Philippe Prugniel$^{3}$, Sven De Rijcke$^{4}$, Werner~W. Zeilinger$^{5}$,\\
  $^{1}$Instituto   de
      Astrof\'{\i}sica  de Canarias,  La Laguna,  E-38200  Tenerife, Spain\\
  $^{2}$Departamento de  Astrof\'{\i}sica, Universidad  de  La Laguna,
      E-38205  La  Laguna,   Tenerife,  Spain \\
  $^{3}$Universit\'e Lyon~1,
  Villeurbanne, F-69622, France; CRAL, Observatoire de Lyon, St. Genis
  Laval, F-69561, France ; CNRS, UMR 5574 \\ 
  $^{4}$Dept. of Physics \& Astronomy, Ghent University, Krijgslaan 281, S9, B-9000 Ghent,
  Belgium\\
  $^{5}$Institut
  f\"{u}r Astronomie, Universit\"{a}t Wien, T\"{u}rkenschanzstrasse
  17, A-1180 Wien, Austria 
  }
\date{Accepted 2011 XXX XX.  Received 2010 XXX XX; in original form 2010 XXX XX}

\pagerange{\pageref{firstpage}--\pageref{lastpage}} \pubyear{XXXX}

\maketitle 

\label{firstpage}

\begin{abstract}
We studied the stellar populations of 40 early-type galaxies using
medium resolution long-slit spectroscopy along their major axes (and
along the minor axis for two of them). The sample, including
elliptical and lenticular galaxies as well as dwarf galaxies, is
combined with other previously published data in order to discuss the
systematics of the radial gradients of age and metallicity over a
large mass range, from $10^{7} M_\odot$ to $10^{12} M_\odot$ ($-9.2 >
M_B > -22.4$ mag). The well-known mass-metallicity relation is
continuous throughout the whole mass range, in the sense that more
massive galaxies are more metal-rich. The age-mass relation is
consistent with the idea of downsizing:~smaller galaxies have more
extended star-formation histories than more massive ones.  The
transition type dwarfs (intermediate between dwarf irregular and
dwarf elliptical galaxies) deviate from this relation having younger
mean age, and the low-mass dwarf spheroidals have older ages, marking
a discontinuity in the relation, possibly due to selection effects.  

In all mass regimes, the mean metallicity gradients are approximately
$-0.2$ and the mean age gradients $+0.1$~dex per decade of radius. The
individual gradients are widely spread: $ -0.1 < \nabla_{\rm Age} <
0.4 $ and $-0.54 < \nabla_{[{\rm Fe/H}]} < +0.2 $. We do not find
evidence for a correlation between the metallicity gradient and
luminosity, velocity dispersion, central age or age
gradient. Likewise, we do not find a correlation between the age
gradient and any other parameter in bright early-type galaxies. In
faint early-types with $M_B \gtrsim -17$~mag, on the other hand, we
find a strong correlation between the age gradient and luminosity:~the
age gradient becomes more positive for fainter galaxies. Together with
the observed downsizing phenomenon this indicates that, as time
passes, star formation persists in dwarf galaxies and becomes more
centrally concentrated. However, this prolonged central star formation
is not reflected in the metallicity profiles of the dwarfs in our
sample. 


We conclude that various physical mechanisms can lead to
similar gradients and that these gradients are robust against
the environmental effects.
  In particular, the gradients observed in
dwarfs galaxies certainly survived the transformation of the
progenitors through tidal harassment or/and ram-pressure
stripping. The diversity of metallicity gradients amongst dwarf
elliptical galaxies may  reflect a plurality of progenitors'
morphologies. The dwarfs with steep metallicity gradients could have
originated from blue compact dwarfs and those with flat profiles from
dwarf irregulars and late type spirals.
\end{abstract}


\section{Introduction}

The internal inhomogeneities of galaxies offer a valuable insight into
their chemical and dynamical evolution, and their star formation
histories. The radial age and metallicity gradients in a galaxy are
the result of its star formation and metal enrichment history. The
stellar population may have originated from pristine gas and continued
forming from self-enriched gas flowing to the galaxy centre, creating
negative gradients in the mean metallicity and positive gradients in
the mean age \citep{larson1974,pipino2008}.  To persist and be
observed, the gradients must survive the galaxy's evolution. The
ageing of the stellar population erases the observational signatures
of age gradients and fades the mean metallicity gradients. Orbital
mixing and merger events will also alter them
\citep{white1980,kobayashi2004,dimatteo2009}.
Thus, the gradients are likely related to other properties of the
galaxies and should reflect the diversity of the formation and
evolution processes. We can a priori expect the mass to play the first
role, and we may expect different trends in different mass regimes, in
fact reflecting different formation mechanisms.

The monolithic collapse models of \citet{pipino2010} predict gradients
in the range $-0.5 < \Delta[{\rm Fe/H}]/\Delta\log(r) \equiv
\nabla_{[{\rm Fe/H}]} < -0.2$\,dex per decade of radius (computed
within the effective radius, \reff). Those gradients appear to be
  independent of the galaxies' masses (their fig.4). The dispersion
results from initial conditions and reflects the competition between
feedback and star formation. As an alternative mechanism, 
smoothed-particle hydrodynamics (SPH) simulations
\citep{bekki1999,hopkins2009} suggest that the dispersion of the
metallicity gradients (between $-0.6$ and $0$) is due to the diversity
of (wet) mergers. Mergers may destroy the gradients but can also
induce star formation to re-generate them. These gradients do not
depend on the progenitor's mass.

A hybrid formation scenario, involving collapse, merging histories
\citep{ogando2005,kormendy1989} and ram-pressure quenching of the star
formation (for the smallest galaxies) may possibly account for the
diversity of histories and characteristics of early-type galaxies.

Gradients of age and metallicity have been observed in normal
elliptical and S0 galaxies ($M > 10^{10} M_\odot$).  They are
determined comparing spectra or colours to models of single stellar
populations (SSPs). These SSP-equivalent parameters are dominated by
the most-luminous stars and hence by the youngest star formation
event. They are closer to luminosity-weighted than to mass-weighted
characteristics, but still they should not be confused with the
former. Colour gradients were observed in E galaxies back in the
1960's \cite[e.~g.][]{devauc1961} and were generally interpreted as
metallicity gradients.  Similar conclusions were reached by studying
the radial distribution of spectral line strengths
\citep{mcclure1969,carollo1993}.  In the recent years, more elaborated
approaches were used to determine simultaneously the age and the
metallicty. \citet{sanchez-blazquez2007} conclude that the
line-strength gradients are essentially due to metallicity and that
the age of the populations is generally homogeneous. Analogous results
are also reached in \citet{mehlert2003} and \citet{rawle2010}:~the
metallicity gradients are spread around $\nabla_{[{\rm
      Fe/H}]}\approx-0.2$ ($-0.5 < \nabla_{[{\rm Fe/H}]} < 0.2$).
\citet{sanchez-blazquez2006} find no, or at most a marginal,
correlation between the metallicity gradients and the mass of the
galaxies or their large-scale environment.  However,
\citet{forbes2005} support a correlation between the metallicity
gradient and the mass, in the sense that more massive galaxies have
steeper negative gradients. \citet{annibali2007} studied intermediate
and massive early-type galaxies (120\,\kms$ < \sigma_{\rm central} <
$ 300\,\kms) and find negative metallicity gradients ($-0.2$), which
decrease with the increase of the mass.

Until recently,
less information was available concerning the low-mass,  early-type
galaxies (M$_B > -18$~mag; $L < 10^{9} L_\odot$). Most of them are dwarf, 
diffuse elliptical galaxies (dEs),  with smooth elliptical
isophotes, exponential surface brightness profiles and in general
quiescent.  They are predominantly observed in galaxy clusters or as
satellites of massive galaxies. They  differ from ``normal'' ellipticals
(E) by their lower concentration \citep{kor1985,kormendy2009}, although
a small population of objects intermediate between dEs and Es exists
\citep[e.g.][]{prugniel1994}.

Some recent studies present evidences for a diversity of age and
metallicity gradients in dEs \citep{chi2009,paperII}.  On average,
negative metallicity gradients, $\nabla_{[{\rm Fe/H}]} = -0.26$ and
positive ages gradients $\Delta\log({\rm Age})/\Delta\log(r) \equiv
\nabla_{{\rm Age}} = +0.1$ were found, but with large
dispersions. They are in the range $-0.5 < \nabla_{[{\rm Fe/H}]} <
0.4$ and $-0.1 < \nabla_{{\rm Age}} < 0.3$.  About 1/3 of the galaxies
have flat profiles that may be connected with rotation or with the
presence of a disc \citep{paperII}, but otherwise, no significant
correlation with any galaxy property was found.

Descending through the mass sequence of the early type galaxies
($M \leq 10^{8-9} M_\odot$), 
a diversity of gradients was found. \citet{makarova2010} found no gradient in
two dwarf spheroidals (dSphs) of the M~81 group, KDG~61 and 64 ($\mathrm{M_B} \approx
-13$~mag).  
\citet{lianou2010} confirmed the homogeneity of these two galaxies, but
found negative photometric metallicity gradients ($\nabla_{[{\rm Fe/H}]}
\approx -0.1 \sim -0.2$ within \reff{}) in two other dwarfs of this
group, out of the eight they studied.  In Centaurus~A group dE/dSph
($-10 < \mathrm{M_B} < -14$), \citet{crnojevic2010} detected weak
metallicity gradients, typically of the order of $\nabla_{[{\rm
      Fe/H}]} \sim -0.07$.


\citet{harbeck2001} searched for population gradients at even 
lower masses ($M < 10^8$) in nine 
Local Group dSphs through horizontal-branch morphologies and
distributions of red giant branch and red clump stars.  Negative
metallicity gradients are suggested in Sextans, Sculptor, Tucana, and
Andromeda~VI, and possibly a positive age gradient in the Carina
dSph. No population gradient is found in Andromeda~I, II, III, and
V.  Positive age gradients were found in the Fornax
\citep{stetson1998} and Sextans \citep{okamoto2008} dSphs and in the
transition-type galaxy Phoenix \citep{martinez-delgado1999}.  Further
spectroscopic studies of individual stars confirmed the metallicity
gradient in Sextans \citep{kleyna2004,battaglia2010} and Sculptor
\citep{tolstoy2004,kirby2009} and revealed gradients in Fornax
\citep{battaglia2006} and probably in Leo~I \citep{gullieuszik2009}.
A negative metallicity gradient is also detected in the disrupted Sagittarius
dwarf \citep{kel2010}.  A general feature is that whenever population
gradients are found in dSphs, they are negative in metallicity and
positive in age. In the three galaxies with a large number of
spectroscopic measurements, Fornax, Scultptor and Sextans, the
metallicity gradients are $\nabla_{\rm [Fe/H]} \approx -0.3$.

The star forming Sm and dIrr galaxies may be the progenitors 
of the quiescent dwarfs. However, the
metallicity of their interstellar medium, estimated from H{\sc ii}
regions, appears homogeneous \citep{kobu1997,vanzee2006}, but the
metallicity of their old stellar populations is generally not
constrained by the present observations.  The stellar population of
the LMC shows only a shallow negative metallicity gradient, and the
SMC seems to be chemically homogeneous \citep{cioni2009,feast2010}.

All these studies suggest that the metallicity gradients in early-type
galaxies are spread around a negative value, without any strong
correlation with the mass of the system.  However, a recent study by
\citet{spo2009}, including one dE/dS0, presents a tight relation between
the metallicity gradient and the central velocity dispersion or mass,
leading from strong gradients in normal E/S0 to flat profiles in dEs.

In this paper, we address questions related to stellar population
gradients in medium to low mass early-type galaxies. Our aim is to
describe the systematics of the gradients in early-type systems.  We
will study various sample of low-mass galaxies, including those of
\citet{paperII} and \citet{spo2009}.  We analyse with the same method
data from different instruments and we compare the results obtained by
different authors and using different approaches.

The paper is organised as follow. In Sect.\,\ref{sec:obsred}, we
introduce the samples and describe the data processing.  In
Sect.\,\ref{sec:radpop} we analyse the data and derive the gradients,
and in Sect.\,\ref{sec:comp} we will compare our results with previous
studies. In Sect.\,\ref{sec:disc} are our discussions, and
Sect.\,\ref{sec:conc} draws the conclusions.

\section{Observations and data reduction} \label{sec:obsred}

To investigate the systematics of the population gradients in low-mass, 
early-type 
galaxies, we gathered a sample of galaxies with good quality
spectroscopic data available in the telescope archives.
The main sample was presented in \citet{paperI} and was focused on dE
galaxies. We also included the sample of \cite{spo2009}.
To verify the robustness of
the analysis, we used the observations of S0 galaxies used in
\citet{bedregal2006} and those of E galaxies presented in
\citet[hereafter VAKU]{vaku}, containing several objects in common
with the previous samples.  Finally, we include NGC~205 \citep{sp02}
and NGC~404 \citep{bouchard2010} which are two nearby galaxies that have
been studied in detail.

For all these galaxies, we use spectroscopic long-slit data acquired
along the major axis, and for NGC~205 and 404 also along the minor
axis. To refine the the morphological classification of the galaxies
we use photometric studies from the literature (the corresponding references
can be found in Table~\ref{table:char} and  Appendix~\ref{appendix:indgal}). 
For the \citeauthor{spo2009}
sample, we gathered images and did our proper analyses 
(see Appendix~\ref{appendix:photometry}).

Though all the galaxies can be broadly qualified as ``early-type'',
the full sample contains a mix of morphologies. We re-examined these
objects and separated them in four groups: (1) ``normal'' elliptical
galaxies ($M_B < -18$ mag), (2) S0 ($M_B < -18$ mag), (3) dE and dS0
and (4) transition type dwarf galaxies (TTDs).  The distinction
between the former two classes and the latter two is made on the basis
of luminosity and concentration. To separate E and S0 galaxies we examined the
isophotes shape, the ${\rm V_{max}}/\sigma_0$ ratios (where 
${\rm V_{max}}$ is the maximum projected velocity of rotation and $\sigma_0$
the central velocity dispersion) and the signatures of discs or dust
lanes on the images.  Finally, to distinguish dEs/dS0s from TTDs we
considered their star-formation histories (a galaxy with constant 
star formation rate, until present will be classified as a TTD) and 
the H{\sc i} or H$_\alpha$ detections.  

Even if NGC\,205 is detected in H{\sc i} and H$_\alpha$ we consider it
as a dE/dS0 because its H{\sc i} is only 5.3$\times$10$^5$M$_\odot$
\citep{young1997}, while the detection limit at the distance of Virgo
is 3$\times$10$^6$M$_\odot$ \citep{conselice2003}\footnote{ Note, that
  with a detection limit of 2$\times 10^7$M$_\odot$ the ALFALFA survey
  \citep{alfalfa} is less deep than the observations used in
  \citet{conselice2003}.}.  The spectrum extracted in a 100\,pc
aperture does not show prominent emission in the Balmer lines (though
these emission lines are clearly visible after subtracting the stellar
component).  We associated the compact elliptical galaxy (cE),
NGC~4486B = VCC~1297, to the E group, as cEs are more related to Es
than to dEs \citep{kor1985,np87}.

The characteristics of the 40 galaxies are presented in Table~\ref{table:char}.
Below, we give details about the data and processing.

\begin{table*}
  \centering
  \begin{minipage}{180mm}
    \caption{Structural properties of our sample. The columns are as follows:
      name of the galaxy in Virgo and Fornax catalogues, other name, type, 
      apparent magnitude in $B$, absolute magnitude in $B$, central velocity 
      dispersion in \kms, ellipticity, effective radius in arcsec, surface 
      brightness at the effective radius in $B$ , S\'ersic index, 
      samples: 1 - \citeauthor{paperI}, 2 - \citeauthor{spo2010}, 
      3 - \citeauthor{bedregal2006}, 4 - VAKU, 5 - NGC\,205, 6-NGC\,404, 
      comments. The measurements are ours, except where indicated. 
      The \citeauthor{paperI} galaxies properties are taken from their paper.}
    \begin{tabular}{|l|l|c|c|c|c|c|c|c|l|l|l} 
      \hline
      F/VCC   & name & type & $m_B$\footnote{values for sample 4 (without repetitive 
        observations) are taken from HyperLeda}  & $M_B$\footnote{computed using 
        the following distance modulus: Virgo - 30.8 , Fornax - 31.6, NGC\,5898 group - 32.6, 
        NGC\,5044 - 32.87, NGC\,3258 - 33.00, NGC205 - 24.5, for NGC404 we adopt distance of
        3.4\,Mpc \citep{tonry2001}. All measurements where corrected for galactic 
        extinction and reddening.} &  $\sigma$ & $\epsilon$\footnote{$\epsilon = 1-b/a$, 
        with b\slash a being the minor-to-major axis ratio of the isophotes. For 
        samples 3,4,5,6 we list values from HyperLeda} & $R_{\rm e}$ &  $\mu_{{\rm e},B}$ &
      S\'ersic \footnote{values for sample 4 were taken from \citet{kormendy2009}, 
        for sample 6 - from \citet{seth2010}} & s &comment \\  
      &      &      & (mag) & (mag) & ({\kms}) &   & (arcsec) & (mag arcsec$^{-2}$) & $n$\\
      \hline
      \multicolumn{12}{c}{E}\\
      \hline
      FCC277 &NGC1428    &E-S0 &13.62&-17.7& 82&0.50& 13.7&20.55&1.93&2&-\\
VCC0575&NGC4318    &E    &14.07&-17.6& 95&0.32&  7.8&20.10&1.66&2&disky\\
VCC0731&NGC4365    &E    &10.39&-20.4&256&0.11& 35.1&22.11&7.11&4&boxy, triaxial, KDC\\
VCC0828&NGC4387    &E    &12.97&-18.3&115&0.41& 26.0&20.17&2.33&2,4&boxy\\
VCC1025&NGC4434    &E    &12.96&-18.8&120&0.04& 10.6&20.05&2.26&2&elliptical isophotes\\
VCC1146&NGC4458    &E    &13.07&-18.2&103&0.07& 14.8&20.84&2.16&2&elliptical isophotes, KDC\\
VCC1178&NGC4464    &E    &13.45&-17.4&129&0.25&  6.0&19.55&2.31&2,4&disky\\
VCC1231&NGC4473    &E    &10.91&-19.9&179&0.41& 12.5&20.39&4.00&4&KDC\\
VCC1279&NGC4478    &E    &12.16&-19.0&144&0.19& 17.0&19.82&2.07&4&-\\
VCC1297&NGC4486B   &cE   &14.22&-16.8&170&0.03&  2.7&18.38&2.63&2&disky\\
VCC1475&NGC4515    &E-S0 &13.42&-17.8& 86&0.21& 10.5&20.27&2.71&2&disky innerparts\\
VCC1630&NGC4551    &E    &12.87&-18.3&107&0.26& 15.0&20.43&1.78&2&slightly boxy\\
VCC1903&NGC4621    &E    &10.52&-20.3&225&0.26& 27.7&21.72&5.36&4&KDC\\
\hline
\multicolumn{12}{c}{S0}\\
\hline
FCC021 &NGC1316    &S0   & 9.29&-22.3&226&0.42& 46.3&21.61&2.90&3&dust, merger?\\
FCC055 &ESO358-006 &S0   &13.90&-17.5& 47&0.48&  5.3&21.50&3.00&3&-\\
FCC148 &NGC1375    &S0   &13.05&-18.2& 69&0.56&  9.1&20.59&1.69&2,3&peanut bulge\\
FCC153 &IC0335     &S0   &12.84&-18.5& 79&0.69&  5.8&20.63&1.83&2,3&thin edge-on disk\\
FCC167 &NGC1380    &S0   &10.80&-20.8&221&0.51& 15.9&20.79&3.30&3&-\\
FCC170 &NGC1381    &S0   &12.02&-19.0&150&0.62&  4.3&19.18&3.10&2,3&big bulge, thin disk\\
FCC177 &NGC1380A   &S0   &13.21&-18.1& 67&0.72&  8.7&21.92&3.70&3&-\\
FCC301 &ESO358-059 &E-S0 &13.95&-17.4& 49&0.17&  3.4&20.60&4.36&2,3&-\\
\hline
\multicolumn{12}{c}{dE/dS0}\\
\hline
       &[FS90]029  &dE5   &15.70&-17.5& 60&0.54&  8.9&22.44&1.79&1&-\\
       &[FS90]075  &dE1,N &16.87&-16.3& 49&0.10&  6.8&23.03&1.74&1&-\\
       &[FS90]076  &dE1   &16.10&-17.1& 57&0.07&  4.4&21.41&2.02&1&KDC\\
       &[FS90]131  &dE5,N &15.30&-17.9& 87&0.54&  8.1&21.83&2.35&1&Peanut\\
       &NGC5898 DW1&dE3   &15.66&-17.6& 44&0.34&  8.7&22.35&1.53&1&-\\
FCC043 &IC1919     &dE3   &13.91&-17.8& 56&0.26& 16.9&22.05&2.17&1&-\\
FCC136 &           &dE2   &14.81&-16.9& 64&0.21& 14.2&22.57&1.71&1&-\\
FCC150 &           &dE4,N &15.70&-16.0& 64&0.19&  5.7&21.48&1.65&1&-\\
FCC204 &ESO358-043 &dS0   &14.76&-17.0& 67&0.61& 11.5&22.06&1.29&1&Spiral\\
FCC245 &           &dE0,N &16.00&-15.7& 40&0.11& 11.4&23.28&1.35&1&-\\
FCC266 &           &dE0,N &15.90&-15.8& 42&0.11&  7.1&22.15&1.08&1&-\\
FCC288 &ESO358-056 &dS0   &15.10&-16.6& 49&0.72&  9.5&21.99&1.11&1&Spiral\\
FCC335 &ESO359-002 &dS0,N &14.50&-17.1& 44&0.34& 15.4&21.99&1.55&2,3&-\\
       &NGC205     &dE5   & 8.61&-16.0& 20&0.39&147.9&21.82&    &5&H{\sc i}, H$\alpha$ emission\\
\hline
\multicolumn{12}{c}{TTD}\\
\hline
       &NGC404     &dS0   &11.19&-16.5& 30&0.00& 42.0&21.05&2.43&6&H{\sc i}, H$\alpha$ emission, LINER\\
       &[FS90]373  &dE3   &15.60&-17.8& 73&0.23&  7.9&22.03&2.71&1&KDC\\
       &NGC5898 DW2&dS0,N &16.10&-17.2& 44&0.57&  5.9&21.95&1.26&1&H$\alpha$ emission\\
FCC046 &           &dE4,N &15.99&-15.7& 61&0.36&  6.7&22.12&1.24&1&H{\sc i}, H$\alpha$ emission\\
FCC207 &           &dE2,N &16.19&-15.5& 61&0.15&  8.4&22.81&1.32&1&H$\alpha$ emission\\

      \hline
    \end{tabular}
  \end{minipage}
  \label{table:char}
\end{table*}

\subsection{\citeauthor{paperI} sample}\label{subsec:derijcke}

This sample, of 16 dE/dS0 galaxies, was observed with the FORS1 and FORS2
spectrographs attached to the VLT. These observations were used in 
\citet{ddzh01,deRijcke03a,d03,d04,mdzpdr04,paperI,michielsen08,paperII} 
to study the kinematics and stellar population of dE galaxies.
For the present study, we use the reduced spectroscopic data 
described in \citet{paperII}.

The FORS1 data have a resolution $\Delta\lambda = 2.6$\,\AA\  
and cover the wavelength range 3300--6200\,\AA.
The FORS2 data have a resolution of $\Delta\lambda = 3.0$\,\AA\ 
and cover the wavelength range 4355--5640\,\AA.
All the data have a central signal-to-noise values (S/N) of typically
30 per wavelength element.

To analyse the data, and in particular the kinematics, the
instrumental broadening of the observations with respect to the model
needs to be accurately determined.  This is characterised by the
line-spread function (LSF) and depends on the wavelength.  The LSF is
determined from twilight and stellar template spectra using the full
spectrum fitting package \ulyss{} \citep{ulyss}. It compares
observations to models having a finite resolution. Thus, it measures
the excess of broadening of the former with respect to the latter;
this is why we speak of the {\it relative} LSF.

We adopted LSFs with mean broadening of $\sigma_{\rm LSF} = 63$ and
$73$\,\kms, for the FORS1 and FORS2 observations, respectively.  Note,
that due to the relatively wide slit of 1\,arcsec, the FORS2 have a
lower spectral resolution and the centres of the galaxies are
marginally spatially resolved within the slit (i. e., the total
broadening is the combination of the intrinsic resolution of the
spectrograph, the width of the slit, the spatial profile and the
internal kinematics of the galaxies). Since, the luminosity profiles
are known, it is in principle possible to take them into account and
recover reliable kinematics. However, we do not focus on the 
galaxies dynamics and we will not attempt these complicated corrections.
For these reasons, the velocity dispersion profiles must be
interpreted with care, as well as the velocities in the central
arcsec.  The internal kinematics of these galaxies was separately
determined using higher resolution observations presented in
\citet{paperI}.

\subsection{Spolaor et al. sample} \label{subsec:spolaor}

The \citeauthor{spo2009} sample consists of 14 galaxies, 6 in the
Fornax cluster (FCC\,148, FCC\,153, FCC\,170, FCC\,277, FCC\,301,
FCC\,335) and 8 in the Virgo cluster (VCC\,575, VCC\,828, VCC\,1025,
VCC\,1146, VCC\,1178, VCC\,1297, VCC\,1475, VCC\,1630). Most of the
Fornax cluster galaxies are S0s. Two of them, FCC\,153 and FCC\,170,
are seen edge-on and display a prominent disc component. Another
galaxy, VCC1327 = NGC~4486A was also observed in the Virgo program.
This low-luminosity E galaxy was set apart from the present sample
because its spectra, highly contaminated by the light of a 12$^{th}$
magnitude foreground star, located 2~arcsec from the nucleus, deserved
a special treatment. Its kinematical and population analysis is
presented in \cite{prugniel2011}.  The surface photometry obtained
from  image taken in the GMOS archive is presented in
Appendix~\ref{appendix:photometry}.

The GMOS long-slit spectra of the Fornax and Virgo sample were
extracted from the Gemini data archive.  The Virgo and Fornax data
sets have the same characteristics with exception of the spectral
resolution as a result of different slit widths (0.5\,arcsec and 1.0\,arcsec
respectively). The details of the two setups and the observational log
are given in Table\,\ref{table:setup},\,\ref{table:obslog}, respectively. At
least two spectra were obtained for each target galaxy. All the
spectra were observed by aligning the slit with the optical major axis
of the galaxy. The long-slit spectra cover the full extent of the 3
detectors including two small gaps. In order to cover these gaps
additional spectra with a shift of the central wavelength by 50 \AA\ 
were obtained in the case of FCC\,277 and the whole Virgo sample. Since
the S/N of the individual data sets was found to be sufficient for the
present investigation, we preferred to keep the data sets separate
instead of stacking them. We used the second, slightly shifted data
set for an independent analysis in order to check the reproducibility
of the results.

%

\begin{table}
\begin{minipage}{\columnwidth}
  \caption{\label{table:setup}
Setup of observations }
  \begin{tabular}{lcc}
    \hline
    \multicolumn{2}{c}{GMOS $@$ Gemini South -- Fornax cluster} \\
\hline
CCDs, EEV\#             & 2037-06-03 / 8194-19-04 / 8261-07-04 \\
\# of pixels            & 3 $\times$ 2048$\times$4068 chips        \\
pixel size [$\mu$m$^2$] & 13.5$\times$13.5  \\
image scale [arcsec pix$^{-1}$]  & 0.0727 \\
readout noise [e$^-$ pix$^{-1}$] & 3.98 / 3.85 / 3.16     \\
gain [ADU (e$^-$)$^{-1}$]        & 2.372 / 2.076 / 2.097  \\
grism                      & B600\_G5323  \\
slit width [arcsec]        & 1.0         \\
spectral range [\AA]       & 3690 -- 6450 \\
FWHM $\delta\lambda$ [\AA] & 4.7         \\
$\sigma_{instr}$ [km s$^{-1}$] & 114     \\
dispersion [\AA\,pix$^{-1}$]   & 0.46    \\
\hline
\hline
\multicolumn{2}{c}{GMOS $@$ Gemini South -  Virgo cluster}\\
\hline
CCDs, EEV\#             & 2037-06-03 / 8194-19-04 / 8261-07-04 \\
\# of pixels            & 3 $\times$ 2048$\times$4068 chips        \\
pixel size [$\mu$m$^2$] & 13.5$\times$13.5  \\
image scale [arcsec pix$^{-1}$]  & 0.0727 \\
readout noise [e$^-$ pix$^{-1}$] & 3.20 / 3.50 / 3.10    \\
gain [ADU (e$^-$)$^{-1}$]        & 2.000 / 1.900 / 1.900 \\
grism                      & B600\_G5323  \\
slit width [arcsec]        & 0.5         \\
spectral range [\AA]       & 3640 -- 6500 \\
FWHM $\delta\lambda$ [\AA]     & 2.5     \\  
$\sigma_{instr}$ [km s$^{-1}$] & 61      \\
dispersion [\AA\,pix$^{-1}$]   & 0.46    \\
\hline
\hline
\multicolumn{2}{c}{FORS2 $@$ VLT -  Fornax cluster}\\
\hline
CCDs, MIT/LL mosaic\#   & CCID20-14-5-3 \\
\# of pixels            & 2 $\times$ 4096$\times$2068 chips \\
pixel size [$\mu$m$^2$] & 15 \\
image scale [arcsec pix$^{-1}$]  & 0.126      \\
readout noise [e$^-$ pix$^{-1}$] & 2.7 / 3.60 \\
gain [ADU (e$^-$)$^{-1}$]        & 0.8 / 0.8  \\
grism                      & GRIS\_1400V  \\
slit width [arcsec]        & 0.51        \\
spectral range [\AA]       & 4560 -- 5900 \\
FWHM $\delta\lambda$ [\AA]     & 1.1   \\
$\sigma_{instr}$ [km s$^{-1}$] & 19.9    \\
dispersion [\AA\,pix$^{-1}$]   & 0.32  \\
\hline
\hline
    \multicolumn{2}{c}{ISIS @ 4.2 WHT -- VAKU sample} \\
\hline
\# of pixels                             & 3471 -- 5501\\      
image scale (arcsec pix$^{-1}$ )          &  0.439 \\
slit width (arcsec)                      & 1.6  \\         
spectral range (\AA)                     & 3978 -- 5501 \\      
FWHM $\Delta\lambda$ (\AA)               & 2.4 \\     
$\sigma_{\rm instr}$ (\kms)                & 58.8 \\             
\hline
\hline
    \multicolumn{2}{c}{CARELEC @ 1.93 OHP -- NGC205 and 404} \\
\hline
\# of pixels                             & 2048\\      
image scale (arcsec pix$^{-1}$ )         &  0.54 \\
slit width (arcsec)                      & 1.5 \\         
spectral range (\AA)                     & 4700 -- 5600\\      
FWHM $\Delta\lambda$ (\AA)               & 1.4 \\     
$\sigma_{\rm instr}$ ({{\kms}})           & 35 \\             
\hline
  \end{tabular}
\end{minipage}
\end{table}

The data reduction was performed using the IRAF-based GMOS standard
pipeline software for bias and flat field correction and wavelength
calibration. Cosmic rays were identified and corrected by the IRAF
task {\sc crblaster}. A sky spectrum was derived for each galaxy spectrum
from regions in the spectral image empty of objects. The sky spectrum
was then subtracted from the object spectrum.

In the case of the Fornax data, the Gaussian width of the LSF
increases from $\sigma_{LSF}$ = 100 to 150\,\kms\ along
wavelength range. This broadening is larger than the velocity
dispersion of most of the target galaxies and a close examination of
the night sky emission line profiles reveals that the spectral
resolution is determined by the the slit profile. Therefore, as for
the FORS2 observations described in Sect.~\ref{subsec:derijcke}, 
the centre of the
galaxies are spatially resolved in the slit. Hence, the precision of the
velocity dispersion measurements will be limited. However, except for
one galaxy, additional FORS2 spectra (see Sect.~\ref{subsec:bedregal}) 
can be used to
accurately constrain the velocity dispersion. We adopted an average
between the LSF derived from the twilight and the template spectra. We
checked that the corresponding velocity dispersion agree with that
measured on the FORS2 spectra and found that the velocity dispersion
is overestimated in the external regions and more importantly, it is 
underestimated in the centres, which produces fake sigma-drops 
(see Appendix~\ref{appendix:indgal}). 
This problem is not discussed in \citet{spo2010a}, 
where kinematical analyses of these data are also presented. 
 In the case of the Virgo data, the slit width was narrower
and the spectral resolution sufficient to derive reliable
kinematics. We adopt for the Fornax and Virgo galaxies a LSF of 114
and 61\,\kms{} respectively.


\begin{table}
  \centering
  \begin{minipage}{\columnwidth}
    \caption{\label{table:obslog}Observation log}
\begin{tabular}{lcrrr} \hline 
Instrument & object  & $\lambda_0 $ [\AA] & exposure (s) & slit PA \\
\hline
GMOS-S  & FCC 148  & 5075 & 4 $\times$ 2040.5 &  90 \\
(Fornax)& FCC 153  & 5075 & 4 $\times$ 2040.5 &  83 \\
        & FCC 170  & 5075 & 3 $\times$ 2040.5 & 139 \\
        & FCC 277  & 5075 & 4 $\times$ 2040.5 & 116 \\
        &          & 5075 & 4 $\times$ 2040.5 & 122 \\
        & FCC 301  & 5075 & 4 $\times$ 2040.5 & 155 \\
        & FCC 335  & 5075 & 2 $\times$ 2040.5 &  47 \\
\hline
GMOS-S  & VCC 0575 & 5250 & 2 $\times$ 1886.5 &  65 \\
(Virgo) &          & 5300 & 2 $\times$ 1886.5 &  65 \\
        & VCC 0828 & 5250 & 2 $\times$ 1886.5 &   3 \\
        &          & 5300 & 2 $\times$ 1886.5 &   3 \\
        & VCC 1025 & 5250 & 3 $\times$ 1886.5 &  15 \\
        &          & 5300 & 2 $\times$ 1886.5 &  15 \\
        & VCC 1146 & 5250 & 2 $\times$ 1886.5 & 174 \\
        &          & 5300 & 2 $\times$ 1886.5 & 174 \\
        & VCC 1178 & 5250 & 2 $\times$ 1886.5 &  10 \\
        &          & 5300 & 2 $\times$ 1886.5 &  10 \\
        & VCC 1297 & 5250 & 2 $\times$ 1886.5 & 100 \\
        &          & 5300 & 2 $\times$ 1886.5 & 100 \\
        & VCC 1327 & 5250 & 2 $\times$ 1886.5 &  18 \\
        &          & 5250 & 1 $\times$ 1500.5 &  18 \\
        &          & 5300 & 4 $\times$ 1200.5 &  18 \\
        & VCC 1475 & 5250 & 2 $\times$ 1886.5 &  15 \\
        &          & 5300 & 2 $\times$ 1886.5 &  15 \\
        & VCC 1630 & 5200 & 2 $\times$ 1886.5 &  58 \\
        &          & 5300 & 2 $\times$ 1886.5 &  58 \\
\hline
FORS2   & FCC 021  & 5200 & 3 $\times$ 1200   &  50 \\
        & FCC 055  & 5200 & 2 $\times$ 2400   &  32 \\
        & FCC 148  & 5200 & 2 $\times$ 1800   &  90 \\
        & FCC 153  & 5200 & 2 $\times$ 1600   &  83 \\
        & FCC 167  & 5200 & 2 $\times$ 1200   &   3 \\
        & FCC 170  & 5200 & 2 $\times$ 1600   &  42 \\
        & FCC 177  & 5200 & 2 $\times$ 1700   &   4 \\
        & FCC 301  & 5200 & 1 $\times$ 2550   &  24 \\ 
        & FCC 335  & 5200 & 1 $\times$ 2250   &  47 \\
\hline
\end{tabular}
  \end{minipage}
\end{table}

\subsection{Bedregal et al.  sample}
\label{subsec:bedregal}

The ESO data archive contains FORS2 long-slit spectra of 5 of the 6
Fornax cluster galaxies of the \citeauthor{spo2009} sample. An
analysis of these data has been published by \citet{bedregal2006}.
The spectral resolution of the FORS2 data is higher than the GMOS
observations, allowing therefore accurate measurements of the internal
stellar kinematics.

The FORS2 setup is described in Table\,\ref{table:setup}. The data
reduction was performed using the ESO FORS pipeline software for bias
and flat field correction, distortion correction and wavelength
calibration. Cosmic rays were identified and corrected by the IRAF
task {\sc crblaster}. A sky spectrum was derived for each galaxy
spectrum from regions in the spectral image empty of objects.  The sky
spectrum was then subtracted from the object spectrum.

The pipeline data reduction ignored the correction of the electronic
noise mentioned by \citet{bedregal2006}. This effect, identified as a
15\,Hz hum in the reading of the CCD, produces an aperiodic
residual. This artifact may affect the analysis of the stellar
population, as it clearly dominates the noise in the external
regions. We also analysed the spectra reduced by A.~Bedregal including
the hum correction, that he courteously made available to us
\citep{kolbed}. We found
a perfect consistency between the two analyses, but a noticeable
divergence in the external regions. With the uncorrected data, the
metallicity reaches higher values than with the corrected ones. The
fact that the corrected version is in a better agreement with the
analysis of the GMOS data, may be used as an indication of the
relevance of the correction. However, in order to keep the analyses
independent, we simply chose to limit our analysis to the radial range
where the two data sets give consistent results.

We adopt a wavelength independent Gaussian relative LSF of 19.9\,\kms
determined from template star spectra.

\subsection{Vazdekis et al. sample, NGC~205 and 404}
 
The other data sets included in our discussions are from \citet[six E galaxies
from the Virgo cluster]{vaku},
\citet[NGC~205]{sp02} and \citet[NGC~404]{bouchard2010}. 
The observational setups are presented in Table~\ref{table:setup}, and their
data reduction is described in the corresponding papers. The seeing
during these observations was larger than the slit widths, thus we 
do not expect problems in the derived kinematic profiles.

The reduced \citeauthor{vaku} data were taken from the
HyperLeda\footnote{\url{leda.univ-lyon1.fr}, \citet{HyperLeda}} FITS
archive, where no associated template star is given.  Therefore, the
LSF could not be determined directly, and we derived it in order to
match the velocity dispersion profiles of NGC~4387 and 4464 with those
obtained from the GMOS data. We checked that the central velocity
dispersion of the other four galaxies also agree with the values
compiled in HyperLeda.  We used a Gaussian LSF with $\sigma_{\rm LSF}
= 58.8 $\,\kms, in good agreement with the nominal resolution of 2.4
\AA.

For NGC~205 and 404 we used the reduced data and the LSF quoted in
the corresponding papers. NGC~205 is a close companion of 
M~31 and a prototype of the dwarf elliptical class of galaxies. 
NGC 404 is an isolated dwarf S0.

\section{Analysis of the stellar population and kinematics} \label{sec:radpop}

In this section, we first describe the method for extracting the
internal kinematics and the SSP-equivalent ages and metallicities.
After, we present the stellar population radial profiles of the 40
galaxies.  Then, we measure the population gradients and finally we
discuss possible biases.

\subsection{Analysis method}

The principle of our analysis is to compare an observed spectrum with
a model of a population broadened to account for the internal
kinematics. A least-square minimisation provides the internal
kinematics and the parameters of the population model, age and
metallicity in the case of a SSP.  This is done with the
\ulyss\footnote{\url{http://ulyss.univ-lyon1.fr}} package
\citep{ulyss}.  The main characteristics of the method, described and
validated in \citet{koleva2008a,ulyss}, is to fit the full spectrum,
pixel-by-pixel, while the most common approach uses spectrophotometric
indices.  The SSP-equivalent ages and metallicity for central
extractions of the \citeauthor{paperI} sample were found perfectly
consistent with Lick indices and three times more precise
\citep{michielsen08, kolbed}.  Beside the precision, another advantage of the
method is its insensitivity to the presence of emission lines, that
can be either masked or fitted with additional Gaussians.

We are using stellar population models generated with 
Pegase.HR\footnote{\url{http://www2.iap.fr/pegase/pegasehr/}}
\citep{PEGASEHR}, assuming Salpeter IMF \citep{salpeter} and Padova isochrones
\citep[][and companion papers]{padova1994} and build with the
ELODIE.3.1 library \citep{PS01, ELODIE31}.  These models have a well
calibrated line spread function (i.  e.  instrumental broadening)
of 0.58~{\AA} corresponding to a resolution R $\approx$ 10000 or
instrumental velocity dispersion $\sigma_{\rm instr} = 13$\,{\kms} at
$\lambda = 5500$\,{\AA}.

The ELODIE library, as any other empirical library, consists mostly
of stars of the solar neighbourhood and hence presents the
abundance pattern of this environment. Therefore, a model made with 
such library cannot match a population with non-solar abundances, such
as in globular clusters or massive elliptical galaxies known 
to harbour an excess of Mg with respect to Fe.
However, it was found \citep{koleva08e},
that the measurements of [{\rm Fe/H}] and [{\rm Mg/Fe}] with full-spectrum fitting
are independent. In other words, a mismatch of [{\rm Mg/Fe}] does not affect
the measurement of [{\rm Fe/H}]. For this reason, we
preferred to conservatively use the empirical library instead of the
$\alpha$-elements resolved semi-empirical one \citep{prugniel07} which
is not tested in detail.


During the fit we perform automatic kappa-sigma clipping to exclude
the spikes (un-removed cosmic rays, or artifacts of the subtraction of
strong sky lines).  The Mg triplet, near 5175~\AA, is also masked in
the case of strong $\alpha$-overabundance.  The NaD region ($\lambda
\sim 589$\,nm) is masked by default, since it is strongly affected by
telluric absorption.


The fit is rendered insensitive to the shape of the spectral energy
distribution thanks to the inclusion of a multiplicative polynomial in
the model.  This feature makes the flux-calibration unnecessary. The
possible biases due to this polynomial have been well studied on other
occasions, and we determined the optimal degree as in \citet{ulyss}.
The adopted degrees essentially depend of the wavelength range.  We
used 40 for the GMOS observations, and 25 for all the others.

\begin{figure*}
\includegraphics[width=0.3\textwidth]{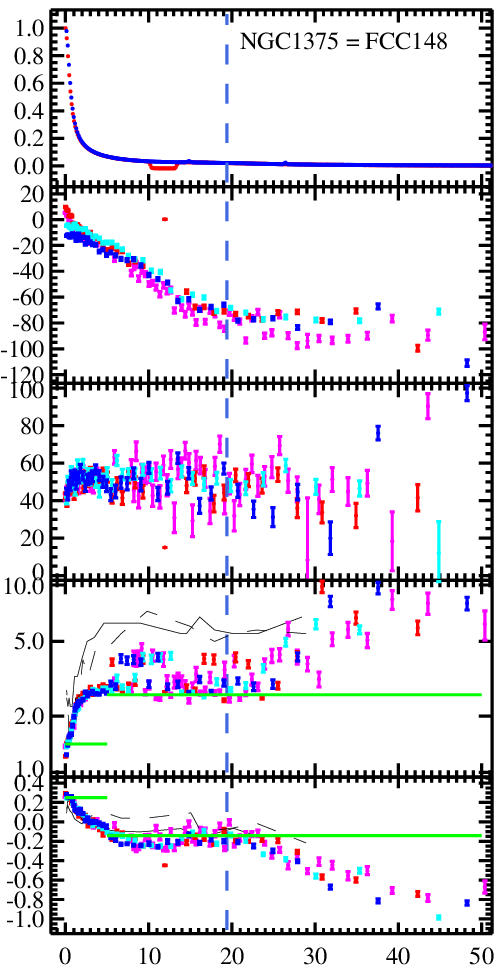}
\includegraphics[width=0.3\textwidth]{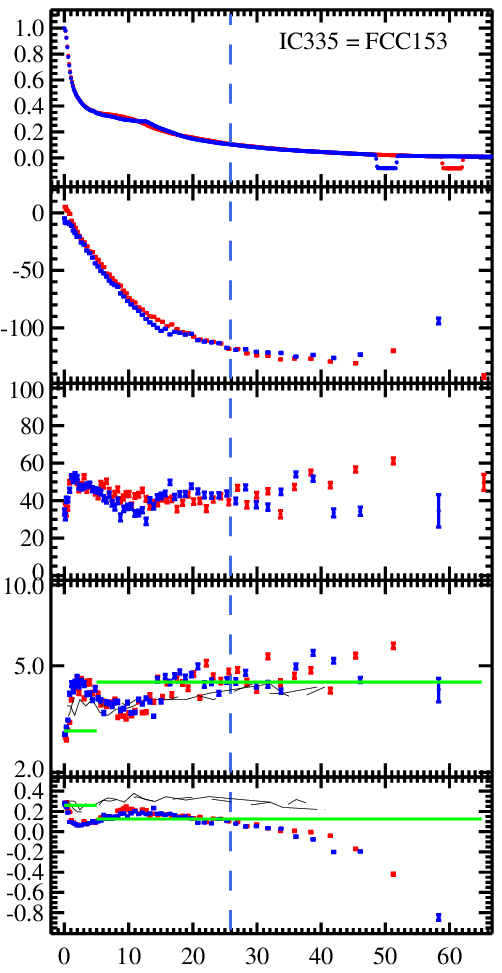}
\includegraphics[width=0.3\textwidth]{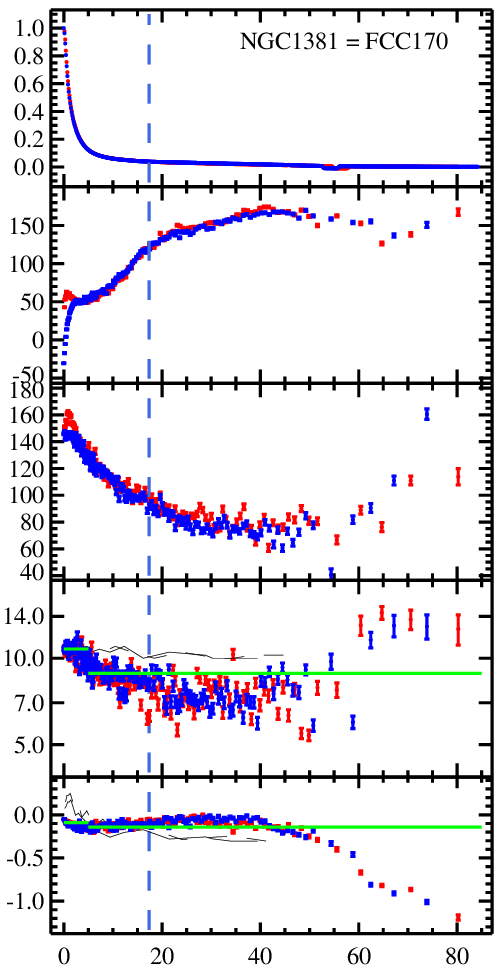}
\includegraphics[width=0.3\textwidth]{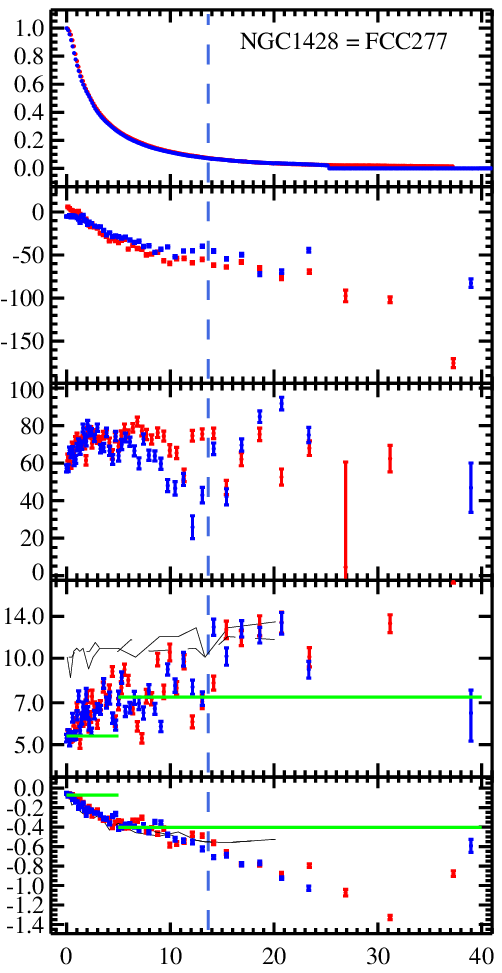}
\includegraphics[width=0.3\textwidth]{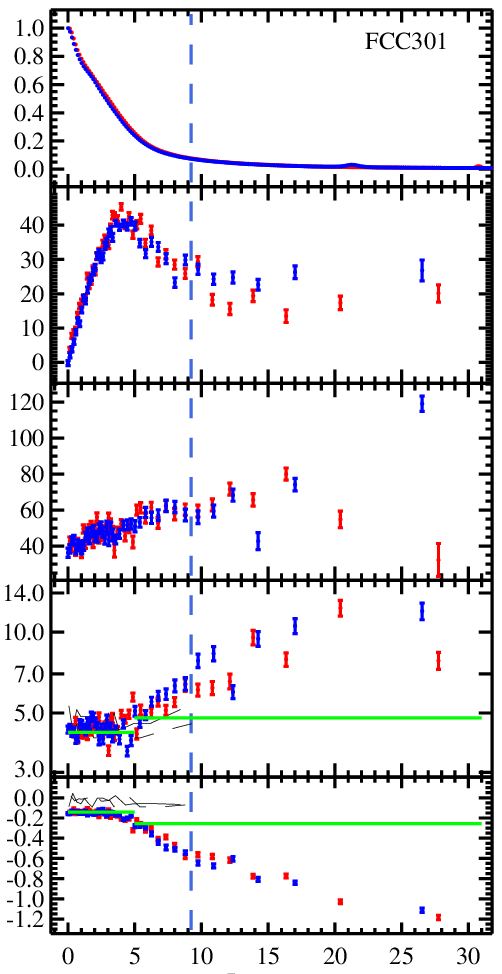}
\includegraphics[width=0.3\textwidth]{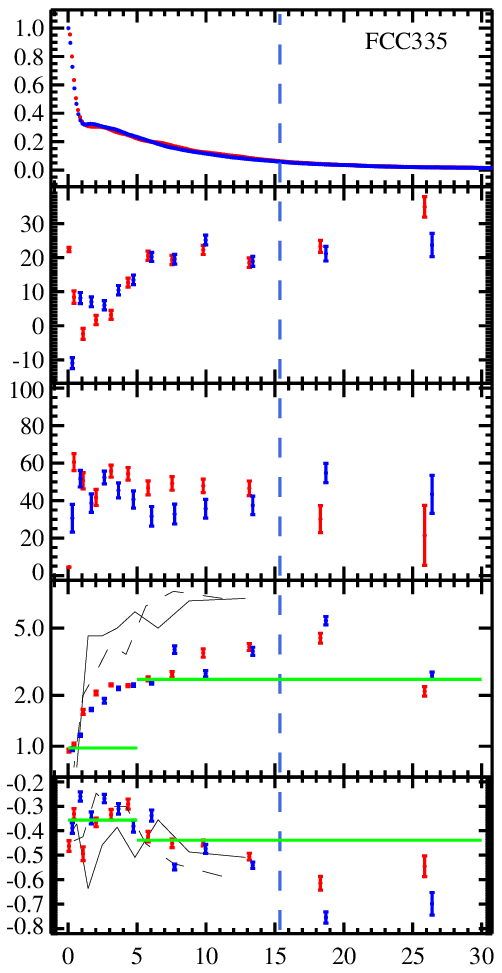}
\caption{Radial profiles of Fornax cluster galaxies from the GMOS 
  observations derived from SSP fits.  The spectra are
  radially  binned for a compromise between S/N and spatial resolution.  
  The profiles are folded around the luminosity peak, the red
  points are for the positive radii and blue for negative (see
  Sect.\,\ref{subsec:radprof}).  
  The semi-major axis of the effective isophotes
  are figured as vertical blue dashed lines. The SSP-equivalent
  metallicities and ages within the core and 1~\reff{} extractions
  (Sect. \ref{sec:1D}) are shown with green lines  (for readability
  they are arbitrarily drawn from 0 to 5 arcsec and from 5
  arcsec outwards). The profiles drawn in black are
  taken from the figures~1,2 and 3 of \citet{spo2010}.
}
\hskip 0cm
\label{fig:radprof1}
\end{figure*}
\addtocounter{figure}{-1}
\begin{figure*}
\includegraphics[width=0.3\textwidth]{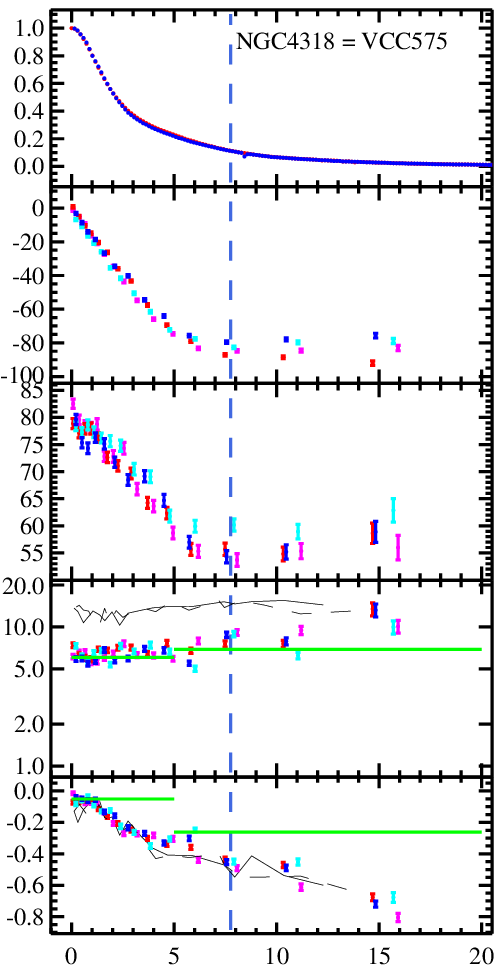}
\includegraphics[width=0.3\textwidth]{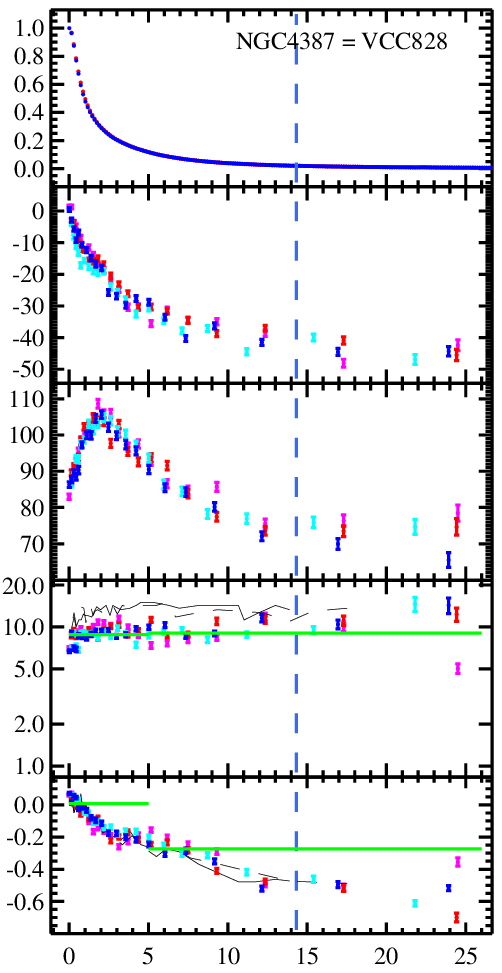}
\includegraphics[width=0.3\textwidth]{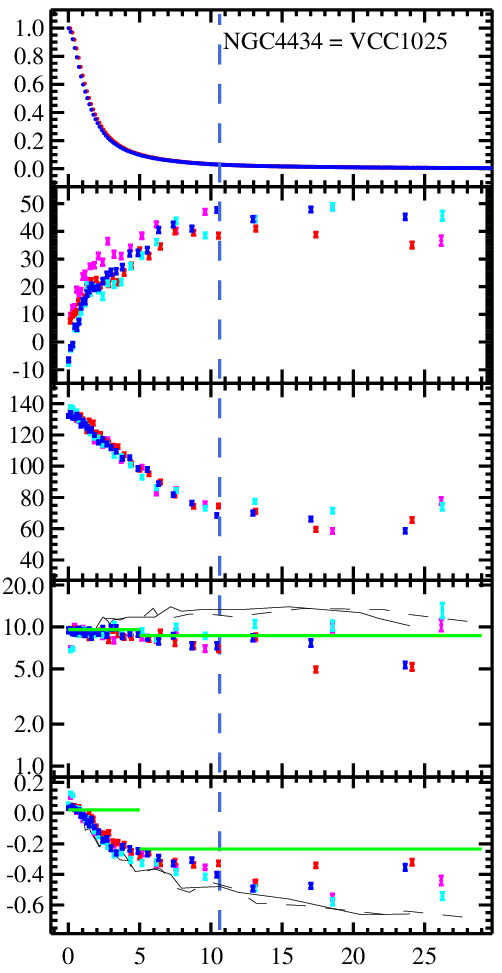}
\includegraphics[width=0.3\textwidth]{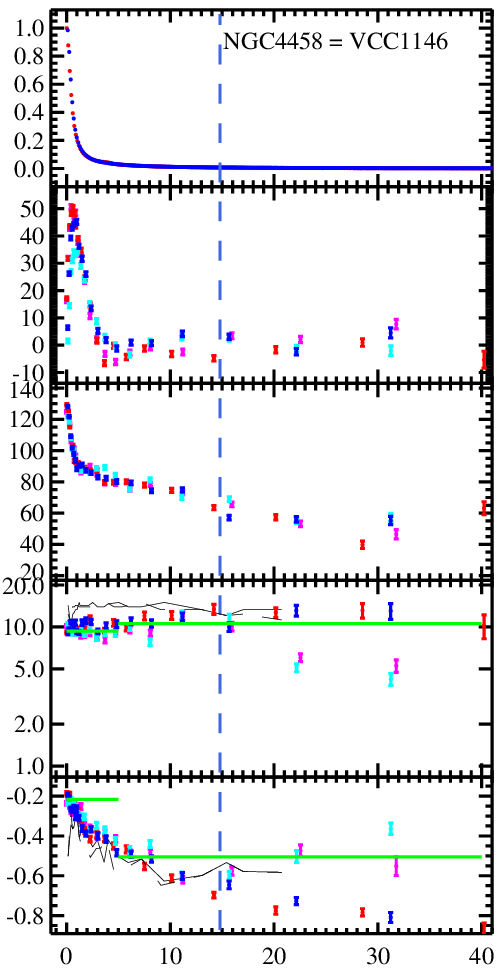}
\includegraphics[width=0.3\textwidth]{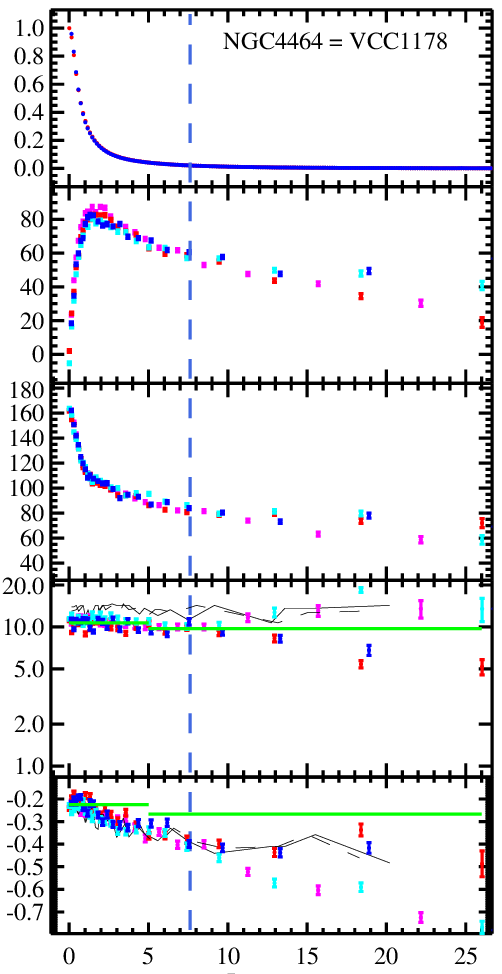}
\includegraphics[width=0.3\textwidth]{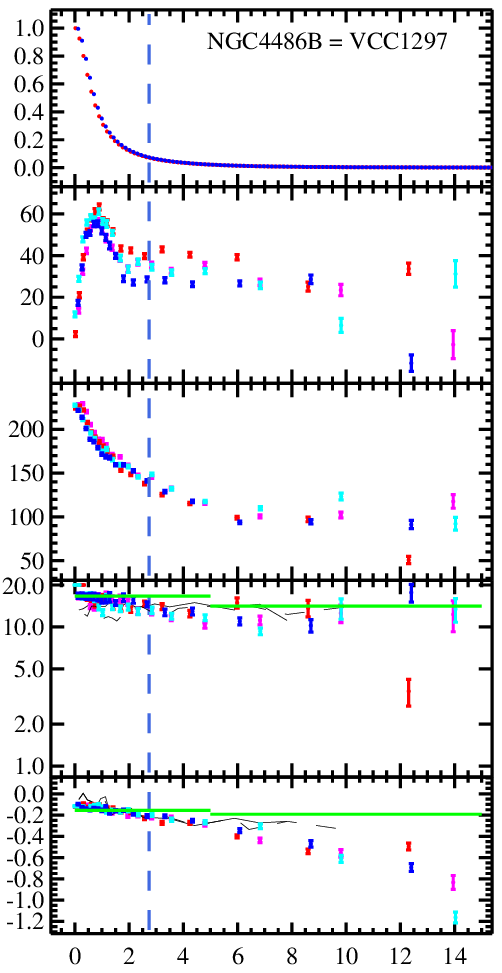}
\caption{Continued (Virgo galaxies).}
\hskip 0cm
\label{fig:radprof2}
\end{figure*}
\addtocounter{figure}{-1}
\begin{figure*}
\includegraphics[width=0.3\textwidth]{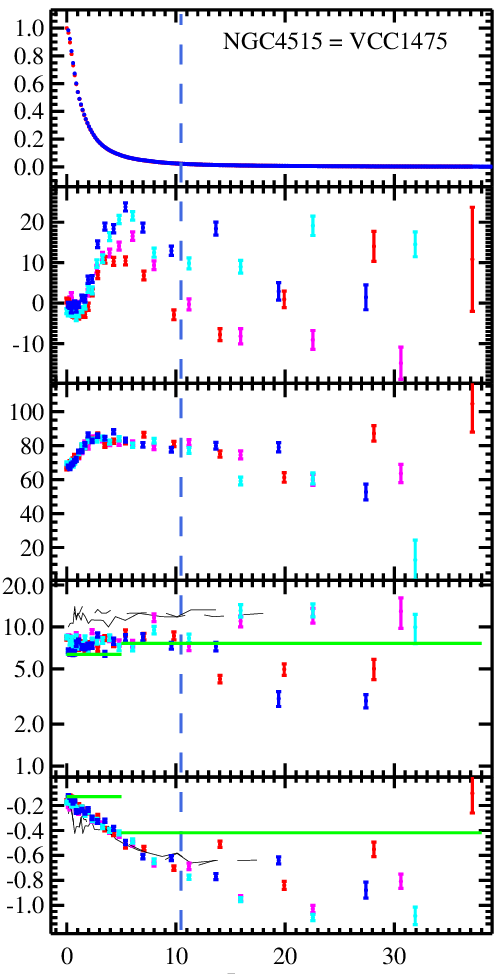}
\includegraphics[width=0.3\textwidth]{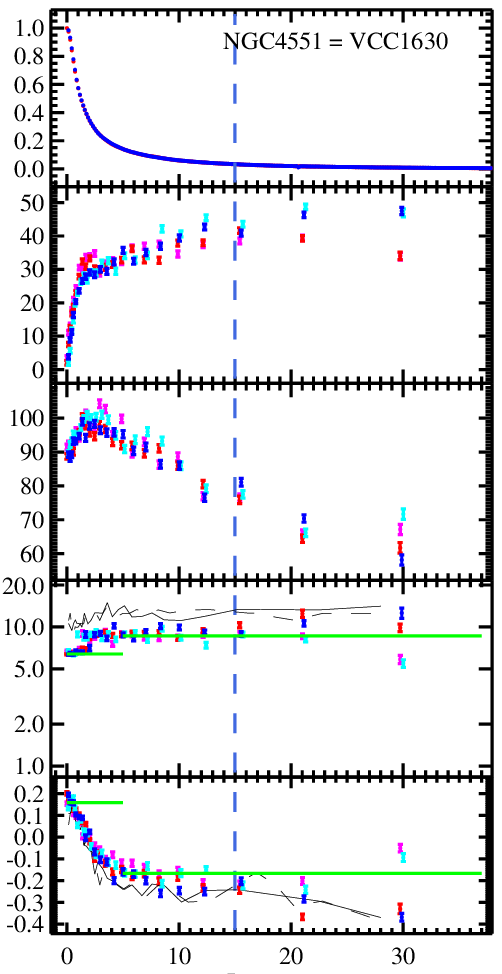}
\caption{Continued}
\hskip 0cm
\label{fig:radprof3}
\end{figure*}

 \begin{figure*}
\includegraphics[width=0.3\textwidth]{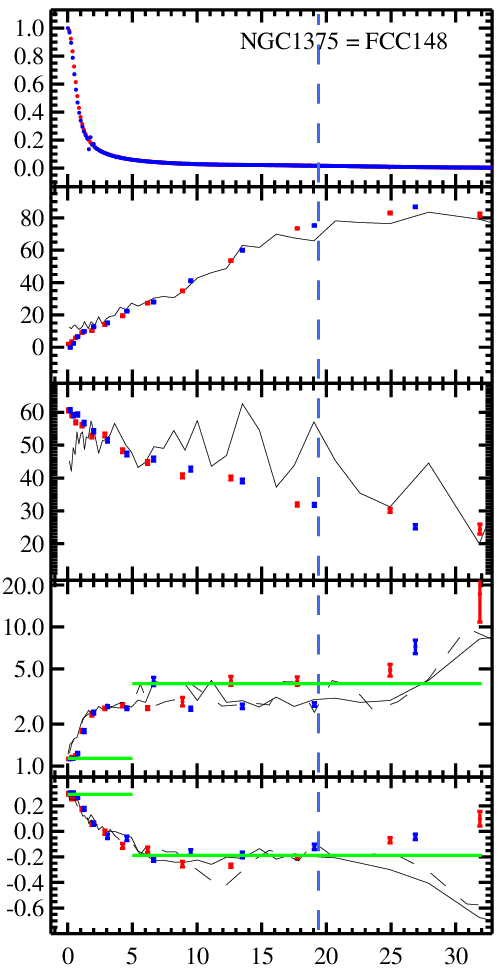}
\includegraphics[width=0.3\textwidth]{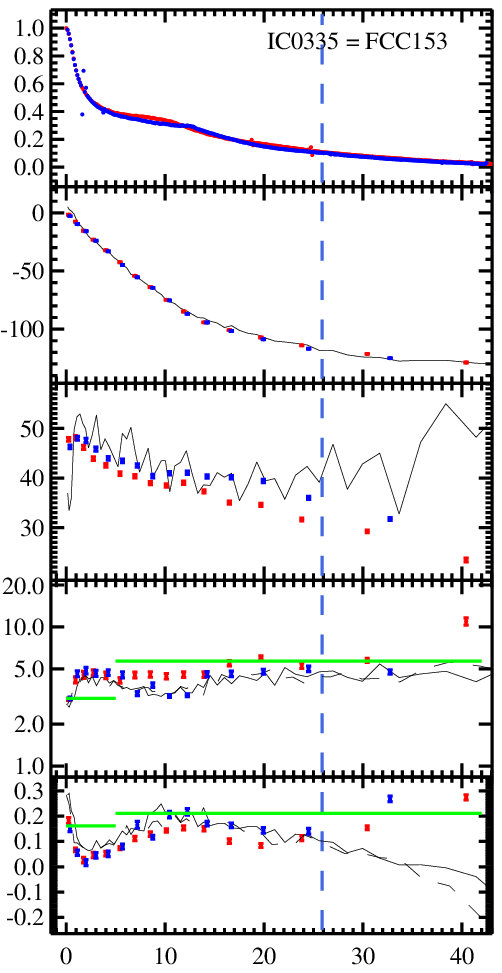}
\includegraphics[width=0.3\textwidth]{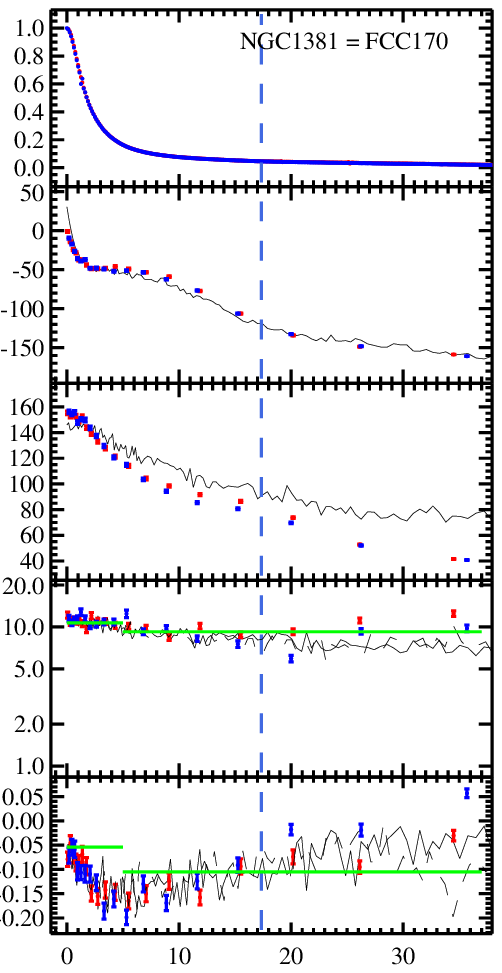}
\includegraphics[width=0.3\textwidth]{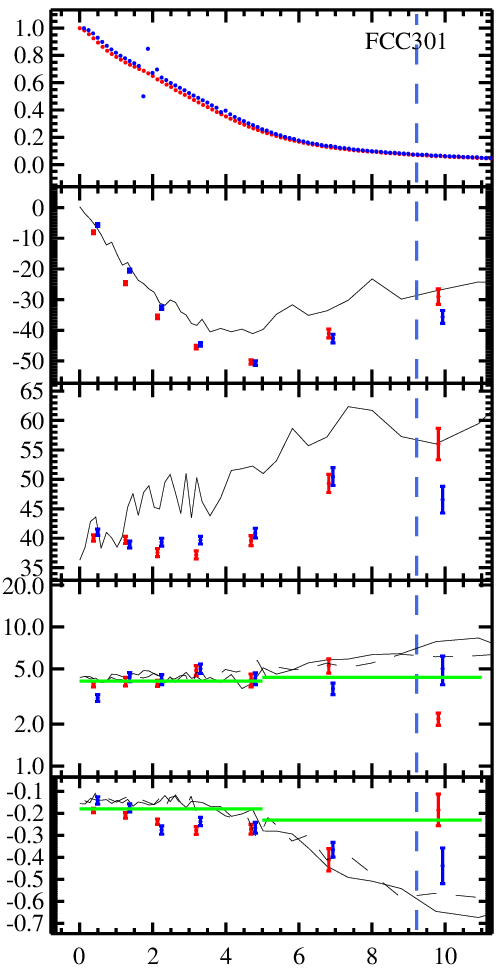}
\includegraphics[width=0.3\textwidth]{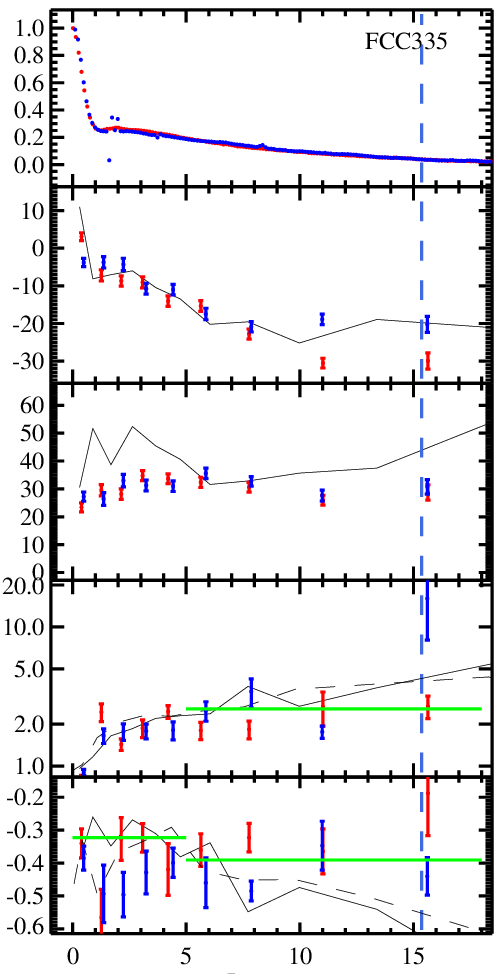}
\includegraphics[width=0.3\textwidth]{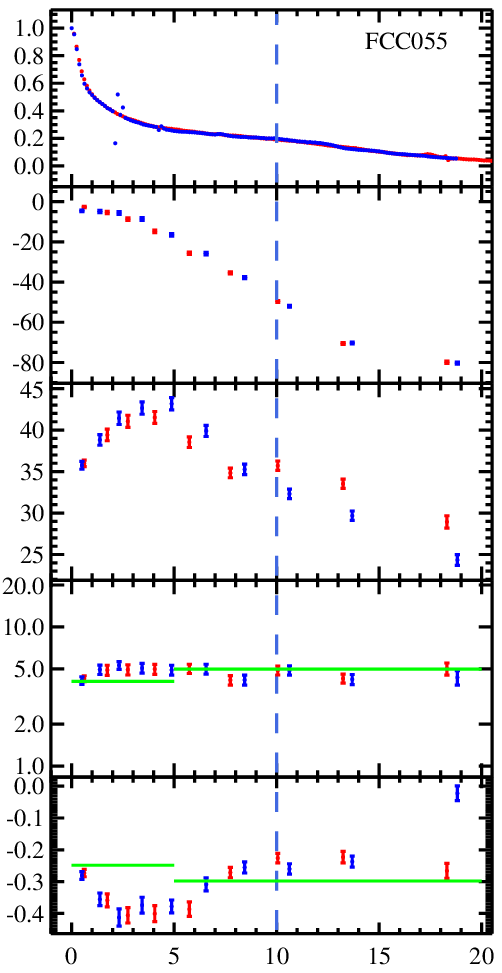}
\caption{Radial profiles of Fornax cluster galaxies from the FORS observations. 
The radial binning of the
spectra and the figure caption is the same as for Fig.~\ref{fig:radprof1}.
The profiles drawn in black are those from the GMOS observations presented
in Fig.~\ref{fig:radprof1}.
}
\hskip 0cm
\label{fig:radprof_fors1}
\end{figure*}
\addtocounter{figure}{-1}
\begin{figure*}
\includegraphics[width=0.3\textwidth]{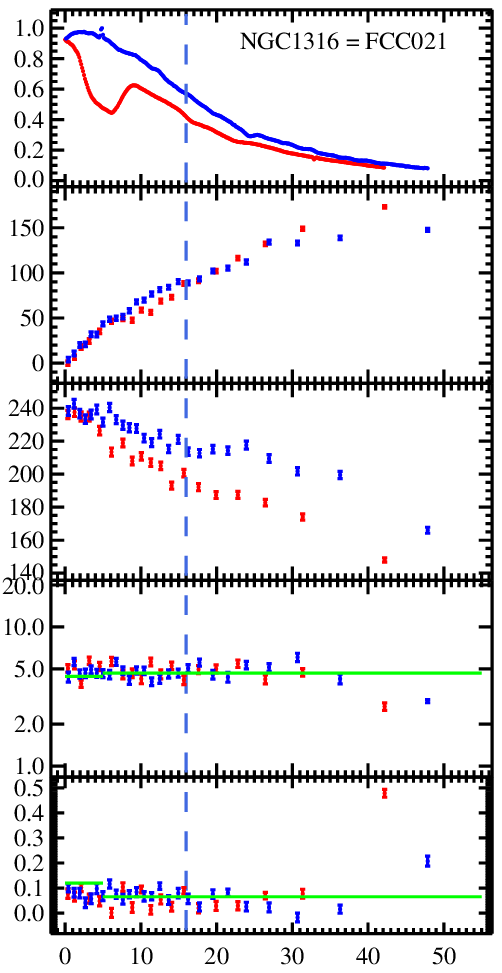}
\includegraphics[width=0.3\textwidth]{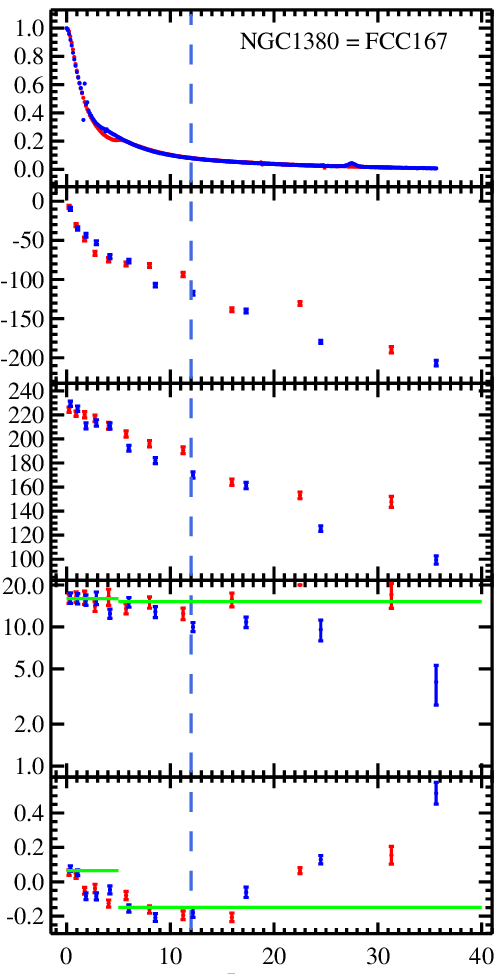}
\includegraphics[width=0.3\textwidth]{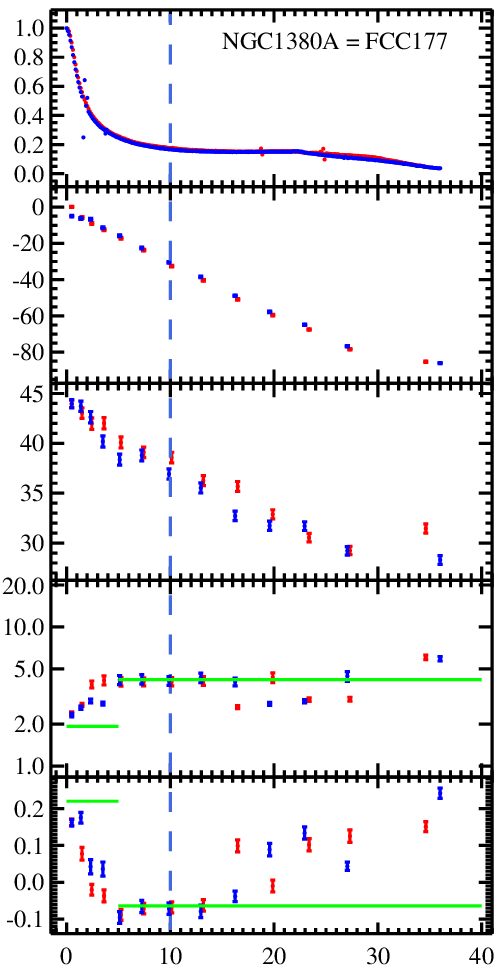}
\caption{Continued.}
\hskip 0cm
\label{fig:radprof_fors2}
\end{figure*}

\begin{figure*}
\includegraphics[width=0.3\textwidth]{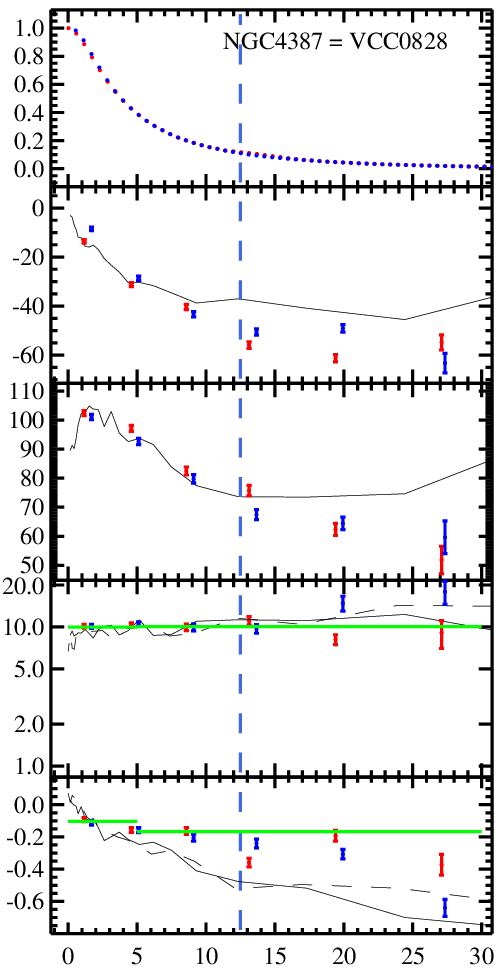}
\includegraphics[width=0.3\textwidth]{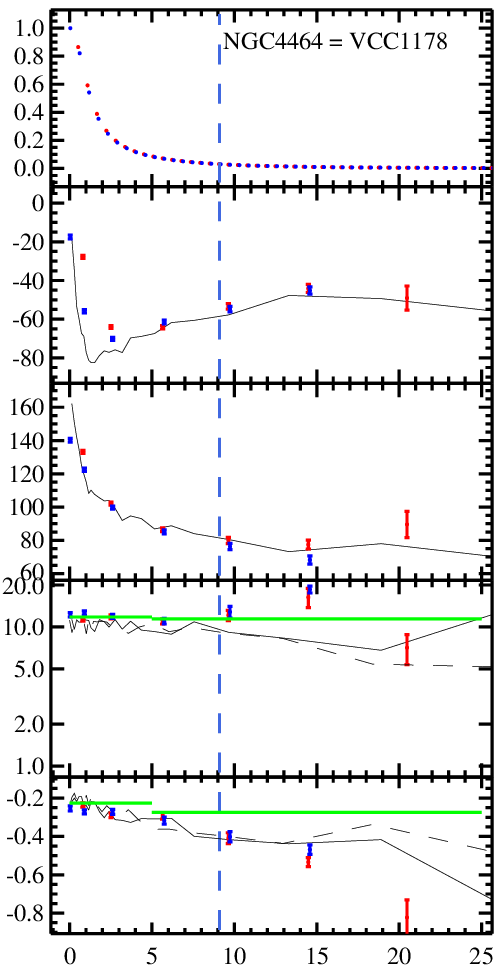}
\includegraphics[width=0.3\textwidth]{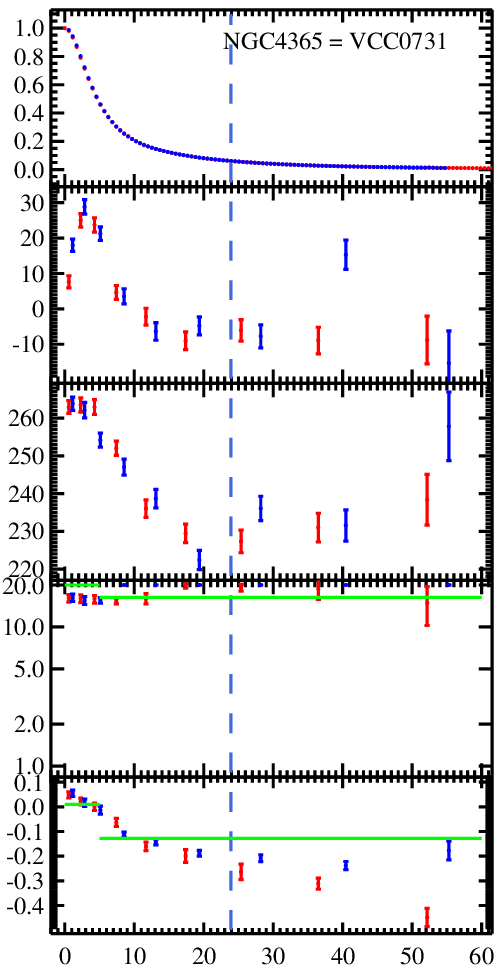}
\includegraphics[width=0.3\textwidth]{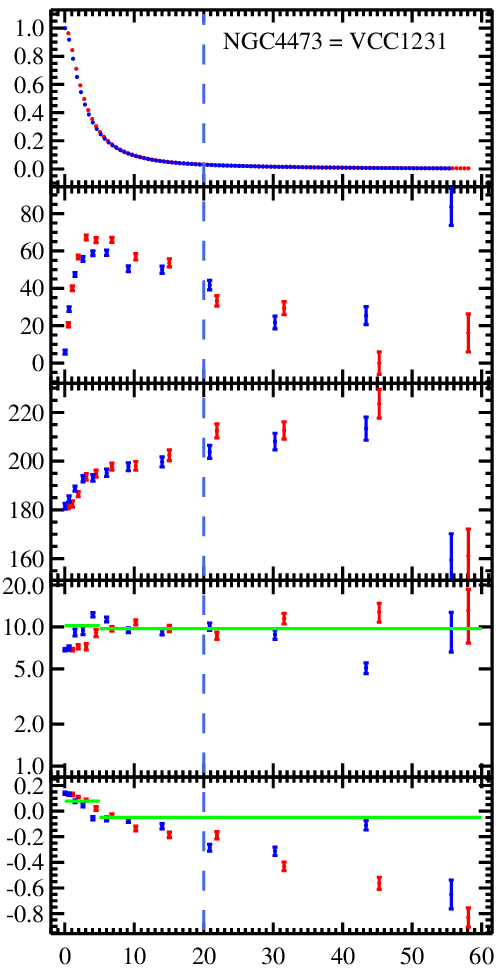}
\includegraphics[width=0.3\textwidth]{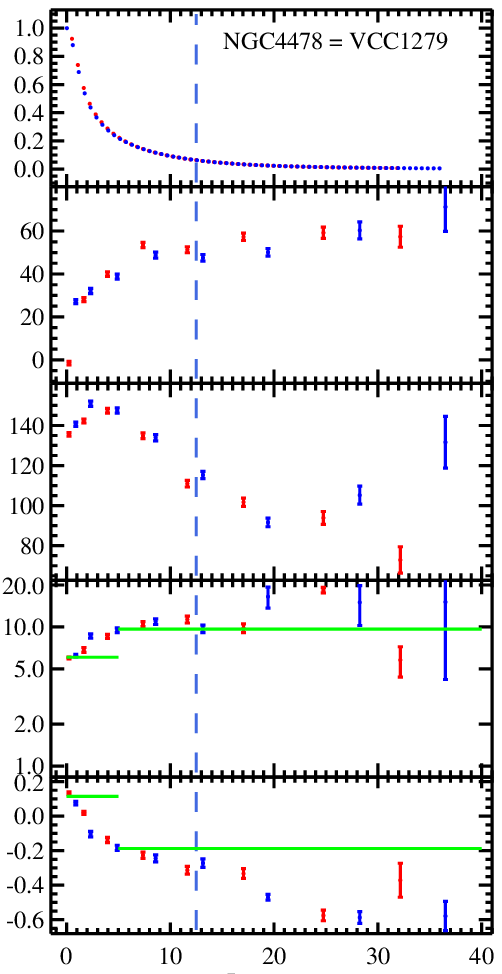}
\includegraphics[width=0.3\textwidth]{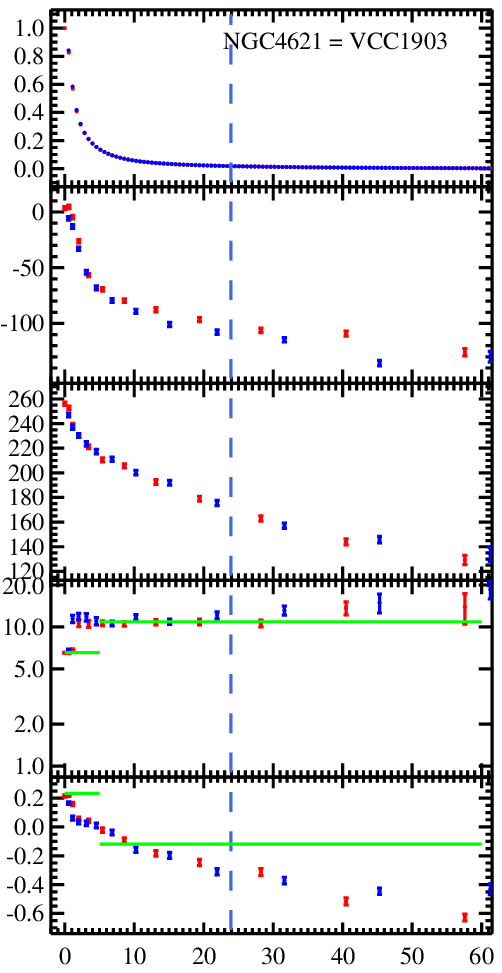}
\caption{Radial profiles of the Virgo galaxies of the VAKU sample. 
The radial binning of the
spectra and the figure caption is the same as for Fig.~\ref{fig:radprof1}.
The profiles drawn in black are those from the GMOS observations presented
in Fig.~\ref{fig:radprof1}.
}
\hskip 0cm
\label{fig:radprof_vaku}
\end{figure*}

\begin{figure*}
\includegraphics[width=0.3\textwidth]{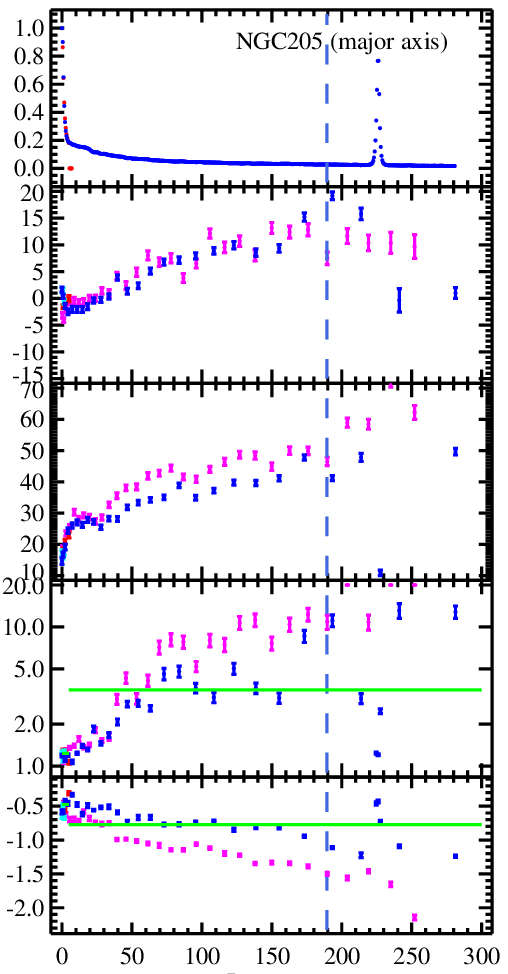}
\includegraphics[width=0.3\textwidth]{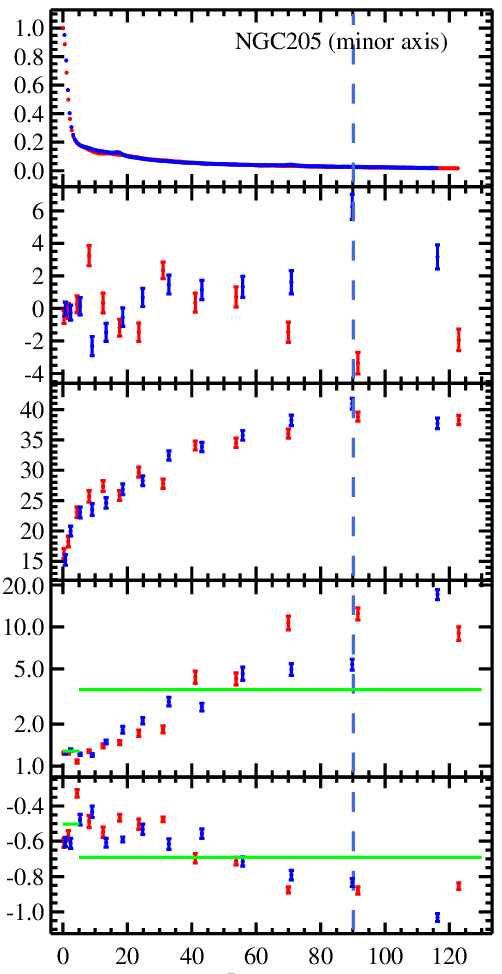}
\caption{Radial profiles for NGC~205. The radial binning of the
spectra and the figure caption is the same as for Fig.~\ref{fig:radprof1}.}
\hskip 0cm
\label{fig:radprof_ngc205}
\end{figure*}

\begin{figure*}
\includegraphics[width=0.3\textwidth]{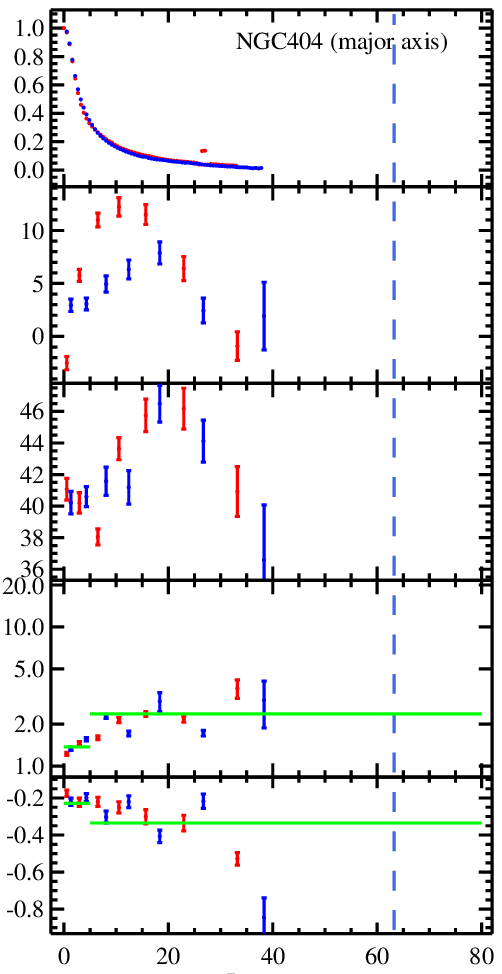}
\includegraphics[width=0.3\textwidth]{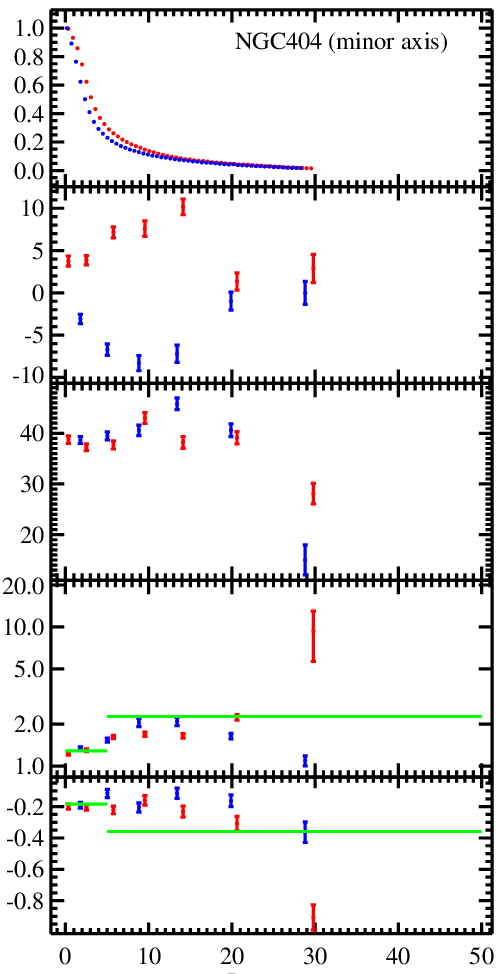}
\caption{Radial profiles for NGC~404. The radial binning of the
spectra and the figure caption is the same as for Fig.~\ref{fig:radprof1}.}
\hskip 0cm
\label{fig:radprof_ngc404}
\end{figure*}

\subsection{Extraction and fit of one-dimensional spectra} \label{sec:1D}

For each object, we extract two one-dimensional (1D) spectra integrated
within simulated apertures of semi-major axes equal to about 50~pc and
one \aeff{} where \aeff{} = \reff/$\sqrt{1-\epsilon}$ is the
semi-major axis of the effective isophote.  The inner extraction is
aimed at characterising the core that may be highly contaminated by a
nucleus.  To make these extractions, each individual spectrum is
weighted by its radius (and consequently the 1\,\aeff{} extractions
are sometimes noisy).

For the Virgo and Fornax galaxies, the core radius was fixed at
0.5\,arcsec, which also corresponds to the seeing disc. The
corresponding physical diameters are 106\,pc in Fornax and 84\,pc in
Virgo.  For the dEs belonging to more distant groups, we used the same
apparent diameter (a smaller extraction would not make sense, because
of the seeing limitation).  For NGC\,205 we adopted a radius of
14\,arcsec for the core extraction, and 3\,arcsec for NGC~404, to
sample a region of about 100\,pc of diameter.  When computing the
outer extractions, we excluded the core region.

The results of the analysis of the core 1D spectra are reported
in Table~\ref{table:result}. The quality and details of the fits
can be checked in Appendix~\ref{appendix:fits}, where we
plot the ones from the central regions.

\begin{table*}
  \centering
  \begin{minipage}{180mm}
    \caption{Central populations and gradients.  \label{table:result}
E(Age) and E([{\rm Fe/H}]) are the uncertainties on the gradients rescaled by
$\chi$ (see Sect.~\ref{subsec:popgrad}).
}
    \begin{tabular}{|l|rrrrrrrr} \hline
Name& $\log {\rm Age}_{core}$&$\log {\rm Age}_{\rm eff}$& $\nabla_{{\rm Age}}$& E({\rm Age})&
[{\rm Fe/H}]$_{core}$&$ [{\rm Fe/H}]_{\rm eff}$& $ \nabla_{[{\rm Fe/H}]}$& E([{\rm Fe/H}])\\
\hline
FCC277   &  0.73$\pm$0.01&  0.89$\pm$0.01&  0.10$\pm$0.01&0.19& -0.07$\pm$0.00& -0.50$\pm$0.01& -0.33$\pm$0.01&0.11\\
VCC575   &  0.80$\pm$0.01&  1.65$\pm$0.03&  0.04$\pm$0.03&0.20& -0.05$\pm$0.01& -0.76$\pm$0.03& -0.32$\pm$0.02&0.11\\
VCC731   &  1.31$\pm$0.00&  1.27$\pm$0.03&  0.08$\pm$0.04&0.09&  0.01$\pm$0.01& -0.14$\pm$0.03& -0.14$\pm$0.03&0.08\\
VCC828   &  0.96$\pm$0.01&  1.93$\pm$0.03&  0.00$\pm$0.02&0.14&  0.01$\pm$0.00& -0.78$\pm$0.03& -0.30$\pm$0.02&0.11\\
VCC1025  &  0.91$\pm$0.01&  1.84$\pm$0.02& -0.05$\pm$0.02&0.09&  0.06$\pm$0.00& -0.77$\pm$0.03& -0.35$\pm$0.02&0.07\\
VCC1146  &  0.98$\pm$0.00&  2.04$\pm$0.04&  0.05$\pm$0.02&0.11& -0.23$\pm$0.00& -1.17$\pm$0.04& -0.26$\pm$0.02&0.08\\
VCC1178  &  1.03$\pm$0.00&  1.99$\pm$0.02& -0.05$\pm$0.02&0.10& -0.23$\pm$0.00& -0.73$\pm$0.03& -0.14$\pm$0.02&0.07\\
VCC1231  &  1.01$\pm$0.02&  1.07$\pm$0.02&  0.17$\pm$0.02&0.15&  0.08$\pm$0.01& -0.12$\pm$0.02& -0.18$\pm$0.02&0.10\\
VCC1279  &  0.79$\pm$0.01&  1.07$\pm$0.02&  0.24$\pm$0.04&0.08&  0.11$\pm$0.01& -0.31$\pm$0.03& -0.35$\pm$0.04&0.06\\
VCC1297  &  1.22$\pm$0.01&  2.28$\pm$0.05& -0.10$\pm$0.09&0.25& -0.15$\pm$0.00& -0.40$\pm$0.04& -0.12$\pm$0.04&0.09\\
VCC1475  &  0.86$\pm$0.01&  1.85$\pm$0.04&  0.06$\pm$0.03&0.15& -0.17$\pm$0.00& -1.11$\pm$0.04& -0.34$\pm$0.02&0.14\\
VCC1630  &  0.82$\pm$0.00&  1.98$\pm$0.02&  0.12$\pm$0.01&0.11&  0.15$\pm$0.00& -0.54$\pm$0.03& -0.30$\pm$0.02&0.08\\
VCC1903  &  0.82$\pm$0.00&  1.11$\pm$0.02&  0.17$\pm$0.02&0.13&  0.23$\pm$0.01& -0.23$\pm$0.02& -0.27$\pm$0.02&0.08\\
\hline
FCC21    &  0.65$\pm$0.03&  0.67$\pm$0.02& -0.03$\pm$0.03&0.14&  0.12$\pm$0.01&  0.07$\pm$0.01& -0.00$\pm$0.03&0.09\\
FCC55    &  0.61$\pm$0.02&  0.65$\pm$0.03& -0.08$\pm$0.07&0.10& -0.25$\pm$0.01& -0.34$\pm$0.03& -0.03$\pm$0.03&0.11\\
FCC148   &  0.15$\pm$0.00&  0.97$\pm$0.01&  0.10$\pm$0.01&0.12&  0.25$\pm$0.00& -0.58$\pm$0.02& -0.36$\pm$0.01&0.10\\
FCC153   &  0.46$\pm$0.00&  0.57$\pm$0.00& -0.00$\pm$0.00&0.16&  0.26$\pm$0.00&  0.16$\pm$0.00&  0.04$\pm$0.00&0.09\\
FCC167   &  1.21$\pm$0.03&  1.11$\pm$0.04& -0.09$\pm$0.06&0.11&  0.07$\pm$0.01& -0.21$\pm$0.03& -0.22$\pm$0.05&0.09\\
FCC170   &  1.04$\pm$0.00&  0.91$\pm$0.00& -0.11$\pm$0.01&0.08& -0.09$\pm$0.00& -0.13$\pm$0.01& -0.01$\pm$0.01&0.07\\
FCC177   &  0.29$\pm$0.01&  0.61$\pm$0.02&  0.22$\pm$0.04&0.18&  0.22$\pm$0.01& -0.11$\pm$0.02& -0.22$\pm$0.03&0.09\\
FCC301   &  0.63$\pm$0.01&  0.72$\pm$0.01&  0.11$\pm$0.01&0.16& -0.14$\pm$0.00& -0.29$\pm$0.01& -0.17$\pm$0.01&0.20\\
\hline
FS029    &  0.46$\pm$0.00&  0.48$\pm$0.01& -0.09$\pm$0.02&0.31&  0.04$\pm$0.01& -0.28$\pm$0.02& -0.19$\pm$0.02&0.14\\
FS075    &  0.60$\pm$0.00&  0.67$\pm$0.03&  0.13$\pm$0.05&0.36&  0.08$\pm$0.00& -0.30$\pm$0.03& -0.33$\pm$0.03&0.11\\
FS076    &  0.79$\pm$0.00&  0.80$\pm$0.03&  0.01$\pm$0.05&0.44&  0.15$\pm$0.00& -0.17$\pm$0.03& -0.34$\pm$0.04&0.19\\
FS131    &  0.62$\pm$0.01&  0.57$\pm$0.03& -0.12$\pm$0.04&0.40&  0.03$\pm$0.01& -0.28$\pm$0.03& -0.21$\pm$0.03&0.16\\
DW0001   &  0.54$\pm$0.01&  0.69$\pm$0.01&  0.04$\pm$0.02&0.23& -0.23$\pm$0.00& -0.25$\pm$0.02& -0.03$\pm$0.02&0.13\\
FCC043   &  0.49$\pm$0.01&  0.35$\pm$0.01& -0.01$\pm$0.01&0.21& -0.23$\pm$0.01& -0.50$\pm$0.01& -0.38$\pm$0.02&0.24\\
FCC136   &  0.65$\pm$0.01&  0.83$\pm$0.04&  0.16$\pm$0.05&0.31& -0.29$\pm$0.01& -0.65$\pm$0.03& -0.36$\pm$0.04&0.25\\
FCC150   &  0.53$\pm$0.01&  0.62$\pm$0.02&  0.02$\pm$0.04&0.45& -0.38$\pm$0.01& -0.70$\pm$0.02& -0.39$\pm$0.04&0.26\\
FCC204   &  0.35$\pm$0.01&  0.49$\pm$0.01&  0.10$\pm$0.01&0.23&  0.12$\pm$0.00& -0.44$\pm$0.02& -0.40$\pm$0.02&0.19\\
FCC245   &  0.62$\pm$0.01&  0.61$\pm$0.03&  0.30$\pm$0.05&0.43& -0.64$\pm$0.01& -0.59$\pm$0.04& -0.08$\pm$0.04&0.36\\
FCC266   &  0.60$\pm$0.01&  0.57$\pm$0.02&  0.33$\pm$0.04&0.57& -0.48$\pm$0.01& -0.88$\pm$0.03& -0.54$\pm$0.04&0.46\\
FCC288   &  0.58$\pm$0.01&  0.41$\pm$0.01& -0.01$\pm$0.01&0.28& -0.59$\pm$0.01& -0.43$\pm$0.01&  0.08$\pm$0.02&0.23\\
FCC335   & -0.01$\pm$0.01&  0.53$\pm$0.01&  0.30$\pm$0.02&0.15& -0.36$\pm$0.02& -0.52$\pm$0.02& -0.22$\pm$0.04&0.19\\
NGC205 ma&  0.13$\pm$0.01&  1.02$\pm$0.03&  0.25$\pm$0.01&0.22& -0.59$\pm$0.02& -2.04$\pm$0.04& -0.29$\pm$0.02&0.18\\
NGC205 mi&  0.11$\pm$0.00&  0.39$\pm$0.02&  0.22$\pm$0.01&0.23& -0.50$\pm$0.01& -0.70$\pm$0.02& -0.16$\pm$0.03&0.24\\
\hline
NGC404 ma&  0.14$\pm$0.01&  0.42$\pm$0.02&  0.18$\pm$0.02&0.14& -0.23$\pm$0.01& -0.38$\pm$0.05& -0.11$\pm$0.04&0.12\\
NGC404 mi&  0.11$\pm$0.01&  0.33$\pm$0.02&  0.14$\pm$0.02&0.14& -0.18$\pm$0.01& -0.25$\pm$0.07& -0.05$\pm$0.06&0.23\\
FS373    &  0.17$\pm$0.00&  0.50$\pm$0.02&  0.15$\pm$0.02&0.29&  0.08$\pm$0.00& -0.58$\pm$0.03& -0.54$\pm$0.03&0.18\\
DW0002   &  0.23$\pm$0.00&  0.37$\pm$0.01&  0.20$\pm$0.01&0.30& -0.92$\pm$0.01& -0.37$\pm$0.02&  0.36$\pm$0.03&0.28\\
FCC046   &  0.03$\pm$0.00&  0.22$\pm$0.01&  0.22$\pm$0.01&0.20& -1.04$\pm$0.01& -0.71$\pm$0.03&  0.21$\pm$0.04&0.29\\
FCC207   &  0.10$\pm$0.00&  0.66$\pm$0.02&  0.45$\pm$0.04&0.52& -0.59$\pm$0.02& -0.74$\pm$0.03& -0.20$\pm$0.04&0.39\\
\hline

    \end{tabular}
  \end{minipage}
\end{table*}

\subsection{Radial profiles}\label{subsec:radprof}

To build the profiles, independent 1D-spectra were generated by adding
together adjacent observed spectra until S/N~=~40 was reached. The
process was started from the central spectrum and extended
successively on both sides. Typically, the central spectrum has
S/N~$\approx$~80, and therefore the central region is not binned.  In
the external parts, several spectra are grouped. If the bin width
reaches r$_{\rm max}$/r$_{\rm min}$~=~1.4, where r$_{\rm min}$ and
r$_{\rm max}$ are its inner and outer limits, the target S/N is
decreased.  The profile is continued until S/N$> 5$.  The central
spectrum is defined as the peak of wavelength-integrated flux, except
for FCC\,21 = NGC\,1316 which is affected by a strong dust lane. In
this case, we used the centre of symmetry of the rotation profile
(interestingly, it is not the centre of symmetry of the velocity
dispersion profile).

When emission lines were present, H$_\beta$, H$_\gamma$ and [O{\sc
    iii}]$~\lambda5007$\,{\AA}, they were fitted with Gaussians together
with all the other free parameters.

The error bars, estimated from the noise and quality of the fit, give the
independent errors on the age and on the metallicity and do not take
into account the age-metallicity degeneracy (a metallicity offset is
compensated by a bias of the age): some points are deviating from a
smooth profile by more than the error bar, but they lie on the same
age-metallicity degeneracy line (or actually age-metallicity-$\sigma$
surface). The error bars determined from Monte-Carlo simulations
are about twice larger, as they take also into account the degeneracies.

The radial profiles are shown in
Figs.~\ref{fig:radprof1} to \ref{fig:radprof_ngc404}.
The corresponding measurements are available in electronic form
at the Centre de Donn\'ees astronomiques de Strasbourg\footnote{CDS, \url{http://vizier.u-strasbg.fr/}}.
For the \citeauthor{paperI} sample, the analysis is essentially identical
to the one presented in \citet{paperII}, the only difference is that
we presently use the official release of \ulyss, while we were previously using a
development version (the cleaning algorithm, to reject the spikes was improved).
The profiles are indiscernible from the published ones, and we do not 
reproduce them here.

The first point to notice is the high degree of symmetry of the
profiles, which is showing that independent, but presumably similar,
spectra are leading to consistent results (though this does not tell
if those parameters are not biased). When two spectra where analysed,
for example for the eight Virgo galaxies observed with GMOS, the two
profiles are also in agreement (they are represented with different
colours on Fig.~\ref{fig:radprof1}).  Moreover, the profiles derived
using different instruments, resolution and wavelength ranges display
a remarkable consistency.  This can be seen on
Figs.~\ref{fig:radprof_fors1} and \ref{fig:radprof_vaku} where the
analysis of the GMOS data is over-plotted.


\subsection{Population gradients}
\label{subsec:popgrad}

In this section, we describe the computation of the gradients
of the population parameters and we present the results. 

\begin{figure*}
\includegraphics[width=0.45\textwidth]{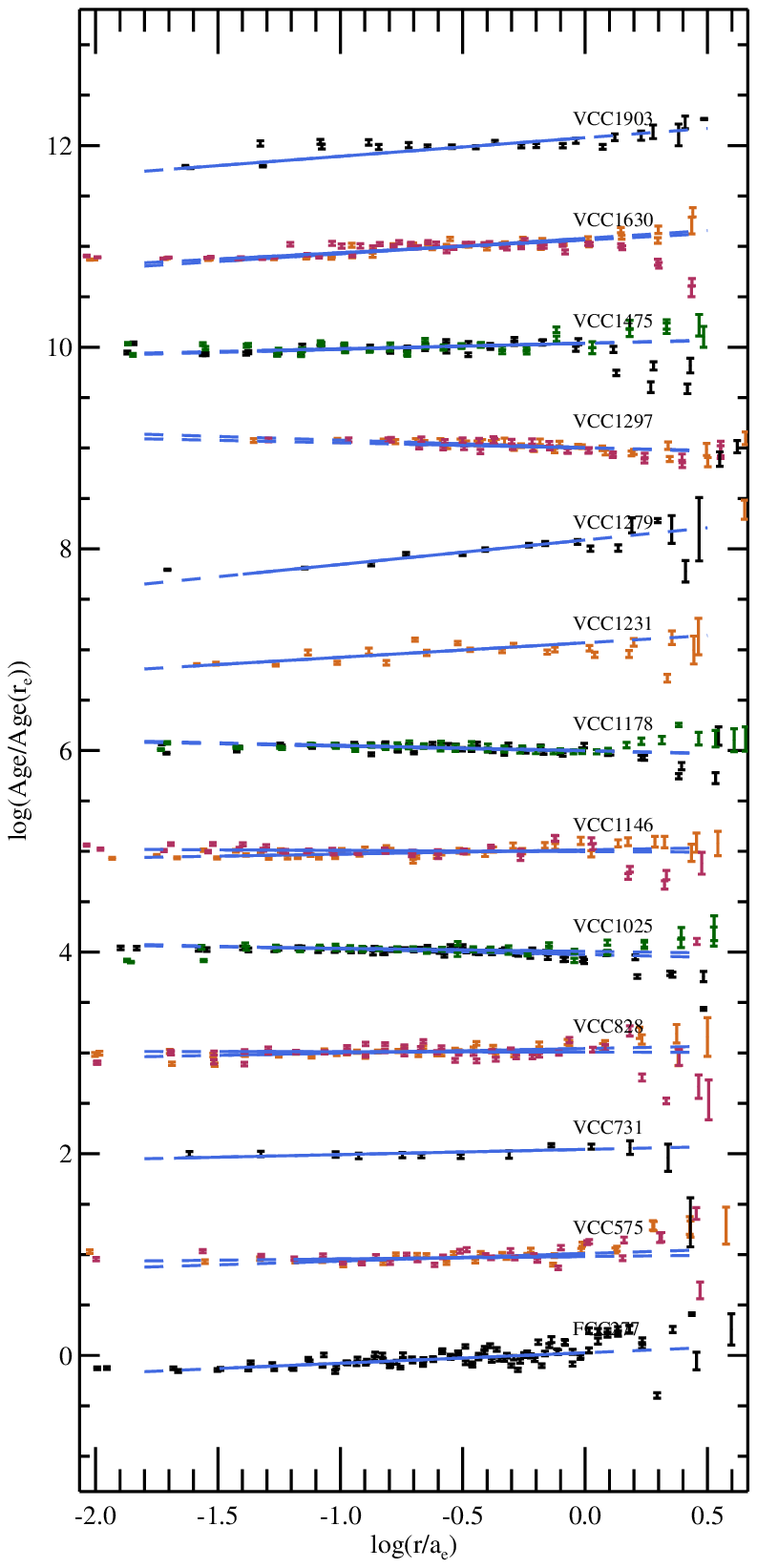}
\includegraphics[width=0.45\textwidth]{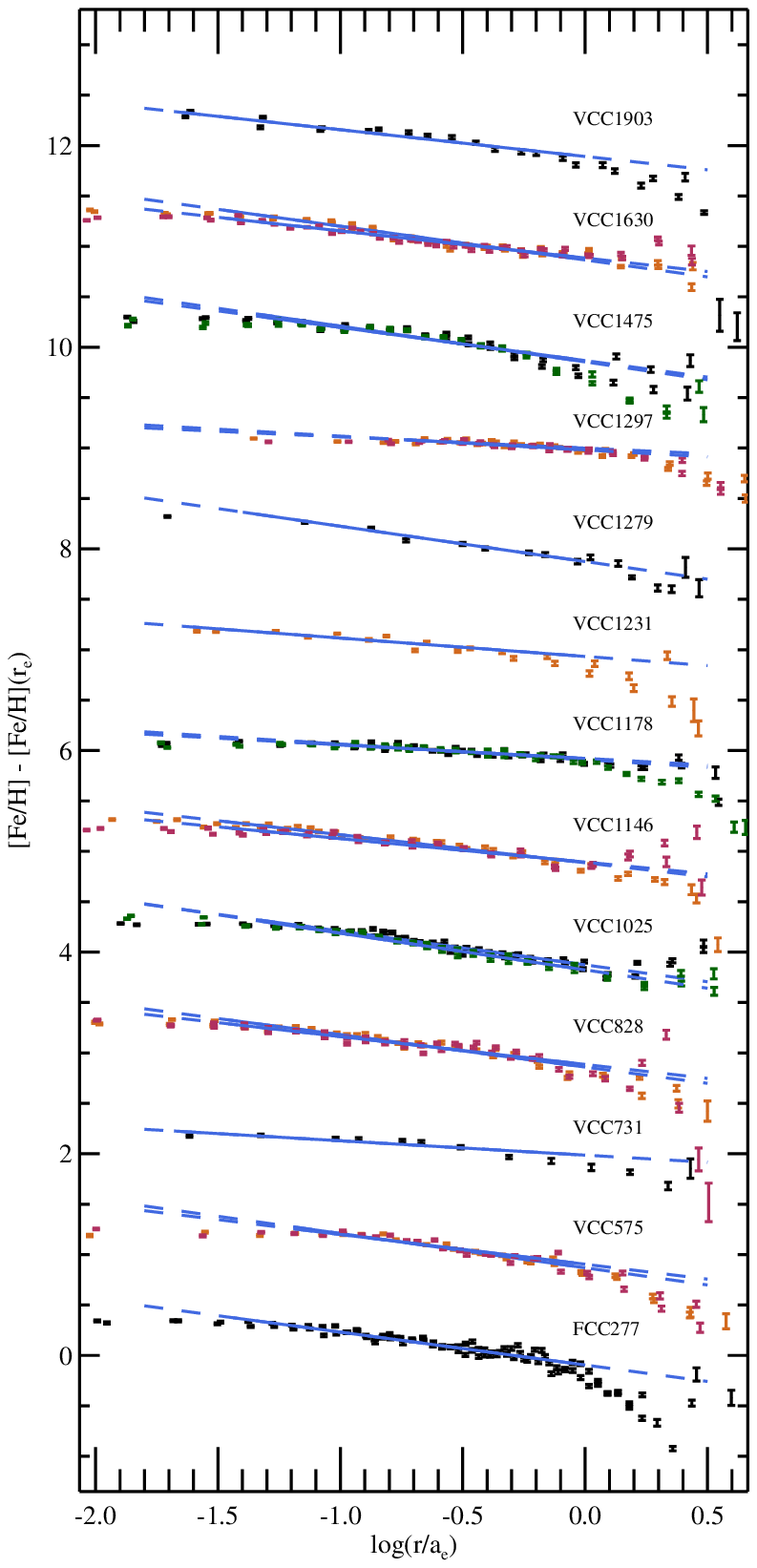}
\caption{Radial gradients of ages and metallicity for the elliptical
  galaxies.  The age profiles are shown on the left panel, and the
  [{\rm Fe/H}] ones on the right one.  The abscissa are the radius
  normalised to the semi-major effective radius (\aeff).  The ordinates are
  the decimal logarithm of age (left) and metallicity (right)
  normalised to the effective aperture extraction.  Each profile is
  labelled by the galaxy name and shifted by 1\,dex.  The blue lines
  are the linear fits of the gradients, and the continuous sections
  represent the region used for the fit (1\,arcsec to 1\,\aeff),  while
  the dashed parts extrapolate the fits to small and large values.  }
\hskip 0cm
\label{fig:grad1}
\end{figure*}

\begin{figure*}
\includegraphics[width=0.45\textwidth]{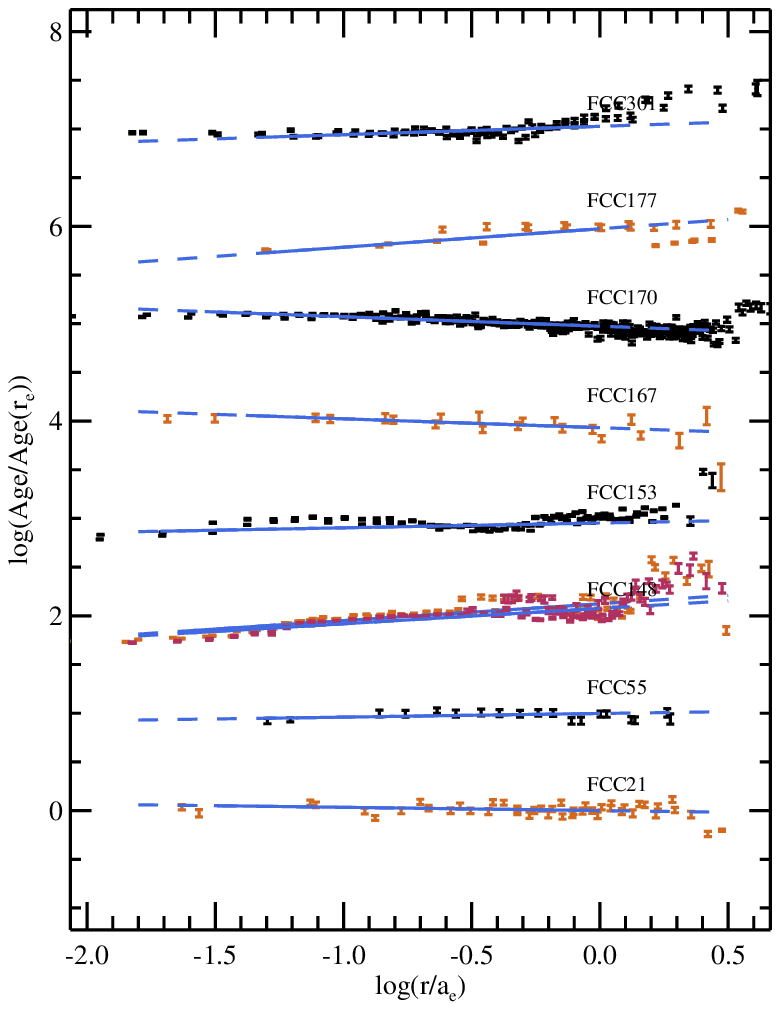}
\includegraphics[width=0.45\textwidth]{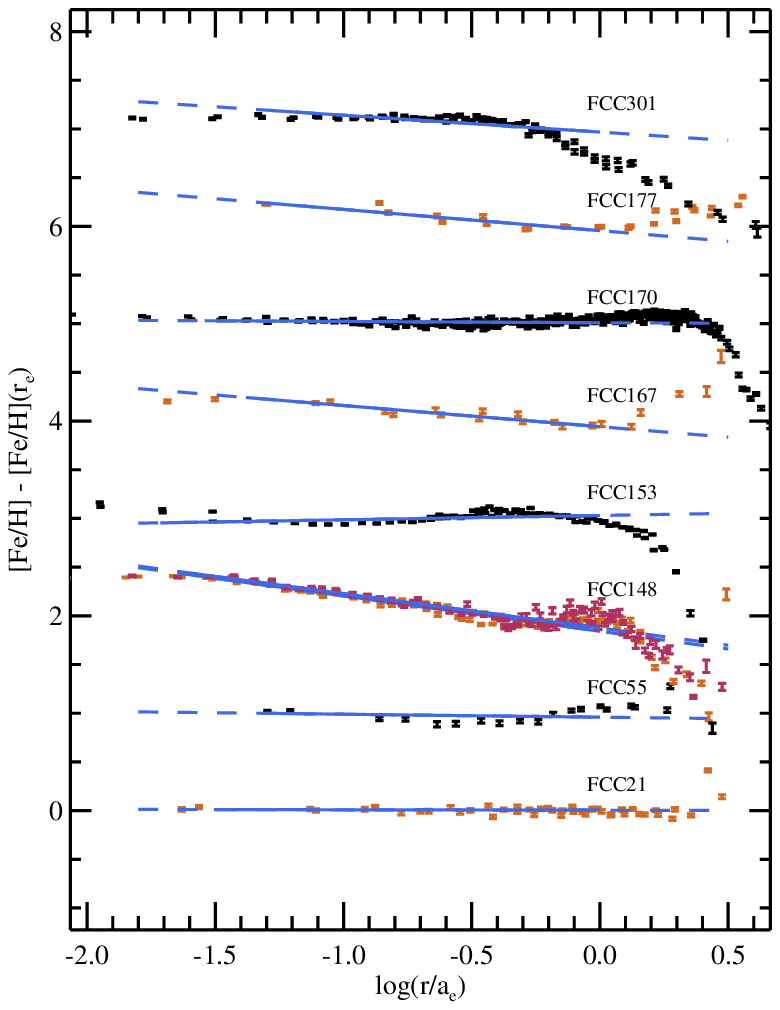}
\caption{Radial gradients of ages and metallicity for the S0
  galaxies. Same conventions as for Fig.~\ref{fig:grad1}}
\hskip 0cm
\label{fig:grad2}
\end{figure*}

\begin{figure*}
\includegraphics[width=0.45\textwidth]{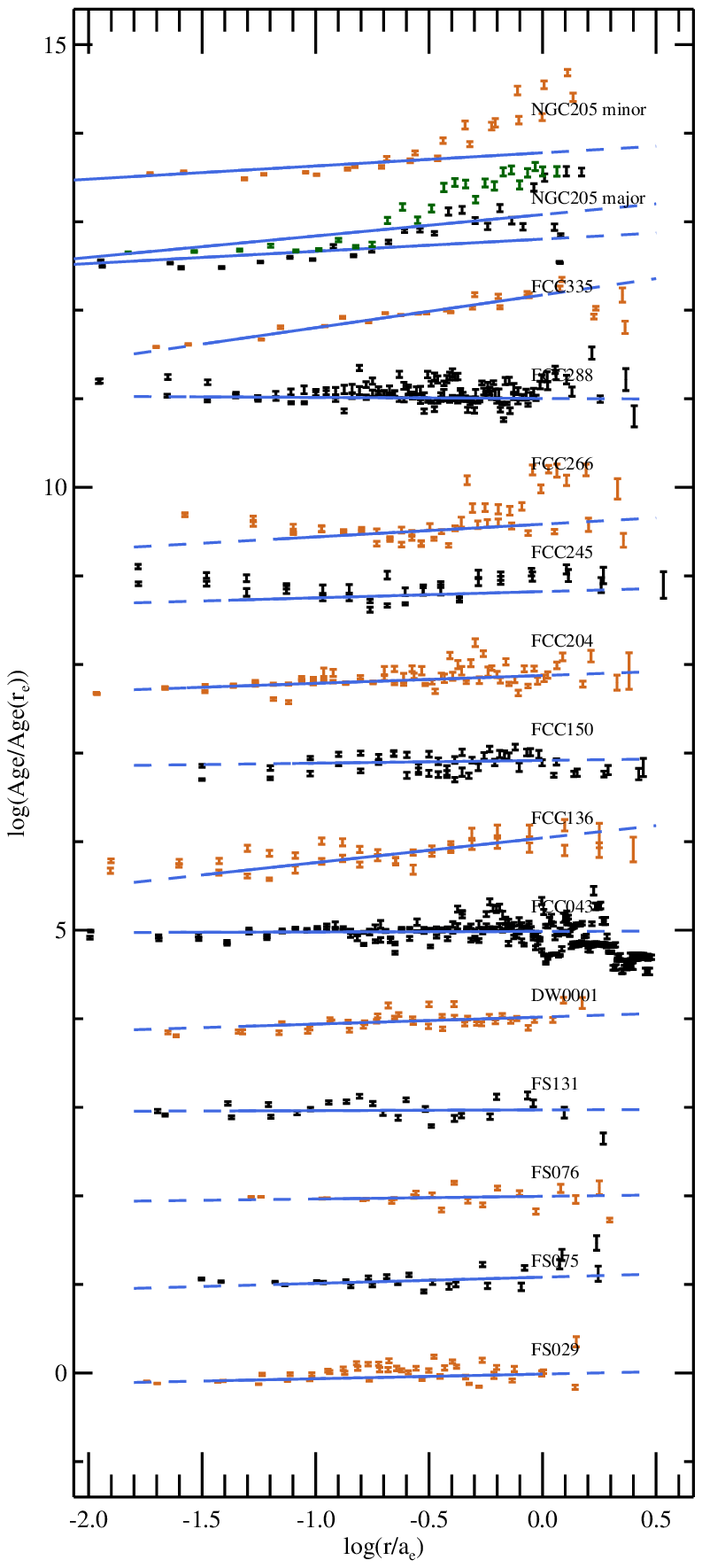}
\includegraphics[width=0.45\textwidth]{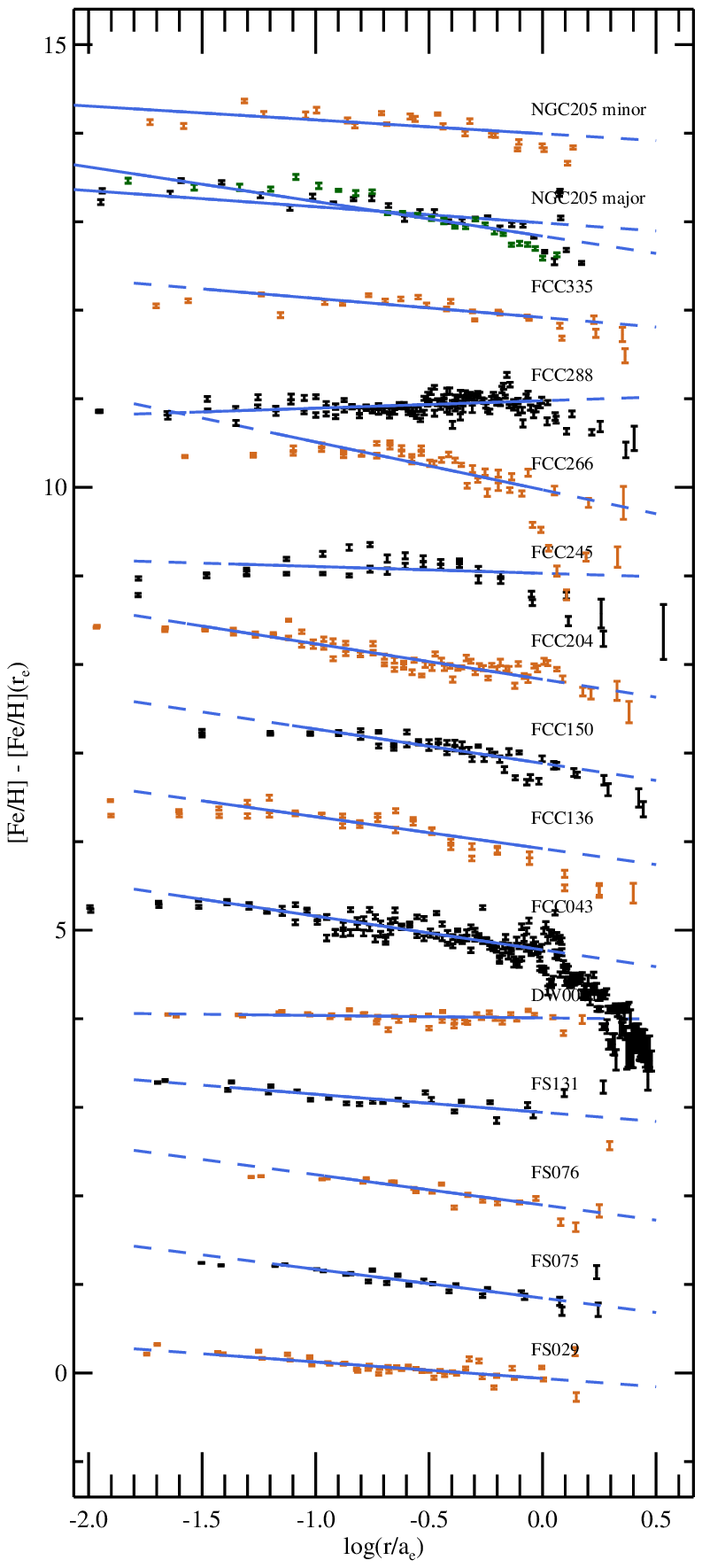}
\caption{Radial gradients of ages and metallicity for the dE/dS0
  galaxies. Same conventions as for Fig.~\ref{fig:grad1}}
\hskip 0cm
\label{fig:grad3}
\end{figure*}

\begin{figure*}
\includegraphics[width=0.45\textwidth]{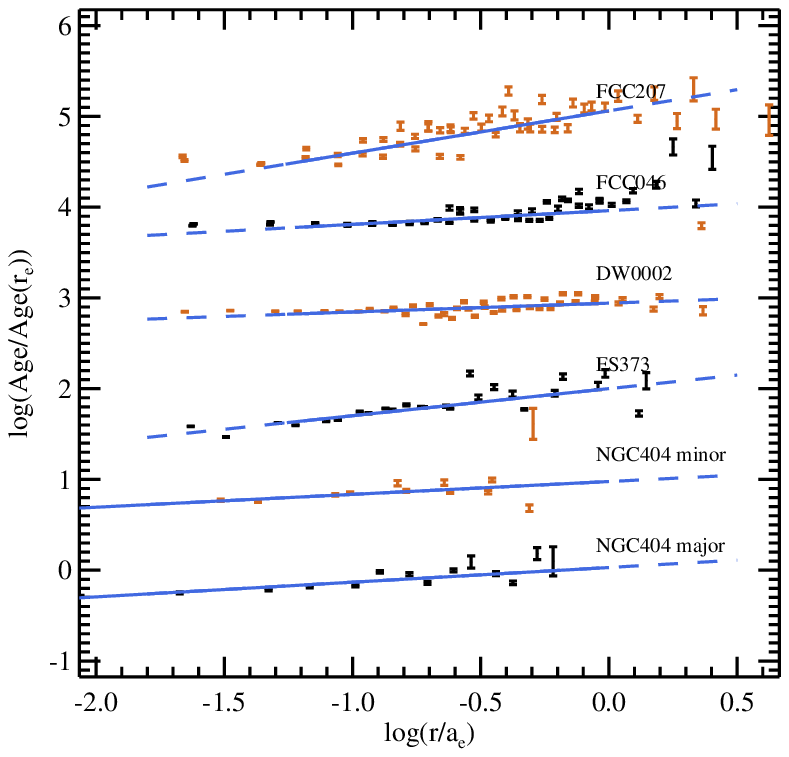}
\includegraphics[width=0.45\textwidth]{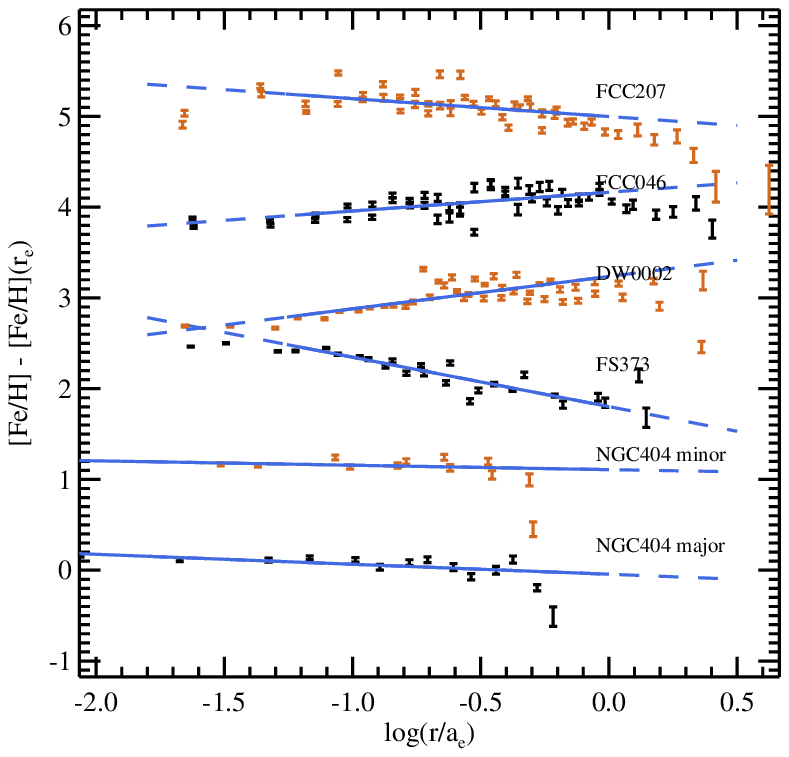}
\caption{Radial gradients of ages and metallicity for the TTD
  galaxies. Same conventions as for Fig.~\ref{fig:grad1}}
\hskip 0cm
\label{fig:grad4}
\end{figure*}

We fitted the gradients assuming 
linear relations between log ${\rm Age}$ or [{\rm Fe/H}] and log(r/\reff):\\
\begin{equation}
\log {\rm Age}(r/r_{\rm eff}) = \log {\rm Age}_{\rm eff} + \nabla_{{\rm Age}}
\log(r/r_{\rm eff})
\end{equation}
and, 
\begin{equation}
[{\rm Fe/H}](r/r_{\rm eff}) = [{\rm Fe/H}]_{\rm eff} + \nabla_{[{\rm Fe/H}]}
\log(r/r_{\rm eff}),
\end{equation}
where $\nabla_{\rm Age}$ and $\nabla_{[{\rm Fe/H}]}$ are the fitted
population gradients and $\log {\rm Age}_{\rm eff}$ and $[{\rm
    Fe/H}]_{\rm eff}$ the fitted age and metallicity at the effective
isophote ($r/r_{\rm eff}$ is the galactocentric distance along the
considered axis, i. e.. along the major axis, it is $a/a_{\rm eff}$).
We used the observed points between a radius of 0.5~arcsec and the
effective isophote, weighted with the observational error.  The fits
of the profiles are presented in Fig.~\ref{fig:grad1} to
\ref{fig:grad4} and the coefficients in Table~\ref{table:result}.

As noted by others \cite[e. g. ][]{baes2007} the profiles are not always well
fitted with these power-laws. This is reflected by large values
of $\chi^2$. In Table~\ref{table:result}, we list also the 
uncertainty on the metallicity gradient multiplied by  $\chi$
(i.e. the formal error obtained if the estimated measurements errors
are rescaled in order to get $\chi^2~=~1$). This gives an idea of
the range of values we would find if we play with the radial limits or
if we sample differently the radius of the galaxy.

\begin{table*}
  \centering
  \begin{minipage}{180mm}
    \caption{Mean characteristics of the sub-samples. \label{table:mean}
The first column identifies the sub-sample and the second is its number
of objects.
The columns labelled (1) give the mean and those labelled (2) the standard deviation.
}
    \begin{tabular}{|l|ccccccccccccc} \hline

Sample& n& \multicolumn{2}{c}{$M_B$}& \multicolumn{2}{c}{$\sigma_0$, \kms} & \multicolumn{2}{c}{$\log {\rm Age}_0$, Myr}& \multicolumn{2}{c}{$[{\rm Fe/H}]_0$} & \multicolumn{2}{c}{$\nabla_{{\rm Age}}$}& \multicolumn{2}{c}{$\nabla_{[{\rm Fe/H}]}$}\\
& &(1)&(2)&(1)&(2)&(1)&(2)&(1)&(2)&(1)&(2)&(1)&(2)\\
\hline
Our full sample & 40&-17.66& 1.52&  93& 59&  3.62& 0.34& -0.15& 0.31&  0.09& 0.13& -0.21& 0.18\\
E            & 13&-18.50& 1.09& 139& 51&  3.94& 0.16& -0.02& 0.14&  0.06& 0.09& -0.26& 0.08\\
S0           &  8&-19.01& 1.56& 113& 70&  3.63& 0.33&  0.05& 0.18&  0.01& 0.11& -0.12& 0.13\\
dE/dS0       & 14&-16.57& 0.63&  53& 14&  3.50& 0.20& -0.24& 0.27&  0.10& 0.14& -0.26& 0.16\\
TTD          &  5&-16.34& 0.82&  53& 15&  3.13& 0.07& -0.54& 0.42&  0.24& 0.11& -0.06& 0.32\\
dSph (literature)& 17&-11.69& 1.18&  12&  7&  3.42& 0.19& -1.39& 0.19&      &     & -0.10& 0.12\\

\hline
    \end{tabular}
  \end{minipage}
\end{table*}

The mean gradients, for the whole sample of 40 galaxies, and
within our four classes are given in Table~\ref{table:mean}.

\begin{table}
\label{table:grad}
\end{table}

\begin{figure}
\includegraphics[width=0.38\textwidth]{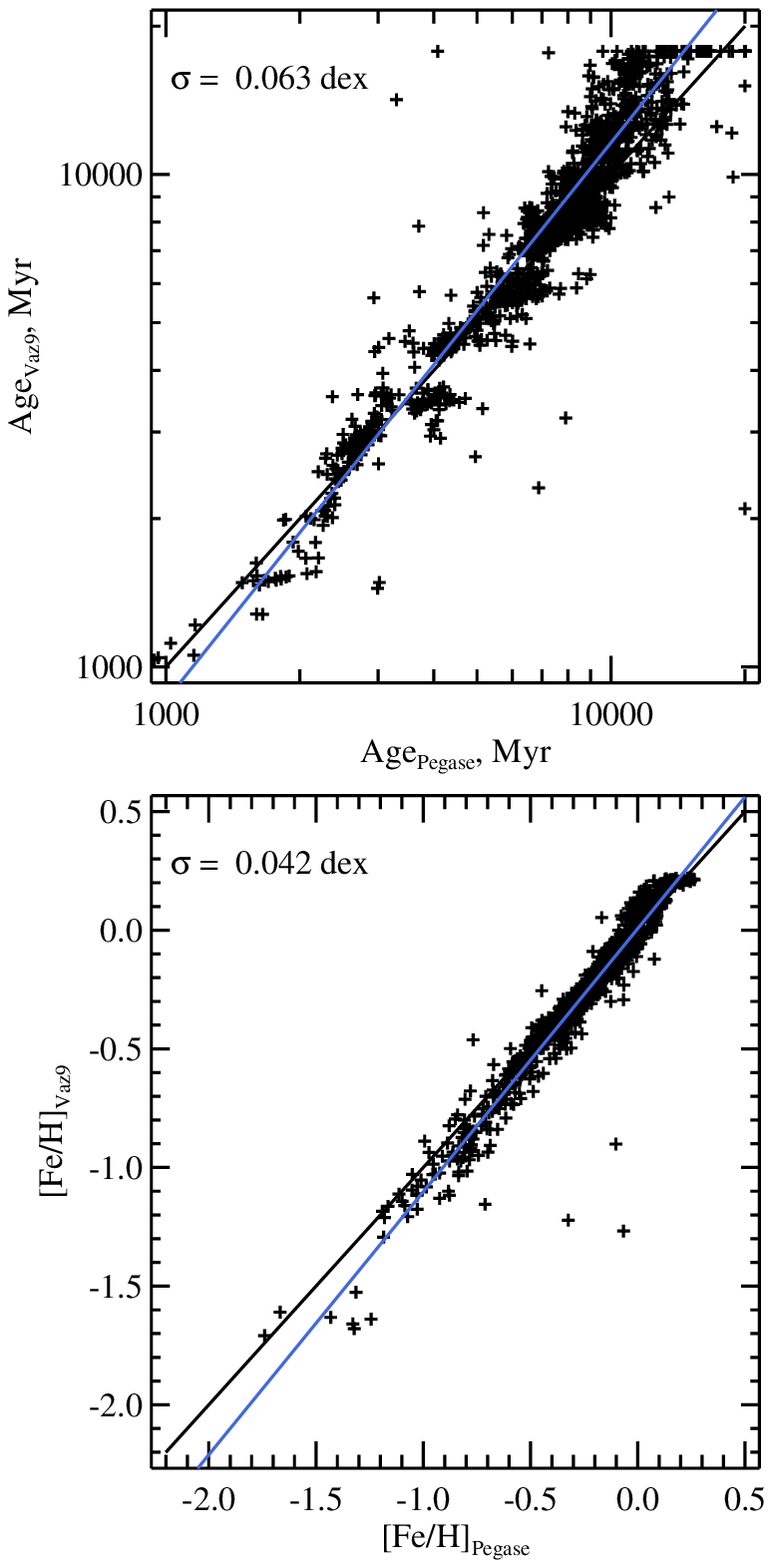}
\caption{Comparison between the SSP analysis with the Vazdekis 9
  models and with the Pegase.HR models based on the ELODIE 3.1 library for the GMOS data. 
  The residual
  dispersions, $\sigma$, labelled on each panel are the dispersions of
  log({\rm Age}) and [{\rm Fe/H}] in dex.  The blue line is a linear fit to the
  data, and the black one is the one-to-one ratio}
\label{fig:compa_vaz9}
\end{figure}
\begin{figure}
\includegraphics[width=0.38\textwidth]{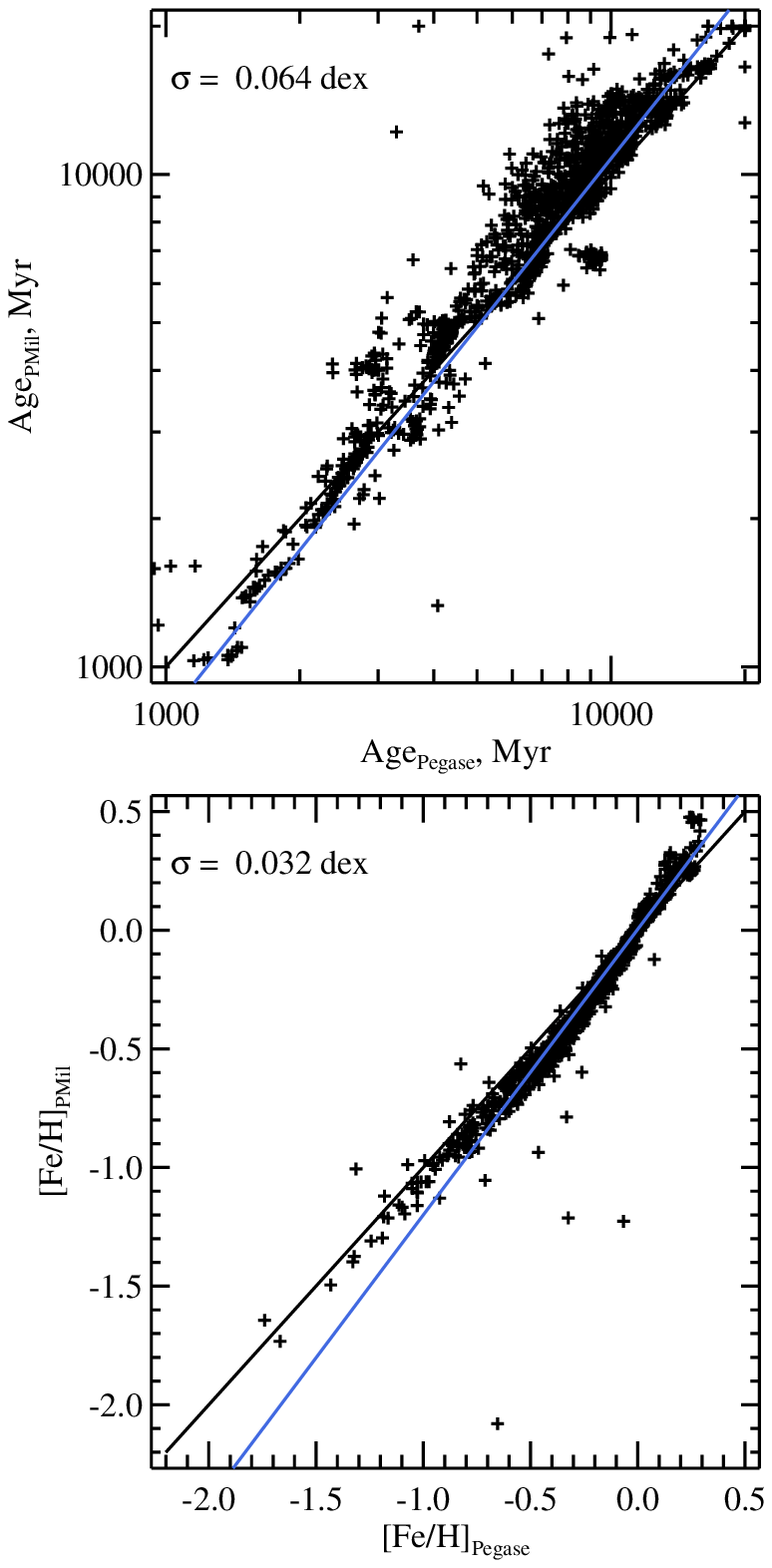}
\caption{Comparison between the SSP analysis with Pegase.HR models
  based on the MILES library (preliminary unpublished version) and the
  ELODIE.3.1 library for the GMOS data. The lines are the same as in Fig.~\ref{fig:compa_vaz9}}
\label{fig:compa_pmil}
\end{figure}

\subsection{Verification}

The SSP-equivalent parameters are sensitive to the uncertainties in the
data reduction and in the population models.  This may not only result
in random errors, but also in systematics.  In \citet{gradan}, we
investigated possible sources of error due to the data reduction and,
in particular, to the sky subtraction. We concluded that this cannot
significantly affect our determination of the gradients.  It was also
established that the fitted parameters do not depend on S/N, at least
down to S/N$\approx 5$ \citep{kol2007,gradan}.

In the present work, we compare the results obtained using our data
reduction and the one from \citet{bedregal2006,bedregal2011}, 
kindly provided to us by the author. We
found a bias of the metallicities associated with the (lack of)
correction of a defect of the CCD amplifier (see
Sect.\,\ref{subsec:bedregal}), but this affects only the external
regions, generally outward of 1\,\reff.  Although this would not
dramatically impact the determinations of the gradients, we rejected
these points.

We tested that our results are independent from whether we mask or fit
(with Gaussians) the emission lines.  In all the cases where
significant emission lines are seen, we fitted H$_\beta$ (and higher
order Balmer lines) and [O{\sc iii}]$~\lambda5007$\,{\AA}.  We also
checked the limitations due to the reliability of the models.  We have
shown that models using sparsely populated empirical libraries can lead
to bias in the population parameters recovery \citep{koleva2008a}, but
the rich and large modern libraries considered here are free of such
effects.  To check the impact of the different physics involved in
different stellar population models, we repeated all the analysis with
other models.  We used the recent models of
\citet[][version\,9]{vazmod} based on the MILES library \citep{miles},
and the unpublished Pegase.HR models also build with MILES.  The
comparisons between the three sets of models are shown in
Fig.\ref{fig:compa_vaz9} and \ref{fig:compa_pmil}. All the models give
consistent results, though some small trends are visible (see 
Prugniel et al. in prep.).

Another caveat is hidden in the interpolation in the meshes of the
stellar population grid. Often, all the points of a profile are
between the same two metallicities of the model grid and sometimes
even between the same two age nodes. In case of \ulyss, the
interpolation is computed with splines, where the variables are
log(age) and [{\rm Fe/H}]. Its accuracy was verified in
\citet{koleva2008a} where the spectra interpolated at metallicities
between those of the grid were compared with those computed by
P\'egase.HR using its internal interpolation. Here, we performed
another test. We generated a grid of models, where the solar
metallicity was removed (the neighbouring metallicities are -0.4 and
+0.4\,dex). We used these models as a grid in \ulyss, to determine the
population parameters of solar metallicity models. We recovered the
metallicity with a precision of about 0.03\,dex (4\,per\,cents of the
mesh size) and the ages to about 0.1 dex. Therefore, the precision of
the interpolation shall not affect the gradients by more than those
values.

\section{Comparison with the literature}\label{sec:comp}

\subsection{Comparison with Spolaor et al.}
\label{subsec:compspo}

One of the goals of this study is to shed some light on the
apparent discrepancy between \citet{spo2009, spo2010} and our own previous
measurements \citep{paperII}. The former study finds a tight
relation between the luminosity or velocity dispersion and the
metallicity gradients, while the latter does not find any relation.
These two papers are based on different samples and use 
different methods, and whether the cause of the discrepancy 
resides in the sample selection, bias in the method or data 
processing could not be clarified before.

A tight relation between the mass and the metallicity gradients is not
immediately expected. First, it could not be anticipated from
any earlier studies, and, second, since the sample is a mix of 
morphologies, it
would imply a new and very general scaling relation. In addition, the
evolution, through environmental effects, should increase the spread,
in particular for the low-mass galaxies. 


We re-analysed the data used in \citet{spo2009},
and we compare the profiles and the gradients of age and 
metallicity obtained in both studies.

\subsubsection{Comparison of the profiles}

We digitised the profiles from figs. 1, 2, and 3 of \citet{spo2010},
and we checked that this process did not introduce significant errors.  We
converted the total metallicity into [{\rm Fe/H}] as: $[{\rm Fe/H}] =
[Z/H] - 0.94\,[\alpha/Fe]$ \citep{thomas03}, consistently with the
models used in the source paper.  We over-plotted these age and
metallicity profiles in Fig.~\ref{fig:radprof1}.

We first remark that our profiles are extending to larger
galactocentric distances. The reason is certainly that our method is
more sensitive (as it uses all the information in the signal).
Looking more closely, we see that while our Virgo profiles are going
typically 1.5 times farther, those of the Fornax galaxies go twice
farther. This difference between the two sub-samples may come from
the fact that in the case of Fornax we co-added the individual
exposures, while we kept them separated for Virgo (the wavelength
setups were different and we wished to avoid an additional step in
the reduction and one more source of uncertainty).

The agreement between the metallicity profiles of the Virgo galaxies
is remarkably good, despite of the differences in the data processing,
analysis and in stellar population models. The ages from
\citet{spo2010} are in general older by $\sim$0.1\,dex 
(the maximum discrepancy is $\sim$0.2\,dex).  This reflects
physical differences in the population models, and in particular in
the isochrones of the stellar evolution.  These systematics will not
be discussed further here, and we note that the shape of the profiles
are similar (i. e. the relative ages are in agreement).

For the Fornax galaxies the agreement in the age and metallicity
profiles is satisfactory, but we note deviations that exceed the error
bars for almost all the galaxies.  The clearest case is FCC301, where
\citet{spo2010} give a flat metallicity profile while our analysis
reveal a gradient extending up to 3\,\aeff.

The reasons for these discrepancies are not clear. We may suspect the
fact that \citeauthor{spo2010} measured the total metallicity and the
$\alpha$-elements enhancement, while we measure only the Fe
metallicity.  However, for the Virgo galaxies, for which
the two analyses agree, the populations are often $\alpha$-enhanced,
while they are not in the Fornax sub-sample.  Still, because of degeneracies,
our [{\rm Fe/H}] could be biased. Nevertheless, the degeneracy between
the [{\rm Fe/H}] and [$\alpha$/Fe] was
examined in \citet[][ see also Prugniel et al., in prep]{koleva08e}
who concluded that they are fairly decoupled.
In addition, in the case of FCC170, \citeauthor{spo2010} do not detect
any enhancement, while our central fit clearly shows one
(Fig. B2), confirmed by \citet{bedregal2008,arrigoni2010}.  

On the basis of the good consistency for the Virgo galaxies,
notwithstanding an age offset, and of the agreement between our analyses
of the GMOS and FORS data for the Fornax galaxies, we tend
to be satisfied by this comparison and trust our measurements.

%

\begin{figure}
\includegraphics[width=0.45\textwidth]{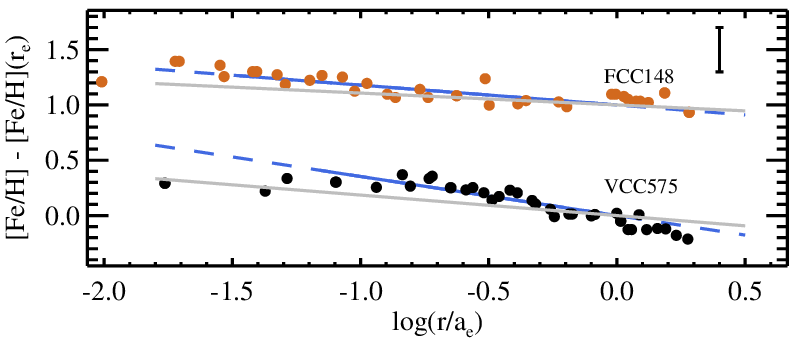}
\caption{Comparison of the fit to the metallicity gradients.
The data points are the measurements from \citet{spo2010}. The grey
lines are their linear fit and the blue lines ours.
The dashed parts of the blue lines correspond to the excluded points,
inside the 0.5\,arcsec radius and outside $a_{\rm eff}$.
The error bar is the median error of the derived parameters.
}
\label{fig:compa_gspo}
\end{figure}

\subsubsection{Comparison of the gradients}

\citeauthor{spo2010} used the same definition as us to 
measure the gradients. Power-law profiles are fitted between 
the seeing disc and the effective radius.

However, the comparison between our gradients and their values
is not consistent, even for the Virgo
sub-sample for which the profiles agree. A quick check for one of the
discrepant Virgo galaxies, VCC~575 reveals the possible
explanation. The hand-fit of the profile on their figure 2,
considering the points between the seeing disc (marked on the graphic)
and the effective radius indicate a total metallicity gradient of
about -0.3, while their tabulated result is -0.11.  The tabulated
value may possibly be reproduced if the central points, within the
seeing disc, are included. To check our initial guess, that the
  discrepancy may result from the erroneous inclusion of the central
  region, we performed unweighted linear fit of all the
  \citeauthor{spo2010} profiles.  We present in
  Fig.\ref{fig:compa_gspo}  the profiles for FCC148 and VCC575,
  together with their and our fits to their data, using the radial
  range between the seeing disc and the effective semi-major axis.
  Our new fits are generally in good agreement with the fits of our
  own profiles. However, another hypothesis shall be examined:
  \citeauthor{spo2010} fits are weighed by the errors on
  individual points, while our test is based on unweighted fits (the
  individual errors in  \citeauthor{spo2010} are not
  available). The data were binned together to reach a minimal S/N/\AA\ of
  35 (near 5200 \AA), and while the errors bars should be nearly
  constant for most of the body of the galaxies, the core region have
  in some cases smaller errors. Nevertheless, this over-weighted
  region is in general within the seeing disc.  Therefore, the
  variation of the weight cannot explain the divergence. A close look
  at the profile of  FCC~148 (Fig.\ref{fig:compa_gspo}) shows that it
  is not possible to reproduce the fit of \citeauthor{spo2010}
  simply by adapting the weight of each point.  Thus, we tend to
  believe that the authors did not exclude the points in the seeing
  disc.   The exclusion of the central region has two strong
justifications:~{\em (i)} the observed profiles are not resolved
within seeing disc and are expected to flatten and {\em (ii)} the
power-law representation does not physically hold at small radii
(i. e. it diverges). Some of the gradients derived in
  \citeauthor{spo2010} may therefore be erroneous. 


%
%
%

\subsection{Comparison with Sanchez-Blazquez et al.}

\citet{sanchez-blazquez2007} analysed KECK observations of 11 early type galaxies
with  $-17 \,{\rm mag}< M_B < -22.4$~mag.
The data have a wavelength coverage of 3110 -- 5617~\AA{} and
a resolution of 8~\AA.  They measured 19 line 
strength indices up to 2\,\reff. As in \citet{spo2009,spo2010}, they
used the $\chi^2$ minimisation technique described in \citet{proctor2002}
and two different population models, a preliminary version of \citet{vazmod}
and  \citet[][TMB]{thomas03}.

We have four galaxies in common with this sample: NGC~4387 = VCC~828,
NGC~4458 = VCC~1146, NGC~4464 = VCC~1178 and NGC~4551 = VCC~1630.  The
comparison between the gradients given by \citet{spo2010},
\citet{sanchez-blazquez2007} and our measurements is given in
Table~\ref{table:compa_san}. For \citet{sanchez-blazquez2007}, 
we used the 19 indices
and TMB models to closely match the analyses made by \citet{spo2010}.
To compare the [{\rm Z/H}] measurements with our of $\nabla_{{\rm Fe/H}}$ we
use the [{\rm Fe/H}] = [{\rm Z/H}] - 0.94 [$\alpha/Fe$].  All of our
measurements are consistent with those from \citet{sanchez-blazquez2007} within the error bars.

\begin{table}
\centering
\begin{minipage}{\columnwidth}
\caption{Comparison between \citet{sanchez-blazquez2007}, \citet{spo2010} and our data.
\label{table:compa_san}
}
\begin{tabular}{lccc} \hline 
\multirow{2}{*}{galaxy} & \multicolumn{3}{c}{$\nabla$[{\rm Fe/H}], dex/\reff} \\
& Sanchez-Blazquez & Spolaor & This work  \\
\hline              
VCC0828  & -0.200$\pm$0.089 & -0.266$\pm$0.039 & -0.30$\pm$0.02 \\
VCC1146  & -0.224$\pm$0.113 & -0.154$\pm$0.039 & -0.26$\pm$0.02 \\
VCC1178  & -0.238$\pm$0.065 & -0.154$\pm$0.039 & -0.14$\pm$0.02 \\
VCC1630  & -0.334$\pm$0.148 & -0.269$\pm$0.039 & -0.30$\pm$0.02 \\
\hline
\end{tabular}
\end{minipage}
\end{table}

\subsection{Comparison with Bedregal et al.}

Five of the six Fornax galaxies observed with GMOS 
were observed by \citet{bedregal2006} with FORS, with higher
spectral resolution, but with lower signal-to-noise ratio. 
We processed and analysed the data from both instruments 
and we compared the results (Fig.\,\ref{fig:radprof_fors1}). 
The consistency is more than satisfactory, so we do not expect any
bias in the radial profiles due to the data processing.

\citet{bedregal2008,bedregal2011} also performed stellar
population analyses of their sample using 10 line strengths (with
different combinations to derive {\rm Fe}, [{\rm MgFe}]$^\prime$ and age) 
and \citet{bc03,vazmod} models. Their age and metallicity
measurements have lower precision, since they use Lick indices, but
are in a good agreement with ours \citep{kolbed}.


\section{Discussion}\label{sec:disc}

We analysed the stellar populations of 40 early-type galaxies along
their major axis (and also along the minor axis for NGC~205 and 404) up to
one to three times the effective radius. 
The sample mixes various morphologies of galaxies and our analysis
revealed (or confirmed) a diversity of kinematical and population
profiles. The individual objects are discussed in 
Appendix~\ref{appendix:indgal}.

In this section, we will enlarge the discussion by comparing our
objects to lower mass dwarf spheroidals and to massive early-type
galaxies.  We include low mass dSph/dE galaxies from the M~81 and
Centaurus~A groups \citep{lianou2010,crnojevic2010}.  We estimated the
gradients from the figures of these papers.  The velocity dispersion
for KDG~64 is taken from \citet{sp02} and the age for KDG~61 and 64
from \citet{makarova2010} (the metallicities from the
\citeauthor{makarova2010} and \citeauthor{lianou2010} are in an
acceptable agreement). We add the Local Group dwarf galaxies with
spectroscopic measurements of a significant number of stars: Sextans
\citep{battaglia2010}, Fornax \citep{battaglia2006} and Sculptor
\citep{kirby2009}. We did not include Leo~I as the gradients
determined by \citet{gullieuszik2009} are based on fewer stars. For
Sextans and Sculptor, the effective radius and luminosity were taken
from \citet{irwin1995} and \citet{walker2010}.  The velocity
dispersion of the Local Group galaxies are from HyperLeda.  Finally,
we also consider samples of more massive, early-type, cluster galaxies
presented in \citet{sanchez-blazquez2007} and in \citet{rawle2010}.

We will first locate our sample in the general mass versus metallicity and
age relations.
Then, we will discuss the systematics of the profiles and of the gradients,
and we will examine the distribution of the gradients as a function
of the mass and of other characteristics.

\subsection{Mass-metallicity and mass-age relations}

\begin{figure*}
  \includegraphics{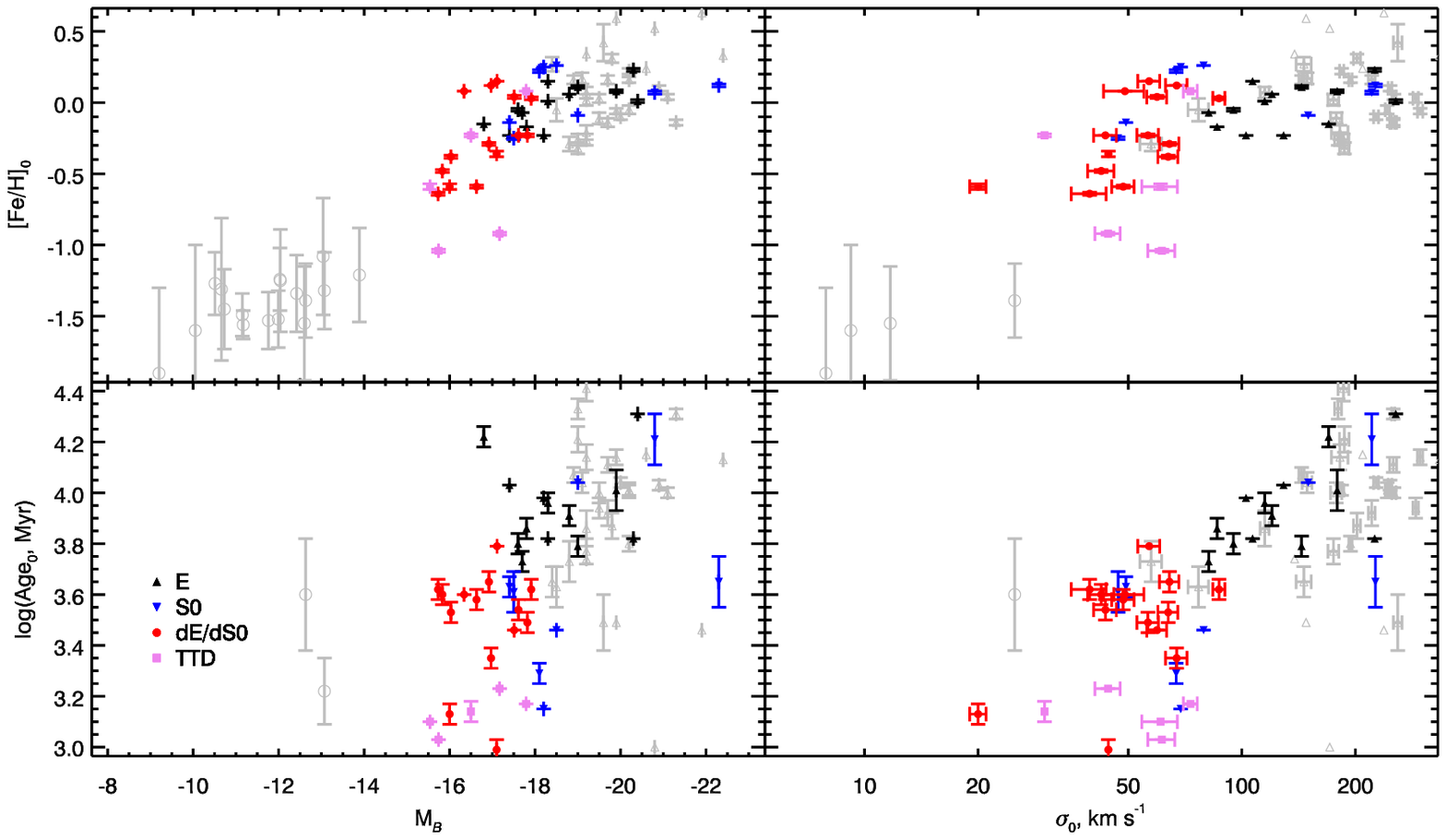}
  \caption{Central age and metallicity vs. absolute magnitude and
    central velocity dispersion.  The different classes of galaxies
    are represented with different colours and symbols according to the
    legend. The error bars are the formal fitting errors.  The grey
    symbols are for the other galaxies compiled from the literature
    (see Sect.~\ref{sec:disc}); for the dSphs, the error bars on the
    metallicity are the cosmic spread.  }
  \label{fig:fcentral}
\end{figure*}

In Fig.~\ref{fig:fcentral}, we show the relations between the central
metallicity and age, and absolute $B$-band luminosity and central
velocity dispersion. There is a clear rising trend of central
metallicity with luminosity and velocity dispersion; an example of the
well-known mass-metallicity relation
\citep[e. g. ][]{henry1999}. Likewise, central age increases with
luminosity and velocity dispersion. This is an example of {\em
  downsizing} \citep{cowie1996}. While these trends appear rather
continuous, we cannot exclude the possibility of a discontinuity, or
at least of a slope change, between the dSphs and the more massive
galaxies. In the luminosity \emph{vs.} metallicity relation, this
discontinuity would be about 0.5\,dex at $M_B \approx -15$~mag, the
dSphs being less metallic than the extrapolation of the relation
derived from the more massive galaxies.  However, the metallicity is
derived using different methods for the two groups. For the dSphs, it is rather 
a mass-weighted value, while for the massive galaxies it is closer 
to luminosity-weighted. The difference qualitatively goes in the 
sense of the observed discontinuity, that may therefore not
be physical.

\subsection{Systematics of the profiles and of the gradients}
\label{subsec:systematics}

The SSP-equivalent profiles, presented in Fig.~\ref{fig:grad1} to
\ref{fig:grad4} in reduced radius (r/\reff) reveal some general trends. 
The mean gradients for each class of galaxies are reported in Table~\ref{table:mean}.

The elliptical galaxies presented in Fig.~\ref{fig:grad1} have mild
and generally positive age gradients ($\langle \nabla \log({\rm
    Age}) \rangle = 0.06$). The only exception is VCC~1297 = NGC~4486B, the
compact elliptical companion of M~87.  Such negative age gradient
$\nabla \log({\rm Age}) = -0.10$ is rare for Es, but may be shared by 
other cEs \citep[][see Appendix~\ref{appendix:indgal}]{chilingarian2010}. The
metallicity gradients of the Es are in general negative, dispersed
around $-0.26\pm0.08$.  This is consistent with other earlier studies
\citep[e.g.][ and the references therein]{sanchez-blazquez2006}.
Both monolithic collapse and hierarchical structure formation (merger
scenario) can explain the diversity of population radial distribution
in Es \citep{pipino2010, white1980}.

In general, the metallicity profiles of our S0  galaxies
(Fig.~\ref{fig:grad2}) are flatter than those of the ellipticals
(dispersed around $\langle  \nabla [{\rm Fe/H}] \rangle \approx
-0.12$). However, we observe  steepening in the external regions
(probably in the disc dominated region). The age gradients are  often
slightly negative (note that when we consider the full errors,
column\,5 of Table\,\ref{table:result}, the negative age gradients are
not detected in 1$\sigma$). We notice that a simple linear relation does
not fit well the age and metallicity profiles of the S0s.

The gradients in age of dEs/dS0s (Fig.~\ref{fig:grad3}) are in general
flat or increasing, but they can be much steeper than the gradients in
the bigger galaxies (Es/S0s) (up to $\nabla_{\rm Age} = +0.3$ for
FCC\,245, FCC\,266 and FCC\,335).  The mean value of the age gradient
in dEs/dS0s is $0.10$ with a cosmic spread of  $0.14$ (standard
deviation), while for Es and S0s the dispersion is $0.10$.  The
metallicity gradient variation is also quite prominent, with a mean
value of $-0.26$ and a dispersion of  $0.16$.  For some of the
galaxies (FCC\,043, FCC\,266, FCC\,288, NGC\,205) we notice the same
metallicity behaviour as for the S0s -  almost flat profile in the
beginning and steeply decreasing after $\sim$1\,\aeff.  In most of the
cases the age is opposite to the metallicity gradient which may
translate into flat colour gradients.  This is confirmed by a simple 
test: we computed a set of colour gradients 
\footnote{in several photometric bands, from \emph{U} to \emph{K} 
using \url{miles.iac.es} web tools.} for FCC\,335 
and found that their values are spread around 0.03\,mag per
1\,\aeff, i.e. of the order of the photometric errors.  

Our five TTDs have strong positive age gradients 
($\langle \nabla_{\rm Age} \rangle = +0.24$ ). The metallicity gradients of 
FCC046 and DW2 are also positive, while for the rest it is flat 
to negative.

We notice that the gradient of the minor axes for NGC\,205 and 404
do not differ from those on their major axes. Thus, these galaxies
seem to have a central (as opposite to axial) symmetry. 

\begin{figure*}
  \includegraphics{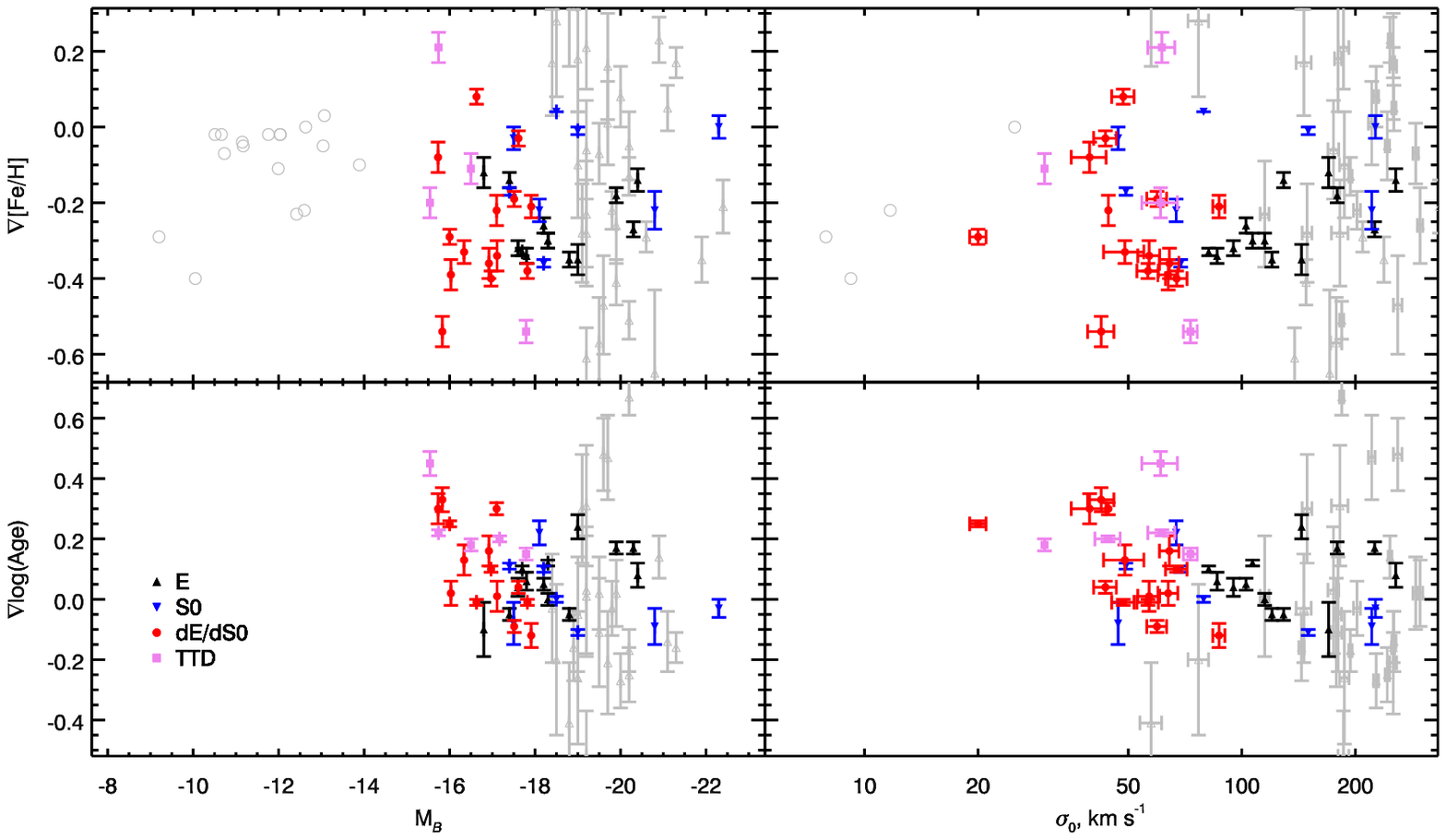}
  \caption{Gradients of age and metallicity vs. absolute magnitude
    and central velocity dispersion.
  The symbols are as in Fig.~\ref{fig:fcentral}.
  The dSph are represented as grey dots without error bars.
  }
  \label{fig:fgradm}
\end{figure*}
\begin{figure}
  \includegraphics{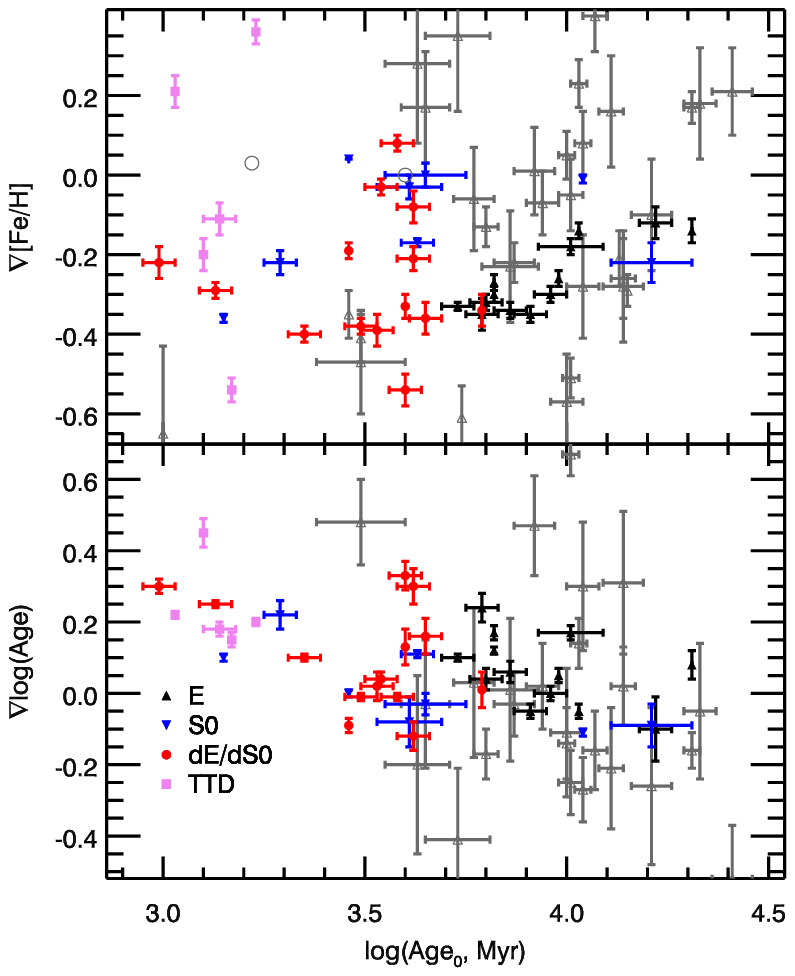}
  \caption{Gradients of age and metallicity vs. central age.
  The symbols are as in Fig.~\ref{fig:fcentral}.
  }
  \label{fig:fgrada}
\end{figure}
\begin{figure}
  \includegraphics{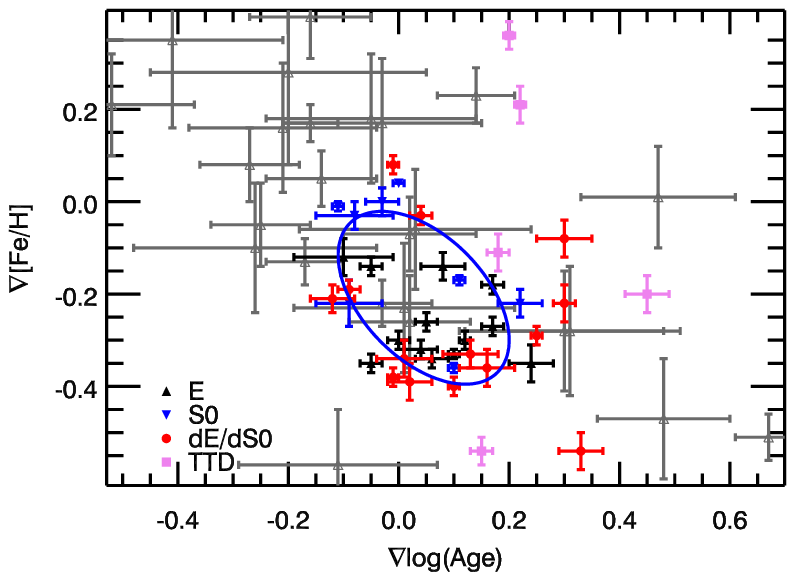}
  \caption{Gradients of age vs. gradients of metallicity.
  The symbols are as in Fig.~\ref{fig:fcentral}.
  The ellipse drawn in blue represent the covariance matrix for
  the E and S0 galaxies.
  }
  \label{fig:fgradgrad}
\end{figure}
\subsection{Gradients versus mass}

In Fig.~\ref{fig:fgradm} we plot the gradients versus the
\emph{B}-band luminosity.  In a wide range of magnitudes ($-22 \,{\rm
  mag}< M_B < -9$~mag) we do not observe any correlation between
metallicity gradient and luminosity (or mass, if we assume constant
mass-to-light ratio). For all masses, the metallicity gradients show a
wide spread in the range $-0.5 < \nabla_{[{\rm Fe/H}]} < +0.1 $. The
most extreme metallicity gradient is observed in FCC~266 ($
\nabla_{[{\rm Fe/H}]} = -0.54$), but, as seen in Fig.~\ref{fig:grad2}
its profile does not follow the power law model, and this large
gradient may not be fully representative.  

For the faintest galaxies in our sample, with $M_B \gtrsim -17$,
the age gradient appears to become more positive with decreasing
luminosity. In other words:~the galaxy centre is increasingly younger
than the outskirts in fainter galaxies. This indicates that star
formation persists over longer timescales in their inner regions than
in their outskirts or that later star-formation events in the smallest
dwarf galaxies occur closer to the galaxy centre \citep{paperII,
  valcke08}. In massive ellipticals, the age gradient shows a wide
spread in the range $-0.2 < \nabla_{{\rm Age}} < +0.4 $. On average,
the galaxies have negative metallicity gradients ($\langle
\nabla_{[{\rm Fe/H}]}\rangle=-0.2$) and positive age gradients
($\langle \nabla_{{\rm Age}}\rangle=0.09$) over the whole mass range.
Replacing $\mathrm{M}_B$ with $\sigma_0$ yields the same conclusion.

The tight relation presented by \citet{spo2009}  is not consistent
  with the strong gradients found in the Local Group dSph galaxies and
  is not confirmed by our present study. To support their empirical
result, \citet{spo2009} mention the concordance with the results of
SPH-simulations with collapse and feedback by \citet{kawata2003}.
These latter authors computed three models of different masses, tuning
the feedback parameters in order to match the colour-luminosity
relation.  The lowest mass model, corresponding to
$\mathrm{M}_B=-17$~mag, displays an homogeneous metallicity. However,
its star formation was quenched too early (at z = 1.7) and the yield
is unrealistically overabundant in $\alpha$~elements ([{\rm Mg/Fe}] =
0.14) to be representative of the observed low-mass galaxies
\citep[see][]{paperII,michielsen08}.   These models have several free
parameters (beside limitations due to the resolution) and the choice
of the feed-back parameters is certainly not a unique solution that
could reproduce the colour-luminosity relation.  With a similar
approach but different parameters, \citet{kawata2006} simulated a
steep metallicity gradient in a low-mass galaxy similar to the Fornax
dSph. Thefore, the  existence of a wide variety of gradients is not
incompatible with the models.

The large scatter on the relations presented so far, suggests that
there is another controlling factor, such as angular momentum.  As
discussed in Sect. \ref{subsec:systematics} and shown in
Table~\ref{table:mean}, the metallicity profiles of the S0 galaxies
are slightly flatter than those of the E galaxies, and, in
\citet{paperII} we suggested that the dEs with flat metallicity
profiles have a discy structure. Moreover, simulations of dwarf
galaxies flattened by rotation show that angular momentum prevents gas
from rapidly sinking to the galaxy centre, thus producing dwarfs
with flat metallicity profiles \citep{schroyen2011}.

We investigate the relation of the gradients with $V_{\rm
  max}/\sigma_0$ (Fig.~\ref{fig:fgradm}). This parameter quantifies
the balance between the ordered rotation and the random
motions. Obviously, there is no significant trend between age or
metallicity gradient and rotation. However, any possible relation
between the galaxies rotation and the gradients will be blurred by
projection effects.

\subsection{Gradients and central ages}

The gradients vs. central age diagrams are presented in
Fig.~\ref{fig:fgrada}, and Fig.~\ref{fig:fgradgrad} presents the
$\nabla_{[{\rm Fe/H}]}$ \emph{vs.} $\nabla_{{\rm Age}}$ diagram.

The dwarfs with a SSP-equivalent
 central ages younger than 4~Gyr show a statistically significant
anti-correlation (95\,per\,cent confidence level) between
$\nabla_{{\rm Age}}$ and ${\rm Age}_0$ (Fig.~\ref{fig:fgrada}). 
The same behavior is 
observed for the S0s, while for our `normal ellipticals' and the 
full sample of dwarfs we do not find such a trend.

Surprisingly, the statistical tests do not show any trend
 in the $\nabla_{[{\rm
      Fe/H}]}$ vs. ${\rm Age}_0$ plane. Consequently, the metallicity
gradient does not appear to be caused only by a young, metal-rich central
stellar population.  This is consistent with our findings discussed in
\citet{paperII}, where we presented a reconstruction of the
star-formation histories of a sample of dS0/dE galaxies as a function
of galactocentric radius. In each radial bin, the stellar population
was decomposed into different age bins using \ulyss. We found
metallicity gradients to be already present in the oldest age bin,
which contains stars that formed 8 to 13\,Gyr ago. Moreover, our
sample does not display a clear relation between $\nabla_{[{\rm
      Fe/H}]}$ and $\nabla_{{\rm Age}}$. Such a relation is observed in
the \citet{rawle2010} data, also represented in
Fig.~\ref{fig:fgradgrad}.  The lack of relation is consistent with the
lack of relation between the metallicity and the central age. However,
if we consider only the massive galaxies (Es and S0s) we can detect
a trend as in 
\citet{rawle2010}.

It should be noted that most of the TTDs have less negative,
i.e. milder, metallicity gradients than dEs/dS0s at the same
luminosity and that two of them have strong positive metallicity
gradients (see Fig.~\ref{fig:fgrada}). dEs/dS0s generally have
extended star-formation histories that were truncated more than a
gigayear ago  \citep{paperII}. TTDs could be surmised to be the latest
generation of such objects that presently on the verge of the
transformation. However, the fact that TTDs belong to the rare class
of early-type galaxies with an almost homogeneous metallicity may be
taken as a sign that they have a different nature.

\subsection{The ancestry of the quiescent dwarfs}

\subsubsection{Building and destroying gradients}

As discussed above, simulations of the formation of dwarf galaxies
show that their chemical evolution is determined by the interplay
between gravity, pulling gas inwards, and supernova feedback, heating
and expelling gas \citep{valcke08}. The outcome of this interplay is
that subsequent star-formation events become more centrally
concentrated, thus producing a positive age gradient and a negative
metallicity gradient.

On the other hand, in dwarf galaxies with high specific angular
momentum the centrifugal barrier prevents gas from rapidly sinking
towards the potential well centre \citep{schroyen2011}. This, together
with differential rotation, decrease star formation
efficiency. Moreover, star formation occurs galaxy-wide, as observed 
in dIrr \citep{vanzee2002}, and not just near the centre. It also is
more continuous and less ``bursty''. Simulations of rotating dwarfs
produce chemically homogeneous systems.

In \citet{paperII}, we presented age and metallicity gradients measured
in a sample of dE/dS0 galaxies. We artificially aged the
stellar populations of these galaxies by 3~Gyr to mimic passive
evolution and remeasured the gradients. We showed that after 3 Gyr of
passive evolution the metallicity gradients are milder while the age
gradients are almost completely removed. Hence, passive evolution
strongly affects the strength of the measured gradients and the
general trend is to weaken them. Moreover, environmental processes
that strip dwarfs from their gas, e.g. ram-pressure stripping and
tidal interactions with massive galaxies, may alter the gradients as
well. Any process that leads to a reshuffling of stellar orbits can be
expected to weaken any existing gradients, although totally erasing
them might be difficult. On the other hand, the same environmental
interactions which remove gas may trigger centrally concentrated star
formation by driving gas to the galaxy centre and\slash or compressing
it. This may reinvigorate the metallicity and age gradients.

All galaxies in our sample, being gasless early-types, must have been
affected by their environments (the smallest ones more so). However,
we still see strong gradients in most of them. 
Since we find metallicity gradients also in the old stellar
populations, they cannot (completely) be explained by recent star
formation triggered by ram pressure and\slash or
interactions. 

\subsubsection{Star forming galaxies and possible progenitors}

\citet{boselli2008a} simulated ram-pressure stripping in Virgo and
found that the photometric properties (colours, concentration, sizes,
luminosities) of the BCDs and late type spirals are consistent with
those of the cluster's early-type dwarfs after being stripped from
their gas and cessation of the star formation. They insist that even
the LMC is a bit too faint to produce dwarfs with $\mu_{e,B} \sim
22$\,mag\,arcsec$^2$, as in our sample. On the other side,
\citet{vanzee2002} predicates that the specific angular momentum of the
BCDs is lower than that of similar mass dwarf irregulars. Therefore,
the star-formation in BCDs is stronger and more centrally concentrated
(which plausibly leads to the emergence of gradients). She argues that
the star formation in dIrrs is randomly distributed along the disc and
that the dIrr have milder and more continuous star-formation histories
than BCDs.

This leads to BCDs as possible progenitors for the cluster dEs.
Assuming that the BCDs are in a burst phase only for 10\,Myr
\citep[e.g.][]{mashasse1999}, \citet{sanchez-almeida2009} argue that
they are quiet for 300\,Myr. In Virgo the dEs outnumber the BCDs by
a factor of 10 \citep{bst1988}, which leads to the conclusion that the
observed number of quiescent BCDs is enough to produce the observed
dEs. \citeauthor{sanchez-almeida2009} measured the metallicity of the
gas and stars of 21493 quiescent BCDs from SDSS and found that their
mean metallicities are around $-0.4$ to $-0.2$~dex, which is
consistent with the mean metallicities of early-type dwarfs. Still, as
a class BCDs have a higher specific angular momentum than do dEs
\citep{vanzee98}. This means that BCDs will not fade passively into
dEs after star formation has ended and that interactions with other
galaxies are required to shed their angular momentum. This is in line
with the observed morphology-density relation for dwarf galaxies.

Unfortunately, as mentioned before, radial studies of the stellar
population in dwarf star forming galaxies (BCDs and dIrrs) are, to our
knowledge, not done up-to-date and we cannot directly compare our
findings.

\section{Conclusions} \label{sec:conc}

We analysed long slit data of 40 early type galaxies using the full
spectrum fitting code \ulyss. We derived the radial distribution of
stellar population and kinematical characteristics. We classified the
galaxies according to their morphological appearance, photometric and
kinematical properties.  We ended up with 14 dE/dS0, 5 TTDs, 13 Es and
8 S0s.  We broadened the discussion by adding nearby dSph galaxies and
massive ellipticals for which similar measurements were available in
the literature.  

We can summarise our main conclusions as follows:
\begin{itemize}
\item All early-type galaxies obey a metallicity-luminosity relation
  in the sense that less massive galaxies contain less metals. This is
  in agreement with previous findings.
\item All early-type galaxies (above $L = 10^7 \times L_{\odot}$) obey an 
  age-luminosity relation in the
  sense that less massive galaxies have younger stellar populations,
  in a  SSP sense. This is the well-known downsizing effect. It
  reflects the fact that the least massive galaxies continue to form
  stars until present, while the most massive galaxies have much
  shorter star-formation histories.
\item For the massive galaxies in our sample we found that the S0s
  have in general milder population gradients than the Es. This 
  maybe due to the fact that angular momentum acts against building
  up gradients, or simply reflects the super-position between a disc and
  a bulge.
\item  We confirm previous studies where metallicity gradients were
  found to correlate with the age gradients, but only for giant galaxies
  (Es, S0s).  For the dwarfs in our sample, we find no evidence for a 
  connection between the metallicity gradient $\nabla_{{\rm Fe/H}}$ and the
  luminosity, central age or age gradient. At every luminosity, there
  is a significant scatter in $\nabla_{{\rm Fe/H}}$. 
\item Galaxies brighter than $M_{\rm B} \sim -17$~mag show no evidence
  for a connection between the age gradient $\nabla_{\rm Age}$ and any
  other parameter. At every luminosity, there is a significant scatter
  in $\nabla_{\rm Age}$.
\item Galaxies fainter than $M_{\rm B} \sim -17$~mag show evidence for
  a relation between the age gradient $\nabla_{\rm Age}$ and luminosity in
  the sense that fainter galaxies have stronger, positive age
  gradients. In other words, star formation persists over longer
  timescales in their inner regions than in their outskirts.
\item Dwarf galaxies with central SSP-equivalent ages younger than 3.5\,Gyr
  show trend of increasing $\nabla_{\rm Age}$ when the central age decreases.
  Consistent with the fact that the newly build age gradients
  disappear after few Gyr of passive evolution. Due to the low, residual
  star-formation rates, the few formed 
  young stars are fastly dissolved into the old population.
\item The majority of our TTDs have positive or only slightly negative
  metallicity gradients, suggesting a different star-formation
  efficiency than in the majority of the dEs.
\end{itemize}
We argue that low angular momentum BCDs may be the progenitors of
early-type dwarfs with gradients, while high angular momentum late
types, i.e. dIrrs or Sm, may be the progenitors of early-type dwarfs without
gradients. In BCDs the star formation is centrally concentrated, while
in discs with higher specific angular momentum it is randomly
distributed \citep{vanzee2002}. BCDs have photometric characteristics
(surface-brightness, effective radii, colours, S\'{e}rsic index)
similar to those of dEs after being ram-pressure stripped
\citep{boselli2008a}. They also have central stellar populations
\citep{sanchez-almeida2009} consistent with the ones of the dEs
\citep{chi2009, paperII, michielsen08}.

\section*{Acknowledgements}
MK has been supported by the Programa 
Nacional de Astronom\'{\i}a y Astrof\'{\i}sica of the Spanish Ministry 
of Science and Innovation under grant \emph{AYA2007-67752-C03-01}.
She thanks CRAL, Observatoire de Lyon, Universit\'{e} Claude
Bernard for an Invited Professorship.
PhP and WZ acknowledge a bilateral collaboration grant between France
and Austria (AMADEUS, collaboration project PHC19451XM).  SdR and PhP
acknowledge a bilateral collaboration grant between the Flander region
and France (Tournesol). We
thank A. Bedregal, C. Gallart, D. Forbes, J. Sanchez-Almeida, M. Monelli
and M. Fioc for useful discussions. 

Based on observations made with: ESO telescopes at Paranal 
observatory under programs 076.B-0196 and 70.A-0332(A); Gemini 
observatory under programs GS-2008A-Q-3 and GS-2006B-Q-74; 
Observatoire de Haute Provence (CNRS), 
France; archival data from WHT, La Palma, Observatorio del Roque de los 
Muchachos of the Instituto de Astrof\'{\i}sica de Canarias.

\bibliographystyle{mn2e}
\bibliography{gmos}   

\appendix

\section[Image analysis and photometry]{Image analysis and photometry} 
\label{appendix:photometry}

We analysed archival GMOS images of the 14 galaxies from
\citet{spo2009}, 6 in the Fornax cluster (FCC\,148, FCC\,153,
FCC\,170, FCC\,277, FCC\,301, FCC\,335) and 8 in the Virgo cluster
(VCC\,575, VCC\,828, VCC\,1025, VCC\,1146, VCC\,1178, VCC\,1297,
VCC\,1475, VCC\,1630). The data are acquisition images, taken before
the spectra. We applied a standard procedure for bias subtraction and
flat-fielding. The sky background was fitted with a low order
polynomial on the edge of the detector and subtracted.

For the photometric analysis, we used our own software.  The code
starts from the light peak and choose a number of isophotal levels
equidistant in mag~arcsecond$^{-2}$.  These isophotes are sampled
along directions separated by 5$^\circ$, and fitted with ellipses. The
errors are computed from the detector characteristics and background
noise.  Fitting the surface-brightness variation along a given
elliptical isophote with a fourth order Fourier series, four more
parameters can be derived that quantify how the isophotes deviate from
pure ellipses.  For instance, the fourth order cosine term quantifies
the boxiness/diskiness of the isophotes. Except for two-component
systems, such as the S0 galaxy FCC~153 which consists of a bulge and a
very extended thin disk, such a model gives an excellent reproduction
of the original image.  The code has been tested by comparing its
results with those derived from MIDAS and has proved its
reliability. The techniques we use have been employed before, e.g. in
\citet{d03} and \citet{derijcke2006}.

The total luminosity and effective radius were measured on the
reconstructed image, and the S\'ersic parameters were fitted on the
equivalent axis (i. e. geometric mean of the major and minor axes).
The error bars were estimated using bootstrapping.  Comparing the
total luminosities measured from these models with the available
HyperLeda values, we derived a mean B-band zero-point of
$m_0=27.252$~mag~arcsec$^{-2}$\slash ADU which was applied to all
data. With this value, we found very good agreement with the
literature values.  The parameters derived from this analysis can be
found in Table~\ref{table:char}.

\section[Fits]{Spectra fits of the 1\,arcsec{} extractions} \label{appendix:fits}

In this section, we present the fits of the core extractions for the 
40 galaxies in our sample, in order to
illustrate the quality of the fits and visualise the misfits.

In the centre of all the Virgo galaxies of our sample, except
in  VCC\,575 and 828, we observe a significant over-abundance of Mg,
and the Mg$_b$ feature,  near 5175 \AA, is masked by the automatic kappa-sigma clipping
({\sc /clean} option of the {\sc ulyss} command). As shown in
\citet{koleva08e}, this has not effect of the derived [{\rm Fe/H}].

For the galaxies for which we fitted separately two observations
(see Sect.\ref{subsec:spolaor}) we present only one fit, as the two
are almost identical. All the figures from this section are available electronically.

\section[Sample description]{Sample description} \label{appendix:indgal}
\subsection{NGC~1375 = FCC~148}
This flat S0 galaxy ($\epsilon \sim 0.56$) has discy isophotes, strong
rotation (${\rm V_{max}} \sim 80$\,\kms) and  ${\rm V_{max}}/\sigma > 1$. On the ACS 
images a clear peanut bulge is seen. A weak nebular emission is detected 
in the core. In the central 7\,arcsec (\aeff$\sim$17\,arcsec) the stellar 
population changes significantly. The metallicity decreases from 0.2 to 
-0.2\,dex and the age rises from 1\,Gyr to 2.5\,Gyr. The $\sigma$-drop 
observed on the GMOS profiles is an artifact due to the wide slit.

 The bulge clearly experienced a recent burst of star formation
(100 to 300\,Myr ago) fed by high metallicity gas.

\subsection{IC~335 = FCC~153}
This galaxy has a prominent edge-on disc, and the isophotes
cannot be fitted with ellipses (thus, the definition and value of
\aeff{} can be debated). The photometric profile reveals the distinct
bulge and disc components.  The maximum rotation, ${\rm V_{max}}$, reaches
130\,\kms, and ${\rm V_{max}}/\sigma > 1$.  In the central region, out to
1-2\,arcsec,  the GMOS profile shows an important $\sigma$-drop, which
is less marked on the FORS2 profile. Due to a lower S/N the FORS2 data
were binned in the central region, but to check the hypothesis of a
$\sigma$-drop we analysed the non-binned data: The 2\,arcsec inner
profile is consistent with a flat $\sigma$. So, we conclude that the
GMOS drop is an artifact due to the wide slit (the effective spectral
resolution is higher in the core).  From the centre to a radius of
2\,arcsec,  the SSP-equivalent age raises from 3 to 5\,Gyr and  the
metallicity drops from 0.25 to 0.05\,dex.  The flat  $\sigma$ profile may
hide a central cold component that may be a young stellar disc.

The metallicity raises between  2 and 12\,arcsec and decreases at larger distance.
The position of this upturn corresponds
to a minimum age and change of slope of the  velocity dispersion
profiles. At
the peak, the metallicity reach super-solar values ([{\rm Fe/H}] =
0.2\,dex).  The upturn corresponds also to the point where the
disc stars to dominate the light.  This complex profile likely
reflects the combination between  a bulge and a more metal-rich disc with a negative
$\nabla_{\rm [Fe/H]}$.  Except in the
centre, where young stars are present, the system has an intermediate
age around 5\,Gyr.

\subsection{NGC~1381 = FCC~170}
This is a S0 galaxy with a prominent edge-on disc
in fast rotation (${\rm V_{max}} \approx$170\,\kms).
The isophotes are not elliptical and the use of \aeff{}
as a spatial scale may be questioned.
The kinematical GMOS profiles displays a $\sigma$-drop
and a displacement of the kinematical centre in the central
2\,arcsec. As this is not seen in the FORS2 profiles,
this is an obvious artifact due to the wide slit.

The metallicity decreases from [{\rm Fe/H}]~$\sim$~0.05\,dex to -0.15\,dex
within 3.5\,arcsec. It remains constant
out to 2\,\aeff, and then decreases to -0.7\,dex at 60\,arcsec. The centre 
of FCC170 is older (12\,Gyr) than the outskirts (7.5\,Gyr).
This is probably consistent with the decomposition into an
old bulge and a younger disc.
 
\subsection{NGC~1428 = FCC~277}
The isophotes of this galaxy are boxy and the flattening is
$\epsilon \sim 0.5$. A foreground star, located 1\,arcsec in the NE
direction from the centre, was removed before
performing the photometry.  The galaxy rotates with ${\rm V_{max}}=$90\,\kms, and
${\rm V_{max}}/\sigma >$1. We classify it as E-S0 and assign it to the E group,
though \citep{merluzzi1998} classified is as S0.
This galaxy was not observed with FORS2, and the $\sigma$-drop
seen in the central 3\,arcsec may be an artifact due to
the wide slit used for this GMOS observing run
(such an effect is observed in the other galaxies for which
a comparison with narrow-slit observations is available).
We extracted the stellar population
parameters and kinematics out to 2\,\aeff. The metallicity drops from 0
to -0.7\,dex in 1\,\aeff, while at 2\,\aeff it is -1.3\,dex. The age raises
from 5 (centre) trough 10 (1\,\aeff) to 14\,Gyr at 2\,\aeff.

\subsection{FCC~301}
The isophotes of FCC301 are slightly discy.
We classify it as E-S0 and we assign it S0 group.
Its ellipticity ($\epsilon$) changes from 0.1 to 0.5 at
16\,arcsec and then again descends to 0.14 at 150\,arcsec. Out to
4\,arcsec the velocity rises to 50\,\kms, as a solid body and 
decreases outwards to a plateau of  20\,\kms{} reached at a
radius of 10~arcsec ($\sim$ 1\,\aeff).
The velocity dispersion is constant within the solid-body region
($\sigma = 40$\,\kms) and it increases by 10-20\,per\,cent outwards.
The stellar population is homogeneous in the solid body region
(5~Gyr and [{\rm Fe/H}]\,$\approx\,-0.2$\,dex), and the metallicity decreases
and the age increases outwards.

\subsection{FCC~335}
It is the only diffuse/dwarf elliptical galaxy in the sample of
\citeauthor{spo2010}
Its prominent nucleus is not resolved in our spectroscopic
observations (we measure FWHM $\sim$ 0.5\,arcsec). The 
central isophotes are irregular, which is caused by the
presence of dust, clearly seen on the composite image\footnote{
The ACS Fornax Cluster Survey,
\url{https://www.cfa.harvard.edu/~ajordan/ACSFCS/Gallery.html}}.

We could fit the spectra a little bit further than 1\,\aeff, where
the rotation velocity is ${\rm V}$\,=\,30\,\kms. 
The central velocity dispersion, $\sigma = 25$\,\kms, rises to
30\,\kms{} at 6~arcsec (on the FORS2 profile; the kinematics on the GMOS
profile are less precise due to the lower resolution).
Such $\sigma$-drops are frequent in the dE galaxies; we note
that it extends at a significantly larger scale than the nucleus.
As other nucleated dwarfs (see further), it is suspected to possess 
lower metallicity in its centre.

\subsection{NGC~4318 = VCC~575}
An elliptical (E3) galaxy with discy isophotes, a S\'{e}rsic 
index typical of low-luminosity galaxies (1.66) and a moderate 
flattening ($\epsilon_{max}$\,=\,0.33 at {\rm r}\,=\,65~arcsec). The rotation 
velocity increases to ${\rm V_{max}}$\,=\,85\,\kms, reached at 8\,arcsec 
($\sim$ 1\,\aeff). Then, it remains constant or decreases slightly. 
Over the same radial range where {\rm V} rises, the velocity dispersion 
decreases from $\sigma_0$\,=\,82\,\kms\ to $\sigma_{a_{eff}}$\,=\,55\,\kms, and 
remains constant outwards. The age is mildly raising  from 6 to 
8\,Gyr, while the metallicity decreases outwards from a Solar centre.


\subsection{NGC~4387 = VCC~828}
This E4 galaxy has boxy isophotes. It has dynamically colder centre.
The velocity dispersion rises from $\sigma_0$ = 82\,\kms\ to
$\sigma$ = 105\,\kms{} at 2\,arcsec, further out it decreases to 75\,\kms.
The rotation velocity reaches $V_{max} = 47$\,\kms{} at {\rm r}\,=\,15--20\,arcsec
(i. e. 1\,\aeff). The profiles derived from the GMOS and VAKU
observations are consistent.

While the age is roughly constant (10\,Gyr) along the body of the galaxy  
(out to the limit of the data at almost 2\,\aeff), the metallicity steeply 
decreases from super-solar to [{\rm Fe/H}]\,=\,-0.43\,dex at 1\,\aeff.

\subsection{NGC~4434 = VCC~1025}
This E0 ($\epsilon$\,=\,0.04) has perfectly elliptical isophotes. 
The rotation velocity reaches $V_{max} = 45$\,\kms{} at 10 -- 15\,arcsec
($\sim 1$\,\aeff) and remains constant afterward.
In the region with a velocity gradient, the velocity
dispersion decreases from  $\sigma_0$\,=\,135\,\kms\ to $\sigma$\,=\,70\,\kms.
The age is uniformly old at about 10\,Gyr, and the metallicity decreases
monotonically with the radius. The slope of the metallicity
profile changes at r\,=\,4\,arcsec,
being steeper in the inner part of the galaxy.

\subsection{NGC~4458 = VCC~1146}
This slowly rotating E0 galaxy ($\epsilon$\,=\,0.07) harbours a
kinematically decoupled core. In the inner 1.5\,arcsec, 
the rotation reaches  $V_{max} = 45$\,\kms{} at 0.7\,arcsec
and vanishes at about 4\,arcsec. Over this radial range, the 
velocity dispersion decreases from 130 to 85\,\kms, and it 
slowly continues to decline at larger radius. Its kinetically 
decoupled core (KDC) was first noted by \citet{hal2001}.

The stellar population parameters do not reflect the presence of a KDC.
The population is old and the metallicity monotonically decreases.
\citet{mor2004} study (with HST imaging and ground based spectroscopy) 
this objects because of the presence of a nuclear stellar disc. 
They found that the nuclear disc is inclined at about 83\,degree and
has $L = 4.8 \times 10^6 L_{\odot, V}$. They did not find any significant 
gradient in the stellar population, but their data are restricted to 
the central region (up to 1/3\,\reff). Their central metallicity is 
consistent with ours ([Fe/H]$_0 \sim -0.2$\,dex).

\subsection{NGC~4464 = VCC~1178}
It is an E2 galaxy ($\epsilon$\,=\,0.25) with a S\'{e}rsic index
{\rm n}\,=\,2.31 and discy isophotes. The central region rotates fast,
reaching {\rm V}$_{max}$\,=\,80\,\kms{} at 1.5\,arcsec, while in the 
outskirts the rotation slowly decreases to 40\,\kms. The 
central velocity dispersion is $\sigma_0 = 165$\,\kms{} and 
it slowly decreases to 70\,\kms. The age profile of the galaxy is
flat (around 10\,Gyr) and its metallicity slowly decreases from
-0.2 to -0.4\,dex at 1\,\aeff (7.5\,arcsec).

\subsection{NGC~4486B = VCC~1297}
This is the only compact elliptical galaxy (cE) of our sample.
This class of galaxies is characterised by a
high mean surface-brightness, we associate it to the E group.
It has discy isophotes, harbours a double nucleus separated by 0.15\,arcsec
\citep{lauer1996} and hosts a super-massive black hole \citep{kormendy1997}.
This black hole is at the origin of the steep rise of the velocity 
dispersion in the centre: Our central value reaches
$\sigma_0 = 230$\,\kms, while \citet{kormendy1997} measures 
$281\pm11$\,\kms{} under better seeing conditions.
The averaged velocity dispersion within an aperture of 2\,arcsec is
about 150\,\kms; with such a value, the galaxy would fall on the 
Fundamental Plane. While for the mean body of the galaxy
the rotation velocity is constant (${\rm V} = 30$\,\kms), it peaks to
${\rm V_{max}} = 60$\,\kms{} at {\rm r}\,=\,0.8\,arcsec
(\citealp{kormendy1997} measures $76\pm7$\,\kms{} at {\rm r}\,=\,0.6 arcsec).

The age has a mild negative gradient outwards, which is 
untypical for the rest of the E galaxies, but this 
characteristics is shared by at least one more cE \citep[NGC~5846cE][]{chilingarian2010}.  
The metallicity decreases from -0.15\,dex in the centre to 
-0.6\,dex at 4\,\aeff. This steep metallicity gradient was 
noticed first by \citet{davidge1991}. The centre has an 
important over-abundance in $\alpha$-elements.

\subsection{NGC~4515 = VCC~1475}
We classify this galaxy as E-S0 ($\epsilon$\,=\,0.2), based on  the images
and on the slightly discy inner parts. The core, within  one arcsec,
is counter-rotating by about 2\,\kms, with respect the the bulk of the
galaxy peaking at ${\rm V_{max}} = 16$\,\kms{} at {\rm r}\,=\,5.5~arcsec.  
The presence of the KDC is not reflected in the stellar
population.  The metallicity steeply decreases  from 
[{\rm Fe/H}]$_0$\,=\,-0.15\,dex to about -0.7\,dex at one \aeff.

\subsection{NGC4551 = VCC1630}
This is an E2 galaxy ($\epsilon$\,=\,0.26) with a $\sigma$-drop 
in the centre. The velocity dispersion increases from 
$\sigma_0 = 90$\,\kms{} to 100\,\kms{} at {\rm r}\,=\,2\,arcsec. 
The rotation velocity concomitantly reaches 30\,\kms{} to 
continue to raise slowly to 40\,\kms. The velocity dispersion
declines to 80\,\kms{} at 1\,\aeff.

The SSP-equivalent age of the core is about 2\,Gyr younger 
that the bulk, aged about 9\,Gyr. The metallicity decreases 
steeply from [{\rm Fe/H}]$_0$\,=\,0.17\,dex, to -0.15\,dex at 
5\,arcsec. It continues to decline slowly outwards.
\vskip 0.5cm
Since the following galaxies were already described in \citeauthor{paperI},
we will only recall some of their important characteristics.
\subsection{FCC~046}
This is a dS0(dE4) galaxy. It is nucleated and detected in H{\sc i} and H$_\alpha$.
This TTD (according to our classification) is suspected in a central
metallicity drop \citep{paperII}.

\subsection{FCC~136}
This is a dE2 galaxy. We detected central depression in the 
metallicity profile \citep[][table~10]{paperII}.

\subsection{FCC~204}
A dS0(6) galaxy, with spiral arms, visible after subtracting 
an axis-symmetric model.  The fit to the photometric profile shows 
positive C$_4$, an indication of the disciness. In the terminology 
of \citet{lisker2007} it will be classified as dE(di).

\subsection{FCC~207}
A dE2, nucleated galaxy suspected to have a central metallicity drop
\citep[fig.2 of][]{paperII}. H$_\alpha$, H$_\beta$, O{\sc iii}
emission is detected.

\subsection{FCC~245}
This is a nucleated galaxy (dE0). As for FCC~136 we detected a central
depression of the metallicity \citep[][table\,10]{paperII}. We find
a ring of intermediate population, which is not seen in the images.

\subsection{FCC~266}
This is a nucleated galaxy with a central depression in
the metallicity profile \citep[][table\,10]{paperII}.

\subsection{FCC~288}
This is a dS0(7) galaxy with spiral underlying structure and with a positive
C$_4$. As for FCC\,204, it wiould be called dE(di) in the terminology of
\citet{lisker2007}. We detected a central depression of the
metallicity profile \citep[][table\,10]{paperII}. It has a slight Mg$_b$
under-abundance.

\subsection{FS~76 = PGC~83852}
This almost round dwarf has a kinematically decoupled core. Its velocity 
dispersion, which initially 
rises with the radius, starts to decline sharply outside of one half-light 
radius. Its S\'{ersic} index of 2 indicates higher level of compactness than the
other dEs at similar luminosity (M$_B$\,=\,-17.1\,mag). This was interpreted as
evidence for a truncated dark halo and hence, for the occurrence of
tidal striping \citep{ddzh01}. Its metallicity profile
regularly decreases from 0.2 to -0.2\,dex in the central
3\,arcsec. The presence of a KDC is not reflected by any particularity
in the stellar population, thus the KDC should not be a recent event,
but rather a long lived structure, which supports the tidal origin
suggested in \citet{d04}.

\subsection{FS~131 = PGC~83905}
This is a dS0,N (dE5,N) dwarf. The central isophotes
have a peanut shape.

\subsection{FS~373 = PGC~83270}
This dE3 galaxy have a KDC. The ellipticity profile of the
isophotes suggests the presence of a small central
stellar disc \citep{d04}.

\subsection{NGC~5898 DW1 = PGC7~92714}
This dE3 has a positive C$_4$ \citep{d03}. Its flattening is stronger
in the inner 10\,arcsec ($\epsilon \sim$\,0.5) than in the outskirts
($\epsilon \sim$\,0.2). In the terminology of \citet{lisker2007} it
would be classified as a dE(di).

\subsection{NGC5898~DW2 = PGC792560}
DW2 is a nucleated dE6 with H{\sc i} and H$_\alpha$ emission.
The metallicity profile is depressed in the centre
\citep[][table10]{paperII}.
\vskip0.5cm
The next 4 SO galaxies are from the sample of \citeauthor{bedregal2006}

\subsection{FCC~055}
It is a faint S0 with M$_B\approx$-17.5\,mag and $\mu_B\approx$21.5\,mag\,arcsec$^{-2}$.
However, it more concentrated (S\'{e}rsic n\,=\,3) than a
typical dwarf galaxy. In the central part
the isophotes are pointed (as expected from a disc presence),
while in the outer parts they are elliptical. The ${\rm V_{max}} = 80$\,\kms\
is reached at 19\,arcsec. While the ages and metallicities are mostly
constant along the radius, there is a small depression at 2-3\,arcsec
associated with a rise in the velocity dispersion.

\subsection{NGC~1316 (Fornax~A) = FCC~021}
This peculiar galaxy, associated with the Fornax~A radio source, 
has been interpreted since a long time as a merger 
remnant \citep{schweizer1980}, produced from a 
collision with a disc galaxy which occurred one to 
three Gyr ago \citep{bosma1985,mackie1998}.
Despite these peculiarities, we associate it to the S0 group.

Our spectrum does not display significant nebular emission.
Our measured velocity dispersion, $\sigma_0$\,=\,239\,\kms, 
agrees with the values reported in HyperLeda.
It is however significantly higher than the value found by
\citet{nowak2008} using IFU observations with adaptive optic
at very high spatial resolution in the {\rm K} band.
These authors observe a central $\sigma$-drop, from
$218\pm8$\,\kms{} at {\rm r}\,=\,0.8\,arcsec to $226\pm9$\,\kms{} in an 8\,arcsec 
aperture. It is possible that the optical and NIR observations do not
trace the same kinematical population in this region affected by
important dust extinction.
The rotation velocity is monotonically rising to ${\rm V} = 160$\,\kms
at the limit of our data ({\rm r}\,=\,45\,arcsec).

The SSP-equivalent age and metallicity are almost homogeneous
at 4.7\,Gyr and [{\rm Fe/H}]\,=\,0.07. The residuals from the fits do
not indicate any strong $\alpha$-elements enhancement. Our results
differ from the ones of \citet{kuntschner2000}, who finds
age of 2\,Gyr and a very metal rich core with mild decrease of the
metallicity along the radius. 

\subsection{NGC~1380 = FCC~167}
This is a fast rotating S0 galaxy,
(${\rm V_{max}} \approx$\,210\,\kms), with a central velocity dispersion of
230\,\kms. The age is constant along the radius.  The metallicity
first decreases by 0.2\,dex (up to 1\,\aeff) and after increases.
We notice an Mg over-abundance in the centre.

\subsection{NGC~1380A = FCC~177}
This S0 shows the same metallicity behaviour
as NGC~1380. We find ${\rm V_{max}} \approx$\,90\,\kms\ and
$\sigma_0$\,=\,45\,\kms.  The age is slightly younger (by 1\,Gyr) in
the centre than in the outskirts. We do not detect central any strong Mg 
over-abundance.
\vskip 0.5cm
In the next paragraphs we describe the 4 remaining  galaxies from
\citet{vaku}, for which we do not have GMOS observations.
\subsection{NGC~4365 = VCC~0731}
The image of this E3 galaxy reveals boxy, slightly hexagonal isophotes.
This galaxy is known to have triaxial intrinsic shapes
and kinematically decoupled core \citep{wag1988}. Earlier stellar
population studies \citep[see][and references therein]{dav2001} have
shown a kinematical distinct inner region, which hosts the same
stellar population as the outskirts. The galaxy harbours  a
metallicity gradient and constant age along the radius. Our profiles
agree with the previous kinematical and stellar populations studies
\citep[e.g.][]{dav2001}. 


\subsection{NGC~4473 = VCC~1231}
This is an E5 galaxy with discy isophotes. \citet{pinkney2003},
argue that this galaxy underwent a recent merger,
because of its photometric peculiarities.
They also find
photometric and kinematic evidences for a cold stellar disc.
Our analysis reveals a younger and more metal rich core. The
metallicity decreases outwards, changing its slope at
5\,arcsec, where the maximum velocity ($\sim$\,60\,\kms)
is observed. The age is constant outside of this radius, but
drops to about 6\,Gyr inwards.

\subsection{NGC~4478 = VCC~1279}
This E2 galaxy has a nuclear stellar disc \citep{mor2004}. We find
that its central stellar population differs from the outskirts in
sense that it is younger (5\,Gyr old) and more metal rich.

\subsection{NGC~4621 = VCC~1903}
This elliptical galaxy is known to host a counter-rotating core
\citep{ben1990,wer2002}, however its size is only  60\,pc or 1\,arcsec
\citep{wer2002} and it is barely visible in our data. However, we
do observe an younger age and higher metallicity in the central pixel,
in agreement with the central, blue colour observed by \citet{kra2004}
using HST\slash WFPC2 data. It has an extended disc, seen from the isophote
analyses \citep[e.g.][]{ben1990}. Except for the central pixels, the
age is constant along the radius, however its metallicity decreases
steeply. 
\vskip 0.5cm
Last we will focus on the 2 nearby dwarf galaxies.

\subsection{NGC~205}
A satellite of M~31, NGC~205 is a prototype of massive dE,N.
It has a dynamical mass of 4.5\,$M_{\odot} / L_{\odot}$ 
\citep{derijcke2006} and a total magnitude M$_{{\it B}}$\,=\,-16.0.
It has a maximum rotation velocity of 11$\pm$5\,\kms{} \citep{sp02,geha2006}
and its dispersion increases with the radius
\citep{bender1991, sp02}. It contains dust in the central 2\,arcmin
\citep{haas98}, atomic and molecular gas \citep{welch98, young1997}. 
Beyond the inner 1\,arcmin radius it is dust and gas free and it 
has intermediate SSP-equivalent ages \citep{demers03}. There are
evidences for a recent star formation 
\citep{baade44a, peletier93,cappellari99, thesis}.

NGC~205 is suspected of close tidal interaction with M~31 
about 100~Myr ago \citep{cepa88}. We find
that it has strong age\slash metallicity positive\slash negative 
gradients. The central age is 1\,Gyr, while the central metallicity
is  -0.5\,dex. At 1\,\reff the age rises to
10\,Gyr, while the metallicity decreases to -1.3\,dex.

\subsection{NGC~404}

 A nearly face-on dS0 galaxy, placed 3.4\,Mpc \citep[e.g.][]{tonry2001}
from the Local Group and fairly isolated. It displays
ongoing star formation \citep{ho1993} and a
dust lane with complex structure within 5\,arcsec from the
centre \citep[e.g.][]{gallagher1986}. \citet{delrio2004} found a ring of
neutral hydrogen (H{\sc i}) ( $M_{\rm HI} = 1.5 \times 10^8~M_{\odot}$)
surrounding the stellar disc. NGC~404 has also a low-ionisation
nuclear emission line region \citep[LINER][]{pogge2000}.

\citet{bouchard2010} performed spectroscopic studies and
found a complex star-formation history (with more recent star-formation near the centre), nearly flat metallicity and age profiles 
within the inner 20\,arcsec and decreasing\slash increasing ages\slash 
metallicities outside. Here, we used the same observations 
as in \citet{bouchard2010}.
NGC~404 has a very peculiar kinematical structure with two 
inversions of the velocity gradient, and the innermost region 
(${\rm r} <$\,3\,arcsec  or 45\,pc) of the galaxy rotates in the same 
direction as the H{\sc i} \citep{delrio2004,bouchard2010}. 
This inner core is also the region where the youngest 
stars were found \citep{bouchard2010}.

\label{lastpage}
\end{document}


\title[Population gradients in early-types galaxies]{
Age and metallicity gradients in early-type galaxies: A dwarf to giant sequence}
\author[Koleva et
  al.]{Koleva et al.}
\pagerange{\pageref{firstpage}--\pageref{lastpage}} \pubyear{XXXX}

\maketitle 

\label{firstpage}

\appendix
\addtocounter{section}{1}
\section[APPENDIX B: Fit plots]{Fits of the 1\,arcsec{} extractions}

In this section, we present the fits of the core extractions for the 
40 galaxies in our sample, in order to
illustrate the quality of the fits and visualise the misfits.

The most characteristics misfitted region is the Mg$_b$ feature, near
5175 \AA, reflecting non-solar abundance of $\alpha$-elements.
In the centre of all the Virgo galaxies of our sample, except
in  VCC\,575 and 828, we observe a significant over-abundance of Mg,
and the Mg$_b$ feature is masked by the automatic kappa-sigma clipping
({\sc /clean} option of the {\sc ulyss} command). However, this does
not effect of the derived [{\rm Fe/H}].

For the galaxies for which we fitted separately two observations 
we present only one fit, as the two are identical.

All the figures will be available online.

\clearpage

\label{lastpage}